\documentclass[ aps, prc, amsmath,amssymb, nofootinbib, preprint ]{revtex4-2}

\usepackage{graphicx}
\usepackage{dcolumn}
\usepackage{bm}
\usepackage{color}

\begin{document}
\newcommand{\intvar}{\int \frac{d^3p}{(2\pi)^3E_p}}
\newcommand{\proj}{\Delta_{\nu_1...\nu_n}^{\mu_1...\mu_n}}
\newcommand{\non}{\nonumber\\}
\newcommand{\ampeq}{ &=& }
\newcommand{\nl}{\nonumber\\ && {}}
\newcommand{\nlq}{\nonumber\\ && {}\quad}
\newcommand{\nlqq}{\nonumber\\ && {}\qquad}
\newcommand{\nlqqq}{\nonumber\\ && {}\qquad\quad}
\newcommand{\nlqqqq}{\nonumber\\ && {}\qquad\qquad}
\newcommand{\ave}[1]{{\langle {#1} \rangle}}
\newcommand{\calE}{{\cal E}}

\title{
The evolution equation for the energy-momentum moments of 
the non-equilibrium density function \& 
The regularized relativistic third order hydrodynamics
}

\author{Dasen Ye\footnote{Currently at HEC Montr\'eal.}}
\author{Sangyong Jeon}%
\author{Charles Gale}
\affiliation{%
 Department of Physics, McGill University, Montreal, Quebec H3A 2T8, Canada
}%

\date{\today}
\begin{abstract}

In this work, we first
derive the evolution equation for the general energy-momentum moment of $\delta f$,
where $\delta f$ is the deviation from the local equilibrium
phase space density.
We then introduce a relativistic extension of regularized hydrodynamics 
developed in the non-relativistic case by Struchtrup and Torrilhon that judiciously mixes
the method of moments and Chapman-Enskog expansion.
Hydrodynamic equations up to the third-order in gradients are then systematially derived
within the context of a single species system and the relaxation time approximation.
This is followed by a series of linear stability and causality analysis.
For the massless particles without any charge conservation, 
the third-order hydrodynamics is shown to be linearly stable and causal.
\begin{description}
\item[Keywords]
relativistic viscous hydrodynamics, linear stability, linear causality, third-order
relativistic hydrodynamics, regularized hydrodynamics
\end{description}
\end{abstract}

\maketitle

\section{Introduction}\label{sec:intro}

The investigation of the hot and dense matter generated during ultra-relativistic
heavy-ion collisions, commonly referred to as quark-gluon plasma (QGP), constitutes
a prominent area of study within modern high-energy nuclear physics. One of
the most challenging aspects of this study is the difficulty to obtain
an analytic or numerical solution to a microscopic many-body QCD problem using
first-principles calculations. What is accessible is the coarse-grained collective
motion of the fluid-like system once approximate local thermal equilibrium is achieved
\cite{sy_heinz}. Accordingly, relativistic viscous hydrodynamics is an indispensible
theoretical tool for modeling the evolution of QGP in relativistic heavy ion collisions. 

The most intuitive and straightforward way of obtaining a relativistic viscous
hydrodynamics theory is to extend the non-relativistic Navier-Stokes theory to
a relativistic one \cite{Eckart:1940,Landau}. These theories are also commonly referred
to as the ``first-order theories'', which only include terms up to first order in
gradients. However, the Navier-Stokes theory is unstable and acausal when slightly
perturbed around thermal equilibrium in linear regime \cite{Hiscock:1985, Hiscock:1987,
Hiscock:1988, Denicol:2008}, and it has been shown that this instability is in fact
caused by the acausality of the theory \cite{Denicol:2008, Hiscock:1983, Shipu:2010}. 
For this reason, the original Navier-Stokes theory 
has been regarded as
not suitable for relativistic hydrodynamics.
However, recent work (usually referred to as the BDNK theory) 
\cite{Bemfica:2017wps, Kovtun:2019hdm, Bemfica:2019knx, Hoult:2020eho, Bemfica:2020zjp,
Hoult:2021gnb, Bemfica:2023res}
has shown that with some modification of the energy-momentum tensor,
the first order theory can be indeed made causal and stable.
(See also Refs.\cite{Das:2020fnr,Dore:2021xqq} for relationship between BDNK and the
second-order theories.)

The most well-known linearly stable and causal relativistic viscous hydrodynamics theory 
is the M\"uller-Isratel-Stewart (MIS) theory 
\cite{Muller:1967zza,IS:19761, IS:19762,IS:1979} 
that used the method of moments
generalizing Grad's work on non-relativistic hydrodynamics \cite{grad:1949}.
Unlike the first-order theories, the MIS theory contains terms
that are up to second-order in gradients, thus it is also commonly referred to as the
second-order theory. However, it has been shown that even the MIS theory
is not always linearly stable and causal. Their transport coefficients must satisfy a
set of constraints to be so \cite{Hiscock:1983, Shipu:2010, Olson:1990, Denicol:2008,
Brito:2020}. Furthermore, the second-order theory is in fact, not unique. 
The original MIS 
paper derived the second-order theory by considering 
entropy production. More recent approaches start with the Boltzmann equation
and derive hydrodynamic equations either using 
the Chapman-Enskog expansion \cite{Chapman,Jaiswal:2013npa,Dash:2023ppc},
or the method of moments \cite{Denicol:2010xn,Denicol:2012cn,Denicol:2012es}.
These approaches all give slightly different results depending on the truncation scheme.
One goal of this work is to provide a framework where truncation scheme is dictated by the
theory itself.

There have also been several recent works that derived the third-order hydrodynamics.
One of the main motivation to obtain the third-order hydrodynamics 
is the fact that the third-order terms may significantly improve the agreement with the kinetic
theory results when
the value of the specific shear viscosity $\eta/s$ is large 
\cite{Jaiswal:2013npa,3rd:2010,3rd:2015}.
In Refs.~\cite{3rd:2010,Younus:2019rrt} positive entropy prodcution rate argument was used 
to derive third-order hydrodynamic equations.
A Chapman-Enskog approach to the third-order hydrodynamics was advocated in
Refs.~\cite{Jaiswal:20132,Jaiswal:2014raa,3rd:2015}.
Naively, these approaches result in parabolic equations that may violate linear stability
and causality as shown in Ref.~\cite{brito2022} but causality may be restored by promoting
gradients of viscous tensor to an independent variable \cite{Panday:2024hqp}
following the prescription from Ref.~\cite{deBrito:2023tgb}.
In contrast, 
the methods of moments was used to derive the third-order equations in Refs.~\cite{brito2022,deBrito:2023tgb}
which were shown to be linearly stable and causal.
In this work, we will explore a method that 
combines a certain features of the method of moments and the Chapman-Enskog
expansion.
This will allow us to systematically derive 
relativistic viscous hydrodynamic equations up to the third order
starting from the evolution equations of the energy-momentum moments.

This is accomplished by
generalizing the non-relativistic 13-moment regularized hydrodynamics (R13) developed
by Struchtrup and Torrilhon \cite{Struchtrup:2003, Struchtrup:2004, Struchtrup:20042,
Struchtrup:2007}, to the relativistic regularized hydrodynamics. In
short, the regularization method combines both the method of moments and Chapman-Enskog
expansion by applying a Chapman-Enskog-like expansion to the energy-momentum moments
instead of the phase space density function. 
Using this method, we derive the third-order hydrodynamic equations 
followed by a linear stability and causality analysis for the massless case 
with a similar procedure outlined in Ref.~\cite{brito2022}.

This paper is organized as follows: in section \ref{sec:con_laws} we introduce
the conservation laws to mainly set the notations.
In section \ref{sec:gm}, we present the derivation of the evolution equations for general
energy-momentum moments of the phase space density.
The regularization method is also introduced in this section.
In section \ref{sec:Chapman_Enskog}, we obtain the Chapman-Enskog-like expansion of the
energy-momentum moments up to the 4-th momentum rank to prepare for the derivation of
the third-order hydrodynamics.
In section
\ref{sec:rrh} we will first briefly discuss the second-order equations obtained using
regularization. Then, we will
proceed to the derivation of the third-order theory before discussing the special
case of massless particles ($m=0$) in section \ref{sec:n1}.
Section \ref{sec:lin_stab_cau} contains
our linear analysis of the third-order hydrodynamics with $m = 0$.
We demonstrate the linear stability and causality of the theory. Finally,
we conclude this work in section \ref{sec:discussion}.
Appendices \ref{app:projectors}-\ref{app:Fintegrals}
contains mathematical and computational details on the projectors, irreducible
momentum polynomials, some derivative identities, details of the derivation of the general
moment equation, and the integrals with the equilibrium density function.

Throughout this paper, we will consider only one particle species. We use the natural
units $c = \hbar = k_B = 1$, and adopt the mostly-positive Minkowski metric $g_{\mu\nu}
= \hbox{diag}(-1, 1, 1, 1)$. To convert tensorial quantities to the mostly-negative metric,
each subscripted (covariant) index is to be
multiplied by $-1$ except the derivatives which work in the
opposite way. In particular,
for the Navier-Stokes tensor $\sigma_{\mu\nu}$ (which involves derivatives of the flow
velocity), this means that
$\sigma_{\mu\nu} \to -\sigma_{\mu\nu}$, $\sigma^{\mu\nu} \to -\sigma^{\mu\nu}$, but
$\sigma_{\mu}^{\nu}$ remains unchanged. The expansion rate 
defined as $\theta = \partial_\mu u^\mu$ (where $u^\mu$ is the local fluid velocity)
and the local time derivative defined as $D = u^\mu\partial_\mu$
also remain the same.

\section{Conservation Laws}\label{sec:con_laws}

The evolution equations of a hydrodynamics theory can be categorized into
two parts: the conservation laws and the moment equations. The conservation laws are
the continuity equations related to the energy-momentum conservation, and any other
charge conservations.
In this work, we will only consider a single species system that does not possess any
additional conserved charges (for instance, a real scalar $\lambda\phi^4$ theory) for the sake of
simplicity.
Hence, only the energy-momentum conservation is relevant:
\begin{equation}\label{eq:conserv}
\partial_\mu T^{\mu\nu} = 0
\end{equation}
where the energy-momentum tensor is further decomposed as
\begin{equation}
T^{\mu\nu} = \varepsilon u^\mu u^\nu+\left(P + \Pi\right) \Delta^{\mu \nu}+\pi^{\mu \nu}
\end{equation}
The fluid 4-velocity $u^\mu$ is defined by 
\begin{equation}
    T^{\mu\nu}u_\nu = -\varepsilon u^\mu
\end{equation}
where $\varepsilon$ is the local energy density and
the fluid 4-velocity $u^\mu$ is normalized to 
$u_\mu u^\mu = -1$.
The thermal pressure at local equilibrium 
is subject to the equation of state, $P = P(\varepsilon)$, and
$\Pi$ is the bulk pressure.
The local 3-metric,
$\Delta^{\mu\nu} = g^{\mu\nu} + u^\mu u^\nu$,
is the projector that extracts the components of any 4-vector that is transverse 
to $u^\mu$. 
The transverse, symmetric, and traceless rank-2 tensor 
$\pi^{\mu\nu}$ is the shear-stress tensor.

It is convenient to decompose Eq.(\ref{eq:conserv}) into the time-like 
and the space-like
components with respect to the fluid 4-velocity $u^\mu$.
Applying $u_\nu$ to $\partial_\mu T^{\mu\nu} = 0$ yields
%
%
the time-like component
\begin{equation}\label{eq:energy_conserv}
    D{\varepsilon} + (\varepsilon+P+\Pi)\theta +
    \pi^{\alpha\beta}\sigma_{\alpha\beta} = 0
\end{equation}
Applying $\Delta^\lambda_\nu$ to $\partial_\mu T^{\mu\nu} = 0$ yields
the space-like components
\begin{equation}\label{eq:momentum_conserv}
    (\varepsilon+P+\Pi) Du^\lambda 
    + \nabla^\lambda (P+\Pi) 
 + \Delta_{\nu}^{\lambda}\partial_\mu \pi^{\mu\nu}  = 0
\end{equation}
where we defined the relativistic substantial derivative
%
(local time derivative)
$D = u^\mu \partial_\mu$, 
the local spatial derivative $\nabla^\mu = \Delta^{\mu\nu} \partial_\nu$, 
the expansion rate $\theta = \partial_\mu u^\mu = \nabla_\mu u^\mu$,
the Navier-Stokes tensor
$\sigma^{\mu\nu} = \nabla^{\langle\mu}u^{\nu\rangle}$,
and the fluid acceleration $Du^\lambda = a^\lambda$.
The angular bracket around a set of indices
represents the transverse (with respect to $u^\mu$), symmetric, and traceless
combination of the indices. In practice, this can be obtained by applying the
projector:
\begin{equation}
    A^{\langle\mu_1...\mu_n\rangle} =
    \Delta_{\nu_1...\nu_n}^{\mu_1...\mu_n}A^{\nu_1...\nu_n}
\end{equation}
where $A^{\mu_1...\mu_n}$ is an arbitrary rank-$n$ tensor.
Some useful facts about the projectors such as the explicit form for $n=2,3$, and recursive
relationships can be found in Appendix \ref{app:projectors}.

Eqs.(\ref{eq:energy_conserv}) and
(\ref{eq:momentum_conserv}) enforce the energy conservation and momentum conservation,
respectively. Together, they constitute the evolution equations for $\varepsilon$ and $u^\mu$.
However, at this point, the evolution equations
for $\Pi$ and $\pi^{\mu\nu}$ are not yet developed. 
In the following sections, we will do so in the context of a single-species kinetic
theory.

\section{General Methods}\label{sec:gm}
\subsection{Energy-Momentum Moments}
To obtain the evolution equations for the bulk pressure $\Pi$ and the shear tensor
$\pi^{\mu\nu}$,
one can start with the kinetic theory equation
\begin{equation}
p^\mu\partial_\mu f = C[f]
\label{eq:kinetic_theory}
\end{equation}
where $f(x,p)$ is the phase space density, and $C[f]$ is the collision integral.
As stated, we will consider a system with a single particle species.
This is also consistent with having no other conserved
quantities.
The energy-momentum tensor is defined as
\begin{equation}
T^{\mu\nu}
=
\int {d^3p\over (2\pi)^3 E_p} p^\mu p^\nu f
\end{equation}
with $E_p = p^0 = \sqrt{{\bf p}^2 + m^2}$.
This tensor satisfies the continuity equations $\partial_\mu T^{\mu\nu} = 0$ as long as
the collisions conserve energy and momentum.

By further decomposing the phase space density as the local equilibrium part and the
correction
\begin{equation}
f(x,p) = f_{0}(x,p) + \delta f(x,p)
\label{eq:f0_deltaf}
\end{equation}
where $f_0(x,p)$ is the local equilibrium density,
we can further define the ideal fluid part of the energy-momentum tensor
\begin{equation}
T_0^{\mu\nu}
=\int {d^3p\over (2\pi)^3 E_p} p^\mu p^\nu f_0
= \varepsilon u^\mu u^\nu + P\Delta^{\mu\nu}
\end{equation}
and the dissipative part 
\begin{equation}
\delta T^{\mu\nu}
=\int {d^3p\over (2\pi)^3 E_p} p^\mu p^\nu \delta f
= \Pi \Delta^{\mu\nu} + \pi^{\mu\nu}
\end{equation}
The local energy density $\varepsilon$ and the flow velocity $u^\mu$
are defined by the Landau matching
condition
\begin{eqnarray}
T^{\mu\nu}u_\nu &=& T^{\mu\nu}_0 u_\nu = -\varepsilon u^\mu
\label{eq:landau_matching}
\end{eqnarray}

As one can see, various components of $T^{\mu\nu}$ are obtained as the 
energy-momentum moments of $f_0$ and $\delta f$. 
Accordingly, their evolution equations can be obtained from the kinetic theory 
equation Eq.(\ref{eq:kinetic_theory}).
To obtain the evolution equations for $\Pi$ and $\pi^{\mu\nu}$, 
it is convenient to define the energy-weighted rank-$n$ tensor moment
of $\delta f$ as
\begin{equation}
    \rho_r^{\mu_1...\mu_n} = \intvar \delta f \mathcal{E}_p^r 
    p^{\langle \mu_1}p^{\mu_2}...p^{\mu_n\rangle}
\label{eq:rho_r_n}
\end{equation}
where $\mathcal{E}_p = -u_\mu p^\mu$ is the energy of a particle
in the rest frame of a fluid cell,
and $p^{\langle\mu_1}p^{\mu_2}\cdots p^{\mu_n\rangle}
= \proj p^{\nu_1}p^{\nu_2}\cdots p^{\nu_n}$
is the symmetric and traceless combination of $p^{\langle\mu\rangle} =
\Delta^{\mu}_{\nu}p^\mu$.
Here, the integer $n$ is the rank of the tensor, and
$\mathcal{E}_p^r$ is the energy weight in which the integer exponent $r$ indicates the energy
order.
In the fluid-cell rest frame,
the local equilibrium density function $f_0$ takes
the form of  $f_0 = {1\over e^{\beta E_p} - \zeta}$, in which $\beta = 1/T$ is the
inverse temperature, and $\zeta$ could be 1 (Bose-Einstein statistics), 0 (Boltzmann
statistics), or $-1$ (Fermi-Dirac statistics).

Using the decomposition $p^\mu = {\cal E}_p u^\mu + p^{\langle\mu\rangle}$,
the Landau matching condition, Eq.(\ref{eq:landau_matching}),
becomes the following two conditions on the moments
\begin{equation}
\rho_2 = \rho_1^\mu = 0
\end{equation}
In terms of the energy-momentum moments, the bulk pressure is given by
\begin{eqnarray}
\Pi = -{m^2\over 3}\rho_0
\end{eqnarray}
and the shear tensor is given by
\begin{eqnarray}
\pi^{\mu\nu} = \rho_0^{\mu\nu}
\end{eqnarray}

\subsection{Derivation of the General Moment Equation}

The evolution equation for $\rho_r^{\mu_1\cdots\mu_n}$ can be obtained by
applying 
the local time derivative
$D = u^\mu\partial_\mu$ to $\rho_r^{\mu_1\cdots\mu_n}$ and then using 
the kinetic equation Eq.(\ref{eq:kinetic_theory}) 
with $f = f_0 + \delta f$.
In this section, we outline the derivation of the evolution equation for the general
energy-momentum moment $\rho_r^{\mu_1\cdots\mu_n}$.
Full derivation can be found in Appendix \ref{app:derivation}.

Applying 
%
the local time derivative
to $\rho_r^{\mu_1\cdots\mu_n}$ in Eq.(\ref{eq:rho_r_n}),
and then projecting onto the transverse
space, we get
\begin{eqnarray}
     \proj D \rho_r^{\nu_1...\nu_n}
     &=& 
     \proj \intvar (D\delta f){\cal E}_p^r p^{\langle \nu_1}p^{\nu_2}...p^{\nu_n\rangle}
     \nonumber\\ && {}\quad
     - n\intvar \delta f {\cal E}_p^{r+1} p^{\langle \mu_1}p^{\mu_2}...a^{\mu_n\rangle}
     \nonumber\\ && {}\quad 
     - r\proj a_\sigma \intvar \delta f {\cal E}_p^{r-1} p^{\langle \sigma
     \rangle}p^{\langle \nu_1}p^{\nu_2}...p^{\nu_n\rangle}
\end{eqnarray}
where we defined the fluid acceleration $a^\mu = Du^\mu$, and used the fact that
$u_\mu D u^\mu = 0$ so that
$D{\cal E}_p = -a_\sigma p^\sigma = -a_\sigma p^{\langle\sigma\rangle}$, and also 
\begin{equation}
\proj D p^{\langle\nu_1\cdots} p^{\nu_n\rangle}
= -n \calE_p p^{\langle\mu_1\cdots}p^{\mu_{n-1}}a^{\mu_n\rangle}
\end{equation}
which is derived in Appendix \ref{app:Dpn}.
Using the identity
\begin{equation}
p^\ave{\lambda} p^{\langle\mu_1}\cdots p^{\mu_n\rangle}
=
p^{\langle\lambda}p^{\mu_1}\cdots p^{\mu_n\rangle}
+
{n\over 2n+1}(\calE_p^2 - m^2)p^{\langle\mu_1}p^{\mu_2}\cdots p^{\mu_{n-1}}\Delta^{\mu_n\rangle\lambda}
\end{equation}
proven in Appendix \ref{app:polynomials},
we can expand the last term on the right hand side to get
\begin{equation}\label{eq:step2}
\begin{split}
     \proj D \rho_r^{\nu_1...\nu_n} &= \proj \intvar (D\delta f){\cal E}_p^r p^{\langle
     \nu_1}p^{\nu_2}...p^{\nu_n\rangle}\\
     &\quad - n\intvar \delta f {\cal E}_p^{r+1} p^{\langle \mu_1}p^{\mu_2}...a^{\mu_n\rangle}\\
     &\quad - ra_\sigma \intvar \delta f {\cal E}_p^{r-1} p^{\langle
     \sigma}p^{\mu_1}p^{\mu_2}...p^{\mu_n\rangle}\\
     &\quad - r\frac{n}{2n+1}a_\sigma \intvar \delta f {\cal E}_p^{r-1} ({\cal E}_p^2 - m^2)p^{\langle
     \mu_1}p^{\mu_2}...\Delta^{\mu_n\rangle \sigma}\\
\end{split}
\end{equation}
For $D\delta f$, we can use the following form of the
Boltzmann equation
\begin{equation}\label{eq:boltz_eq}
\begin{split}
    p^\mu \partial_\mu f_0 + {\cal E}_pD\delta f + p^{\langle\mu\rangle}\nabla_\mu \delta f
    = C[f]
\end{split}
\end{equation}
where $C[f]$ is the collision term of the relativistic Boltzmann equation, and we used
\begin{equation}
p^\mu\partial_\mu = {\cal E}_p D + p^{\langle\mu\rangle}\nabla_\mu
\label{eq:pmu_partialmu}
\end{equation}
This gives
\begin{equation}
\begin{split}
     \proj D \rho_r^{\nu_1...\nu_n} &= - n\intvar \delta f {\cal E}_p^{r+1} p^{\langle
     \mu_1}p^{\mu_2}...a^{\mu_n\rangle}\\
     &\quad - r a_\sigma \intvar \delta f {\cal E}_p^{r-1} p^{\langle
     \sigma}p^{\mu_1}p^{\mu_2}...p^{\mu_n\rangle}\\
     &\quad - r\frac{n}{2n+1}\intvar \delta f {\cal E}_p^{r-1} ({\cal E}_p^2 -
     m^2)p^{\langle\mu_1}p^{\mu_2}...a^{\mu_n\rangle}\\
     &\quad + \proj \intvar C[f]{\cal E}_p^{r-1}p^{\langle \nu_1}p^{\nu_2}...p^{\nu_n\rangle}\\
     &\quad - \proj \intvar (\partial_\lambda f_0){\cal E}_p^{r-1}p^{\lambda}p^{\langle
     \nu_1}p^{\nu_2}...p^{\nu_n\rangle}\\
     &\quad - \proj \intvar (\nabla_\lambda \delta f){\cal E}_p^{r-1}p^{\langle \lambda
     \rangle}p^{\langle \nu_1}p^{\nu_2}...p^{\nu_n\rangle}\\
\end{split}
\label{eq:Drho_1}
\end{equation}
The first three lines of Eq.(\ref{eq:Drho_1})
can be expressed in terms of the energy-momentum moments.
The term with the collision integral is in general a non-linear functional of $\delta f$
that will not admit a simple expression. In the rest of this work, we will use the relaxation time
approximation so that this term {\em can} be expressed in terms of the energy-momentum
moments.
The line involving the equilibrium density 
$f_0$ will not result in the energy-momentum moments. Instead, it gives the constitutive
relationships. 
The rest of the derivation is then to deal with the last line. Details of transferring
$\nabla_\lambda$ from $\delta f$ to the other factors
can be found in Appendix \ref{app:derivation}.
The final result is
\begin{equation}\label{eq:EOM_final}
\begin{split}
    \proj D \rho_r^{\nu_1...\nu_n} &= \intvar C[f]{\cal E}_p^{r-1}p^{\langle
    \mu_1}p^{\mu_2}...p^{\mu_n\rangle}\\
    &\quad - \intvar (\partial_\lambda f_0){\cal E}_p^{r-1}p^{\lambda}p^{\langle
    \mu_1}p^{\mu_2}...p^{\mu_n\rangle}\\
    &\quad -\frac{n(2n+r+1)}{2n+1}\rho_{r+1}^{\langle\mu_1...\mu_{n-1}}a^{\mu_n\rangle}\\
    &\quad + rm^2\frac{n}{2n+1}\rho_{r-1}^{\langle\mu_1...\mu_{n-1}}a^{\mu_n\rangle}\\
    &\quad -ra_\lambda \rho_{r-1}^{\lambda\mu_1...\mu_n}\\
    &\quad - \Delta_{\nu_1...\nu_n}^{\mu_1...\mu_n} \nabla_\lambda
    \rho_{r-1}^{\lambda\nu_1...\nu_n}\\
    &\quad - \frac{n}{2n+1}\nabla^{\langle\mu_1}\rho_{r+1}^{\mu_2...\mu_n\rangle}\\
    &\quad + m^2\frac{n}{2n+1}\nabla^{\langle\mu_1}\rho_{r-1}^{\mu_2...\mu_n\rangle}\\
    &\quad - \frac{n+r+2}{3}\theta\rho_r^{\mu_1...\mu_n}\\
    &\quad - (r-1)\sigma_{\lambda\alpha}\rho_{r-2}^{\alpha\lambda\mu_1...\mu_n}\\
    &\quad + \frac{(r-1)m^2}{3}\theta\rho_{r-2}^{\mu_1...\mu_n}\\
    &\quad
    -\frac{n(2n+2r+1)}{2n+3}\rho_r^{\lambda\langle\mu_1...\mu_{n-1}}\sigma_\lambda^{\mu_n\rangle}\\
    &\quad -n\rho_r^{\lambda\langle\mu_1...\mu_{n-1}}\omega_\lambda^{\mu_n\rangle}\\
    &\quad -
    \frac{(2n+r)(n-1)n}{(2n-1)(2n+1)}\rho_{r+2}^{\langle\mu_1...\mu_{n-2}}\sigma^{\mu_{n-1}\mu_n\rangle}\\
    &\quad +
    2m^2\frac{(r-1)n}{2n+3}\rho_{r-2}^{\lambda\langle\mu_1...\mu_{n-1}}\sigma_\lambda^{\mu_n\rangle}\\
    &\quad -
    m^4\frac{(r-1)(n-1)n}{(2n+1)(2n-1)}\rho_{r-2}^{\langle\mu_1...\mu_{n-2}}\sigma^{\mu_{n-1}\mu_n\rangle}\\
    &\quad +
    m^2\frac{(2n+2r-1)(n-1)n}{(2n+1)(2n-1)}\rho_r^{\langle\mu_1...\mu_{n-2}}\sigma^{\mu_{n-1}\mu_n\rangle}
\end{split}
\end{equation}
Here, $\omega^{\mu \nu}=\frac{1}{2}\left(\nabla^\mu u^\nu-\nabla^\nu u^\mu\right)$
is the anti-symmetric vorticity tensor. 
For $n=0, 1, 2, 3, 4$, Eq.(\ref{eq:EOM_final}) agrees with the results obtained by Denicol
and others \cite{Denicol:2012es,deBrito:2023tgb} as they should.
This general evolution equation was first derived by one of the authors in
Ref.\cite{Ye:2023wem}.
As far as we know, this was the first time the evolution
equation for a general energy-momentum moment was explicitly derived in literature.
This equation also appeared in a recent paper \cite{deBrito:2024vhm}.
Even though we will eventually use Boltzmann statistics, Eq.(\ref{eq:EOM_final}) is
valid for quantum statistics as well.

\subsection{Regularization Methods}

As one can see in Eq.(\ref{eq:EOM_final}) the time evolution of
$\rho^{\mu_1\cdots\mu_n}_r$ involves
$\rho^{\mu_1\cdots\mu_n}_r$, 
$\rho^{\mu_1\cdots\mu_n}_{r-2}$, 
$\rho^{\mu_1\cdots\mu_{n-1}}_{r\pm 1}$, 
$\rho^{\mu_1\cdots\mu_{n-2}}_{r\pm 2}$, 
$\rho^{\mu_1\cdots\mu_{n-2}}_{r}$, 
$\rho^{\mu_1\cdots\mu_{n+1}}_{r-1}$, 
and
$\rho^{\mu_1\cdots\mu_{n+2}}_{r-2}$. 
As such, Eq.(\ref{eq:EOM_final}) represents an infinite set of coupled 
%
%
partial differential equations.
To get a closed set of equations for a finite number of moments, one must use 
a truncation scheme. 
The two well-known truncation schemes are the method of moments
\cite{grad:1949,IS:19761, IS:19762, IS:1979}, and the Chapman-Enskog method \cite{Chapman}. 
In the method of moments, one assumes that
$\delta f$ is such that all $n$-th rank moments are proportional to each other regardless
of their energy weights \cite{deBrito:2023tgb}.
On the other hand, the Chapman-Enskog method
expands $\delta f$ 
using the Boltzmann equation as the recursion equation to
obtain $\delta f$ as a derivative expansion.

In a series of papers \cite{Struchtrup:2003, Struchtrup:2004, Struchtrup:20042,
Struchtrup:2007}, Struchtrup and Torrilhon developed a novel method they named the
``Regularized Hydrodynamics'' that combines
both the method of moments and the Chapman-Enskog expansion. This technique 
applys a Chapman-Enskog-like expansion directly to the 
%
%
energy-momentum moments instead of $\delta f$, excluding
the moments that serve as the dynamic hydrodynamic variables. 
This technique provides a more systematic way
to produce a set of equations to any given order in the expansion parameter $\epsilon$
without introducing any additional assumptions.

In the usual Chapman-Enskog method, the collision term is scaled as $C[f] \to (1/\epsilon)C[f]$
and the non-equilibrium part of the phase space density is expanded as
\begin{equation}
    \delta f = \sum_{n=1}^\infty \epsilon^n \delta f_{|n}
\end{equation}
Here and here after, the vertical bar in the subscript indicates the relevant 
$\epsilon$-order.
These are then plugged into the Boltzmann equation. Collecting terms having the same
power of $\epsilon$, the $n$-th order piece $\delta f_{|n}$ can be found iteratively
involving a maximum of $n$ spatial derivatives of $\beta$ and $u^\mu$.
The resulting equations are at best parabolic, and hence potentially acausal.
This can lead to instability unless additional evolution equations 
for $\Pi$, $\pi^{\mu\nu}$ and other dissipative currents are postulated using the constitutive
relationships 
%
%
\cite{Hiscock:1985, Hiscock:1987, Hiscock:1988, Denicol:2008, Hiscock:1983, Shipu:2010,Panday:2024hqp}.

In the method of Struchtrup and Torrilhon, instead of $\delta f$, the energy-momentum moments
of $\delta f$ are expanded in powers of $\epsilon$
\begin{equation}\label{eq:rhorneps}
    \rho_r^{\mu_1\cdots\mu_n}
=
\sum_{n=1}^\infty \epsilon^n \rho_{r|n}^{\mu_1\cdots\mu_n}
\end{equation}
Working out the order-by-order solution by putting Eq.(\ref{eq:rhorneps}) in
Eq.(\ref{eq:EOM_final})
would be completely equivalent to the usual Chapman-Enskog method.
What we would like to do differently, however, is {\em not} to expand the
hydrodynamic variables, such as $\Pi$ and $\pi^{\mu\nu}$, 
whenever they occur while expanding all other moments in terms of them.
However, at higher orders of $\epsilon$, there is no guarantee that $\Pi$ and
$\pi^{\mu\nu}$ (which are $O(\epsilon)$) are the only relevant dynamic variables. As we will see below,
we may need to promote some higher moments to be dynamic to get a closed set of equations.

\section{Chapman-Enskog expansion of the moments}
\label{sec:Chapman_Enskog}

In this section, we work out the $\epsilon$-expansion of the energy-momentum moments up to
$n = 4$ within the relaxation time approximation. The results from this section will be
used in the later sections to bulid hydrodynamic equations.

To determine the $\epsilon$-order of each $\rho^{\mu_1\cdots\mu_n}_r$ explicitly, 
we consider the relaxation time approximation for the collision term
\begin{equation}\label{eq:AW}
    C[f] = -{{\cal E}_p\over \epsilon\tau_R} \delta f(x,p)
\end{equation}
where we have explicitly indicated the expansion parameter $\epsilon$. The relaxation
time $\tau_R$ is assumed to be a constant.
The parameter $\epsilon$
is set to $1$ at the end of calculations.
Putting Eqs.(\ref{eq:rhorneps}) and (\ref{eq:AW}) into the general moment
equation Eq.(\ref{eq:EOM_final}) and collecting the $O(\epsilon^0)$ terms, we get
the first order coefficient function
\begin{equation}\label{eq:O1}
    \rho_{r|1}^{\mu_1\cdots\mu_n} = -\tau_R F_{r-1|0}^{\mu_1\cdots\mu_n}
\end{equation}
where we defined the equilibrium density term to be
\begin{equation}\label{F}
\begin{split}
    F_{r}^{\mu_1\cdots\mu_n} & = \intvar \mathcal{E}_p^r p^{\langle\mu_1}\cdots
    p^{\mu_{n-1}}p^{\mu_n\rangle}p^\lambda\partial_\lambda f_0
\end{split}
\end{equation}
Here, $\rho_{r|1}^{\mu_1\cdots\mu_n}$ is the $O(\epsilon)$ part of
$\rho_{r}^{\mu_1\cdots\mu_n}$ and $F_{r-1|0}^{\mu_1\cdots\mu_n}$ is the $O(\epsilon^0)$ part
of $F_{r-1}^{\mu_1\cdots\mu_n}$. 
Using Eq.(\ref{eq:pmu_partialmu}), one can show that
$p^\lambda\partial_\lambda f_0 = -f_0(1+\zeta f_0) p^\lambda \partial_\lambda(\mathcal{E}_p\beta)$ can contain only $1,
p^{\langle\mu_1\rangle}, p^{\langle\mu_1}p^{\mu_2\rangle}$. Hence the orthogonality
%
%
of the irreducible polynomials
$p^{\langle\mu_1}\cdots p^{\mu_n\rangle}$ 
({\it c.f.}~Eq.(\ref{eq:p_poly_ortho}) in Appendix \ref{app:polynomials}
and also Ref.~\cite{Denicol:2012cn}) demands that
\begin{equation}\label{eq:F_zero}
    F^{\mu_1\cdots\mu_n}_{r} = 0 \ \ \ \hbox{for $n \ge 3$}
\end{equation}
For $n = 0, 1, 2$, we get
\begin{equation}\label{eq:FrS}
    F_r = \phi_{r|0}\theta 
    + \phi_{r|1}^{\pi\Pi}(\pi^{\gamma\rho}\sigma_{\gamma\rho} + \theta\Pi)
\end{equation}
\begin{equation}\label{eq:FrV}
    F_r^\mu = 
    \psi_{r|1} \left(\Delta^{\mu}_{\gamma}\partial_{\rho}\pi^{\rho\gamma} 
    + \nabla^\mu  \Pi 
    + a^\mu \Pi\right)
\end{equation}
\begin{equation}\label{eq:FrR2}
    F_{r}^{\mu\nu} = \varphi_{r|0}\sigma^{\mu\nu}
\end{equation}
where the coefficient functions $\phi,\psi$ and $\varphi$ are functions of $\beta$ only. 
Derivations can be found in Appendix~\ref{app:Fintegrals}.
Observe that $F_r$, $F_r^\mu$ and $F_r^{\mu\nu}$ all involve gradients
and time derivatives of the hydrodynamic variables. Consequently, they can be 
described as 
%
%
physical thermodynamic
forces that are driving the evolution of the system. In deriving the
above expressions, we have used Eq.(\ref{eq:energy_conserv}) to express $D\beta$
in terms of spatial derivatives. The acceleration $a^\mu = Du^\mu$ can also be expressed 
in terms of spatial derivatives using Eq.(\ref{eq:momentum_conserv}) but we leave it as it
is for brevity.
Details can be found in Appendix~\ref{app:Fintegrals}.

From Eqs.(\ref{eq:O1}) and (\ref{eq:F_zero}),
it follows immediately that
$\rho_{r|1}^{\mu_1\cdots\mu_n} = 0 $ for $n \ge 3$.
One should also note that $\rho_{r|1}^\mu = 0$ because there is no number (mass)
conservation.
Hence
\begin{equation}\label{eq:rho_order_1}
    \rho_{r}, \rho_{r}^{\mu_1\mu_2} = O(\epsilon)
\end{equation}
\begin{equation}\label{eq:rho_order}
    \rho_{r}^{\mu_1\cdots\mu_n} = O(\epsilon^2) \ \ \ \hbox{for $n=1$ and $n \ge 3$}
\end{equation}
In fact, only $n = 1, 3, 4$ moments are $O(\epsilon^2)$. To see this,
note that in
Eq.(\ref{eq:EOM_final}), the lowest momentum order on the right-hand side is
$n-2$. Hence, for $n = 5,6$, the lowest momentum order appearing on the right-hand side
is $n=3$ and $n=4$ respectively. This implies that the right-hand sides for $n=5,6$ are
at most $O(\epsilon^2)$, which further implies that
$\rho_{r|2}^{\mu_1\cdots\mu_n}/(\epsilon\tau_R)
= 0$ for $n = 5, 6$ since there are no $O(\epsilon)$ terms in the right hand side of
Eq.(\ref{eq:EOM_final}). Equivalently,
\begin{equation}
    \rho_{r}^{\mu_1\cdots\mu_n} = O(\epsilon^3) \ \ \hbox{for}\ \ n = 5, 6
\end{equation}
Continuing this way, it can be established that in general
\begin{equation}
    \rho_{r}^{\mu_1\cdots\mu_n} = O(\epsilon^{\lceil n/2\rceil}) \ \ \hbox{for}\ \
    n \ge 3
\end{equation}
where $\lceil n/2 \rceil$ is the closest integer that is larger than or equal
to $n/2$.

The second-order hydrodynamics theory is based 
on energy density $\varepsilon$, fluid
flow velocity $u^\mu$, shear stress tensor $\pi^{\mu\nu}$, and bulk viscous pressure $\Pi$.
From Eq.(\ref{eq:rho_order_1}) one can see that $\Pi$ and $\pi^{\mu\nu}$ are 
$O(\epsilon)$. Therefore, in this method,
the second-order theory includes the $O(\epsilon^0)$ terms and the $O(\epsilon)$ terms.
To obtain the third-order theory, we need to include the $O(\epsilon^2)$ terms.

Since we have now established the $\epsilon$-order of the energy-momentum
moments, we do not have to carry $\epsilon$ around from here on although we will keep
referring to the $\epsilon$-order of specific terms. For the relaxation time approximation,
the $\epsilon$-order is the same as
the number of $\tau_R$ factors.

As stated, the goal of this section is to work out the $\epsilon$-expansion of the energy-momentum moments up to
$n = 4$.
We start with the scalar moments.
The general equation of motion for an arbitrary scalar moment ($n=0$) is
\begin{eqnarray}
D \rho_r 
&=&
-\frac{\rho_r}{\tau_R}
-F_{r-1} 
+{1\over 3}\left(
(r-1) m^2 \rho_{r-2} - (2+r) \rho_r 
\right)\theta
\nl
-\nabla_\lambda \rho_{r-1}^\lambda
-r a_\lambda \rho_{r-1}^\lambda 
-(r-1) \sigma_{\lambda \alpha} \rho_{r-2}^{\alpha \lambda}
\label{eq:rank0_eq}
\end{eqnarray}
Collecting the $O(\epsilon^0)$ terms, we get
\begin{eqnarray}
\rho_{r|1} = -\tau_R F_{r-1|0} = -\tau_R \phi_{r-1|0}\theta
\end{eqnarray}
The scalar moment up to and including $O(\epsilon^2)$ terms are then
\begin{eqnarray}
\rho_{r} &=&
\tau_R\bigg[
-F_{r-1}
- D\rho_{r|1}
+{1\over 3}
\left((r-1)m^2 \rho_{r-2|1} -(2+r) \rho_{r|1}
\right)\theta
-(r-1) \sigma_{\lambda \alpha} \rho_{r-2|1}^{\alpha \lambda}
\bigg]
\nonumber\\ && {}
+ O(\epsilon^3)
\label{eq:rho_r}
\end{eqnarray}
where we used the facts that $\tau_R = O(\epsilon)$, $\rho_{r-1}^\lambda = O(\epsilon^2)$,
and $F_{r-1}$ contains both the $O(\epsilon^0)$ terms and $O(\epsilon)$ terms.
The time derivative term is
\begin{eqnarray}
D\rho_{r|1} & = &
D(\tau_R \phi_{r-1|0}\theta)
\nonumber\\ 
& = &
\tau_R \left({\partial \phi_{r-1|0}\over \partial\beta}\right) \chi_{\beta|0} \theta^2
+ \tau_R\phi_{r-1|0} D\theta
+ O(\epsilon^2)
\label{eq:Drho_r1}
\end{eqnarray}
where $\chi_{\beta|0}$ is defined in Appendix \ref{app:Fintegrals}.
%
%
To keep the theory from becoming parabolic,
the right hand side of Eq.(\ref{eq:rho_r})
should not contain any derivatives of thermodynamic variables upon using suitable
constitutive relationships.
To deal with
%
$D\theta = D\partial_\mu u^\mu$ that contains second derivatives, we can use 
\begin{eqnarray}
\rho_{0} &=& -{3\over m^2}\Pi
\nonumber\\ 
& = & 
\tau_R\bigg[
-F_{-1}
-\tau_R \left( {\partial\phi_{-1|0}\over \partial\beta}\right) \chi_{\beta|0}\theta^2
-\tau_R \phi_{-1|0}D\theta
-{1\over 3}\left(m^2 \rho_{-2|1} +2 \theta \rho_{0|1}\right)\theta
+ \sigma_{\lambda \alpha} \rho_{-2|1}^{\alpha \lambda}
\bigg]
\nonumber\\ && {}
+ O(\epsilon^3)
\label{eq:Drho_01}
\end{eqnarray}
which will be used only in the context of obtaining the $\epsilon$-expansion of other
moments.
Replacing $D\theta$ in Eq.(\ref{eq:Drho_r1}) with $D\theta$ in
Eq.(\ref{eq:Drho_01}), we get
\begin{eqnarray}
\rho_{r} 
& = &
-{3\over m^2}\Phi_r \Pi
\nonumber\\ && {}
+
\tau_R\bigg[
-(F_{r-1|1} - \Phi_r F_{-1|1})
\nonumber\\ && {} \qquad
-\tau_R  \left({\partial\phi_{r-1|0}\over \partial\beta} - \Phi_r {\partial\phi_{-1}\over
\partial\beta}\right)\chi_{\beta|0}\theta^2
\nonumber\\ && {}\qquad
-\left(\frac{2+r}{3} \rho_{r|1} - {2\Phi_r\over 3} \rho_{0|1}\right)\theta
+{m^2\over 3} \left( (r-1) \rho_{r-2|1} + \Phi_r \rho_{-2|1}\right)\theta
\nonumber\\ && {} \qquad
-\sigma_{\lambda \alpha}
\left( (r-1) \rho_{r-2|1}^{\alpha \lambda} + \Phi_r \rho_{-2|1}^{\alpha\lambda}\right)
\bigg]
+ O(\epsilon^3)
\label{eq:rho_r_upto2}
\end{eqnarray}
where $\Phi_r= \phi_{r-1|0}/\phi_{-1|0}$.
Using
the first order constitutive relationships
\begin{eqnarray}
\label{eq:r_con1}
\Pi & = & {m^2\over 3}\tau_R\phi_{-1|0}\theta + O(\epsilon^2)
\\
\label{eq:r_con3}
\pi^{\mu\nu} &=& -\tau_R\varphi_{-1|0}\sigma^{\mu\nu} + O(\epsilon^2)
\\
\rho_{r|1} &=& -{3\over m^2}\Phi_r \Pi
\label{eq:r_con4}
\end{eqnarray}
$\rho_r$ can then
%
%
be expressed solely in terms of $\Pi$ and
$\pi^{\mu\nu}$ without involving any derivatives or an explicit factor of $\tau_R$.

From Eq.(\ref{eq:EOM_final}), the evolution equation for the general rank-2 moment 
can obtained as
\begin{eqnarray}
\Delta_{\nu_1 \nu_2}^{\mu_1 \mu_2} D \rho_r^{\nu_1 \nu_2} 
&=&
-\frac{\rho_r^{\mu_1 \mu_2}}{\tau_R}-F_{r-1}^{\mu_1 \mu_2} 
\nonumber\\ && {}
+ {2\over 15}
\left(
-(4+r) \rho_{r+2} 
+m^2 (2 r+3) \rho_r 
-m^4 (r-1) \rho_{r-2} 
\right) \sigma^{\mu_1\mu_2}
\nonumber\\ && {}
-r a_\alpha \rho_{r-1}^{\alpha \mu_1 \mu_2} 
\nonumber\\ && {}
+ {2\over 5}
\left(
r m^2 \rho_{r-1}^{\left\langle\mu_1\right.} a^{\left.\mu_2\right\rangle}
-(r+5) \rho_{r+1}^{\left\langle\mu_1\right.} a^{\left.\mu_2\right\rangle} 
\right)
\nonumber\\ && {}
-\frac{2}{5}\left(\nabla^{\left\langle\mu_1\right.}
\rho_{r+1}^{\left.\mu_2\right\rangle}-m^2 \nabla^{\left\langle\mu_1\right.}
\rho_{r-1}^{\left.\mu_2\right\rangle}\right) 
\nonumber\\ && {}
-2 \omega_\lambda^{\left\langle\mu_1\right.} \rho_r^{\left.\mu_2\right\rangle \lambda}
-(r-1) \sigma_{\lambda \alpha} \rho_{r-2}^{\alpha \lambda \mu_1 \mu_2} 
\nonumber\\ && {}
+ {2\over 7}
\left(
-(2 r+5) \sigma_\lambda^{\left\langle\mu_1\right.}
\rho_r^{\left.\mu_2\right\rangle \lambda} 
+2(r-1) m^2 \sigma_\lambda^{\left\langle\mu_1\right.}
\rho_{r-2}^{\left.\mu_2\right\rangle \lambda} 
\right)
\nonumber\\ && {}
+
{1\over 3}
\left(
m^2(r-1) 
\rho_{r-2}^{\mu_1 \mu_2} 
-(4+r) \rho_r^{\mu_1 \mu_2}
\right)\theta
\nonumber\\ && {}
-\Delta_{\nu_1 \nu_2}^{\mu_1 \mu_2} \nabla_\lambda \rho_{r-1}^{\lambda \nu_1 \nu_2}
\label{eq:rank2_eq}
\end{eqnarray}
where $F_{r}^{\mu\nu} = F_{r|0}^{\mu\nu} = \varphi_{r|0}\sigma^{\mu\nu}$.
Following the similar procedure as in the scalar case,
we obtain 
\begin{eqnarray}
\rho_r^{\mu_1 \mu_2}
& = &
\Sigma_r \rho_0^{\mu_1\mu_2}
\nonumber\\ &&
+\tau_R\bigg[
-\frac{\theta}{3}\left(r \rho_{r \mid 1}^{\mu_1 \mu_2}
-(r-1) m^2 \rho_{r-2 \mid 1}^{\mu_1 \mu_2}
- \Sigma_r m^2 \rho_{-2 \mid 1}^{\mu_1 \mu_2} 
\right)
\nonumber\\ && \qquad
+\frac{2}{7}\left(
-2 r \sigma_\lambda^{\left\langle\mu_2\right.}
\rho_{r \mid 1}^{\left.\mu_1\right\rangle \lambda}
+(2 r-2) m^2 \sigma_\lambda^{\left\langle\mu_2\right.} 
\rho_{r-2 \mid 1}^{\left.\mu_1\right\rangle \lambda}
+2 m^2\Sigma_r \sigma_\lambda^{\left\langle\mu_2\right.} 
\rho_{-2 \mid 1}^{\left.\mu_1\right\rangle \lambda}
\right)
\nonumber\\ && \qquad
+\frac{2}{15} \sigma^{\mu_1 \mu_2}\left(-(4+r) \rho_{r+2 \mid 1}+(2 r+3) m^2 
\rho_{r \mid 1}-(r-1) m^4 \rho_{r-2 \mid 1}\right) 
\nonumber\\ && \qquad
-\Sigma_r\frac{2}{15} \sigma^{\mu_1 \mu_2}
\left(-4 \rho_{2 \mid 1} +3 m^2 \rho_{0 \mid 1} + m^4 \rho_{-2 \mid 1}\right) 
\nonumber\\ && \qquad
 +
 \tau_R
\left(
\frac{\partial\left( \varphi_{r-1 \mid 0}\right)}{\partial \beta} 
-\Sigma_r\frac{\partial\left( \varphi_{-1 \mid 0}\right)}{\partial \beta} 
\right)\chi_{\beta|0}\theta \sigma^{\mu_1 \mu_2} 
\bigg]
\nonumber\\ && \qquad
+O\left(\epsilon^3\right)
\label{eq:rho_r_munu_upto2}
\end{eqnarray}
where $\Sigma_r = \varphi_{r-1|0}/\varphi_{-1|0}$ and we used
the $\epsilon$-expansion of
$\pi^{\mu\nu} = \rho_0^{\mu\nu}$ to replace
$\Delta_{\nu_1 \nu_2}^{\mu_1 \mu_2} D \sigma^{\nu_1 \nu_2}$.
Upon using Eqs.(\ref{eq:r_con1}), (\ref{eq:r_con3}), and (\ref{eq:r_con4}),
$\rho_r^{\mu_1\mu_2}$ can be re-expressed solely in terms of $\pi^{\mu\nu}$ and $\Pi$ without
their derivatives or an explicit factor of $\tau_R$.

For the $O(\epsilon^2)$ moments, we start with the vector moments whose evolution
equation is given by
\begin{eqnarray}
\Delta_{\nu_1}^{\mu_1} D \rho_r^{\nu_1}
& = &
-\frac{\rho_r^{\mu_1}}{\tau_R}-F_{r-1}^{\mu_1}
\nonumber\\ && {}
+ {1\over 3}
\left( (r-1) m^2\rho_{r-2}^{\mu_1} -(3+r) \rho_r^{\mu_1} \right)\theta
\nonumber\\ && {}
-r a_\alpha \rho_{r-1}^{\alpha \mu_1} 
-\Delta_{\nu_1}^{\mu_1} \nabla_\lambda \rho_{r-1}^{\lambda \nu_1}
-\omega_\lambda^{\mu_1} \rho_r^\lambda
\nonumber\\ && {}
-(r-1) \sigma_{\lambda \alpha} \rho_{r-2}^{\alpha \lambda \mu_1} 
\nonumber\\ && {}
+ {1\over 3}
\left( r  m^2 \rho_{r-1} -(r+3) \rho_{r+1} \right) a^{\mu_1}
\nonumber\\ && {}
-\frac{1}{3}\left(\nabla^{\mu_1} \rho_{r+1}-m^2 \nabla^{\mu_1} \rho_{r-1}\right) 
\nonumber\\ && {}
+ {1\over 5}
\left( -(2 r+3) \rho_r^\lambda +2(r-1) m^2 \rho_{r-2}^\lambda \right) \sigma_\lambda^{\mu_1} 
\label{eq:rank1_eq}
\end{eqnarray}
Since $\rho^\mu_r = O(\epsilon^2)$,
the $O(\epsilon)$ terms on the right-hand-side must add up to zero, yielding
\begin{eqnarray}
\rho_{r}^{\mu_1}
& = &
-\tau_R \psi_{r-1|1} 
\left(
\Delta_\gamma^\mu \partial_\rho \pi^{\rho\gamma}
    + \nabla^{\mu_1}  \Pi 
    + a^{\mu_1} \Pi\right)
\nonumber\\ && {}
+\tau_R\bigg[
-\Delta_{\nu_1}^{\mu_1} \nabla_\lambda \rho_{r-1|1}^{\lambda \nu_1}
-r a_{|0\alpha} \rho_{r-1|1}^{\alpha \mu_1} 
\nonumber\\ && {}\qquad
-\frac{1}{3}\left(\nabla^{\mu_1} \rho_{r+1|1}-m^2 \nabla^{\mu_1} \rho_{r-1|1}\right) 
\nonumber\\ && {}\qquad
+{1\over 3}
\left( r m^2 \rho_{r-1|1} -(r+3) \rho_{r+1|1} \right)
a_{|0}^{\mu_1} 
\bigg] + O(\epsilon^3)
\label{eq:rho_mu_r2}
\end{eqnarray}
Further details can be found in Appendix~\ref{app:Fintegrals}.
Unlike the $O(\epsilon)$ moments, this cannot be expressed
solely in terms of $\Pi$ and $\pi^{\mu\nu}$ without involving derivatives.

For the rank-3 moments, we have
\begin{eqnarray}
\Delta_{\nu_1 \nu_2 \nu_3}^{\mu_1 \mu_2 \mu_3} D \rho_r^{\nu_1 \nu_2 \nu_3} 
&=&
-\frac{\rho_r^{\mu_1 \mu_2 \mu_3}}{\tau_R} 
\nonumber\\ && {}
+ {1\over 3}
\left(
-(5+r) \rho_r^{\mu_1 \mu_2 \mu_3}
+(r-1) m^2 \rho_{r-2}^{\mu_1 \mu_2 \mu_3} 
\right)\theta
\nonumber\\ && {}
+ {6\over 35}
\left(
-(6+r) \rho_{r+2}^{\left\langle\mu_1\right.} \sigma^{\left.\mu_2 \mu_3\right\rangle} 
+ (2 r+5) m^2 \rho_r^{\left\langle\mu_1\right.} \sigma^{\left.\mu_2 \mu_3\right\rangle} 
- (r-1) m^4 \rho_{r-2}^{\left\langle\mu_1\right.} \sigma^{\left.\mu_2 \mu_3\right\rangle} 
\right)
\nonumber\\ && {}
-3 \omega_\lambda^{\left\langle\mu_1\right.} \rho_r^{\left.\mu_2 \mu_3\right\rangle \lambda} 
\nonumber\\ && {}
+ {1\over 3}
\left(
-(2 r+7) \sigma_\lambda^{\left\langle\mu_1\right.} \rho_r^{\left.\mu_2 \mu_3\right\rangle \lambda} 
+2(r-1) m^2 \sigma_\lambda^{\left\langle\mu_1\right.} \rho_{r-2}^{\left.\mu_2 \mu_3\right\rangle \lambda} 
\right)
\nonumber\\ && {}
-r a_\alpha \rho_{r-1}^{\alpha \mu_1 \mu_2 \mu_3} 
\nonumber\\ && {}
-\frac{3}{7}\left(\nabla^{\left\langle\mu_1\right.} \rho_{r+1}^{\left.\mu_2 \mu_3\right\rangle}
-m^2 \nabla^{\left\langle\mu_1\right.} \rho_{r-1}^{\left.\mu_2 \mu_3\right\rangle}\right) 
\nonumber\\ && {}
+ {3\over 7}
\left(
r  m^2 \rho_{r-1}^{\left\langle\mu_1 \mu_2\right.} a^{\left.\mu_3\right\rangle}
-(r+7) \rho_{r+1}^{\left\langle\mu_1 \mu_2\right.} a^{\left.\mu_3\right\rangle} 
\right)
\nonumber\\ && {}
-\Delta_{\nu_1 \nu_2 \nu_3}^{\mu_1 \mu_2 \mu_3} \nabla_\lambda 
\rho_{r-1}^{\lambda \nu_1 \nu_2 \nu_3}
-(r-1) \sigma_{\lambda \alpha} \rho_{r-2}^{\alpha \lambda \mu_1 \mu_2 \mu_3}
\label{eq:rank3_eq}
\end{eqnarray}
As before, the $O(\epsilon)$ terms on the right hand side must add up to zero, yielding
\begin{eqnarray}
\label{eq:33rd}
\rho_r^{\mu_1 \mu_2 \mu_3}
&=&
-{3\tau_R\over 7}\bigg[ 
\nabla^{\langle\mu_1} \rho_{r+1 \mid 1}^{\mu_2 \mu_3\rangle}
+ (r+7) \rho_{r+1 \mid 1}^{\left\langle\mu_1 \mu_2\right.} a^{\left.\mu_3\right\rangle} 
\nonumber\\ && {}\qquad
-m^2 \nabla^{\left\langle\mu_1\right.} \rho_{r-1 \mid 1}^{\left.\mu_2 \mu_3\right\rangle}
-r m^2 \rho_{r-1 \mid 1}^{\left\langle\mu_1 \mu_2\right.} a^{\left.\mu_3\right\rangle}
\bigg]+O\left(\epsilon^3\right)
\end{eqnarray}
Again, this cannot be expressed solely in terms of $\Pi$ and $\pi^{\mu\nu}$ without any
derivatives.
%
%
One may take this as the first sign that the rank-1 and rank-3 moments
need to be promoted to dynamic variables as we will do so below.

For the rank-4 moments, we have
\begin{eqnarray}
\Delta_{\nu_1 \nu_2 \nu_3 \nu_4}^{\mu_1 \mu_2 \mu_3 \mu_4} D \rho_r^{\nu_1 \nu_2 \nu_3 \nu_4} 
&= & 
-\frac{\rho_r^{\mu_1 \mu_2 \mu_3 \mu_4}}{\tau_R} 
-r a_\alpha \rho_{r-1}^{\alpha \mu_1 \mu_2 \mu_3 \mu_4} 
\nonumber\\ && {}
- {4\over 9}
\left(
(r+9) \rho_{r+1}^{\left\langle\mu_1 \mu_2 \mu_3\right.} a^{\left.\mu_4\right\rangle}
- r m^2 \rho_{r-1}^{\left\langle\mu_1 \mu_2 \mu_3\right.} a^{\left.\mu_4\right\rangle}
\right)
\nonumber\\ && {}
-\frac{4}{9}\left(\nabla^{\left\langle\mu_1\right.} 
\rho_{r+1}^{\left.\mu_2 \mu_3 \mu_4\right\rangle}
-m^2 \nabla^{\left\langle\mu_1\right.} \rho_{r-1}^{\left.\mu_2 \mu_3 \mu_4\right\rangle}\right)
\nonumber\\ && {}
-\Delta_{\nu_1 \nu_2 \nu_3 \nu_4}^{\mu_1 \mu_2 \mu_3 \mu_\lambda} 
\nabla_\lambda \rho_{r-1}^{\lambda \nu_1 \nu_2 \nu_3 \nu_4}
\nonumber\\ && {}
+ {4\over 21}
\left(
-(8+r) \rho_{r+2}^{\left\langle\mu_1 \mu_2\right.} 
\sigma^{\left.\mu_3 \mu_4\right\rangle}
+ (2 r+7) m^2 \rho_r^{\left\langle\mu_1 \mu_2\right.} 
\sigma^{\left.\mu_3 \mu_4\right\rangle} 
- (r-1) m^4 \rho_{r-2}^{\left\langle\mu_1 \mu_2\right.} 
\sigma^{\left.\mu_3 \mu_4\right\rangle}
\right)
\nonumber\\ && {}
-4 \omega_\lambda^{\left\langle\mu_1\right.}
\rho_r^{\left.\mu_2 \mu_3 \mu_4\right\rangle \lambda} 
-(r-1) \sigma_{\lambda \alpha} \rho_{r-2}^{\alpha \lambda \mu_1 \mu_2 \mu_3 \mu_4}
\nonumber\\ && {}
+ {4\over 11}
\left(
-(2 r+9) \sigma_\lambda^{\left\langle\mu_1\right.}
\rho_r^{\left.\mu_2 \mu_3 \mu_4\right\rangle \lambda} 
+2 (r-1) m^2 \sigma_\lambda^{\left\langle\mu_1\right.}
\rho_{r-2}^{\left.\mu_2 \mu_3 \mu_4\right\rangle \lambda} 
\right)
\nonumber\\ && {}
+ {1\over 3}
\left(
(r-1) m^2 \rho_{r-2}^{\mu_1 \mu_2 \mu_3 \mu_4} 
-(6+r) \rho_r^{\mu_1 \mu_2 \mu_3 \mu_4}
\right)\theta
\label{eq:rank4_eq}
\end{eqnarray}
Collecting all $O(\epsilon)$ terms on the right hand side,
the rank-4 moments up to $O(\epsilon^2)$
are given by
\begin{equation}\label{eq:43rd}
\begin{gathered}
\rho_r^{\mu_1 \mu_2 \mu_3 \mu_4}=\tau_R\left[-(8+r) \frac{4}{21} \rho_{r+2 \mid
1}^{\left\langle\mu_1 \mu_2\right.} \sigma^{\left.\mu_3 \mu_4\right\rangle}+(7+2 r)
\frac{4}{21} m^2 \rho_{r \mid 1}^{\left\langle\mu_1 \mu_2\right.} \sigma^{\left.\mu_3
\mu_4\right\rangle}\right. \\
\left.-(r-1) \frac{4}{21} m^4 \rho_{r-2 \mid 1}^{\left\langle\mu_1 \mu_2\right.}
\sigma^{\left.\mu_3 \mu_4\right\rangle}\right]+O\left(\epsilon^3\right)
\end{gathered}
\end{equation}
which can be expressed using only $\pi^{\langle\mu_1\mu_2}\pi^{\mu_3\mu_4\rangle}$ and
without an explicit factor of $\tau_R$.

\section{Relativistic Regularized Hydrodynamics up to $O(\epsilon^2)$}\label{sec:rrh}

Within the relaxation time approximation,
the full evolution equation for the bulk pressure $\Pi = -(m^2/3)\rho_0$ can be obtained
by setting $r = 0$ in Eq.(\ref{eq:rank0_eq}):
\begin{equation}\label{pi_equation}
\begin{split}
    D\Pi &= 
    - \frac{\Pi}{\tau_R}
    + \frac{m^2}{3}\left[\phi_{-1|0}\theta + 
    \phi_{-1|1}^{\pi\Pi}\left(\theta \Pi + \pi^{\gamma\rho}\sigma_{\gamma\rho}\right)\right]
    \\ &
    +\frac{m^2}{3}\nabla_\lambda \rho_{-1}^{\lambda} - \frac23\theta\Pi -
    \frac{m^2}{3}\sigma_{\lambda\alpha}\rho_{-2}^{\lambda\alpha}
    + \frac{m^4}{9}\theta\rho_{-2}
\end{split}
\end{equation}
From this, one can identify the bulk viscosity 
as $\zeta = \tau_R m^2 \phi_{-1|0}/3$.
Similarly, the full evolution equation for $\pi^{\mu\nu} = \rho_0^{\mu\nu}$ is
obtained from Eq.(\ref{eq:rank2_eq}) by setting $r = 0$:
\begin{equation}\label{shearstress}
\begin{split}
    \Delta_{\alpha\beta}^{\mu\nu}D\pi^{\alpha\beta} &= -\frac{\pi^{\mu\nu}}{\tau_R}
    -(\varphi_{-1|0}\sigma^{\mu\nu})
    - \Delta_{\alpha\beta}^{\mu\nu}\nabla_\lambda \rho_{-1}^{\lambda\alpha\beta} 
    \\ &\quad
    + \frac{2m^2}{5}\nabla^{\langle\mu}\rho_{-1}^{\nu\rangle} -
    \frac43\theta\pi^{\mu\nu} + \sigma_{\lambda\alpha}\rho_{-2}^{\alpha\lambda\mu\nu}
    - \frac{m^2}{3}\theta\rho_{-2}^{\mu\nu}\\
    &\quad - \frac{10}{7}\pi^{\lambda\langle\mu}\sigma_\lambda^{\nu\rangle}
    - 2\pi^{\lambda\langle\mu}\omega_\lambda^{\nu\rangle} 
    \\
    &\quad - \frac{4m^2}{7}\rho_{-2}^{\lambda\langle\mu}\sigma_\lambda^{\nu\rangle}
    + \frac{2m^4}{15}\rho_{-2}\sigma^{\mu\nu} 
    -\frac{6}{5}\Pi \sigma^{\mu\nu}
\end{split}
\end{equation}
The shear viscosity can be identified as 
$\eta = \tau_R\varphi_{-1|0}/2$.
In obtaining Eqs.(\ref{pi_equation}) and (\ref{shearstress}), 
we used the Landau condition $\rho_2 = \rho_1^{\mu} = 0$.
These equations are not closed because
the following moments appearing in the above two equations 
\begin{eqnarray}
\rho_{-2},
\rho_{-1}^\mu, 
\rho_{-2}^{\mu\nu},
\rho_{-1}^{\lambda\alpha\beta},
\rho_{-2}^{\alpha\lambda\mu\nu}
\end{eqnarray}
are not $\Pi$ nor $\pi^{\mu\nu}$.
The goal is to use the $\epsilon$-expansion of these moments to re-express  
Eqs.(\ref{pi_equation}) and (\ref{shearstress}) so that the equations are closed,
adding extra dynamic degrees of freedom when necessary.

Before we carry out the $O(\epsilon^2)$ analysis, 
we can first check the $O(\epsilon)$ results.
Using the $O(\epsilon)$ terms from the $\epsilon$-expansions of $\rho_{-2}$ and
$\rho_{-2}^{\mu\nu}$ (Eqs.(\ref{eq:rho_r_upto2}) and (\ref{eq:rho_r_munu_upto2})),
the evolution equation for $\Pi$ can be expressed as
\begin{equation}
\begin{split}
    D\Pi & = -\frac{\Pi}{\tau_R} 
    + \frac{m^2}{3} \phi_{-1|0}\theta 
    - \frac23\theta \Pi 
    + \frac{m^2}{3}
    \phi_{-1|1}^{\pi\Pi}\left(\theta \Pi + \pi^{\gamma\rho}\sigma_{\gamma\rho}\right)
    \\
    &\quad 
    - \frac{m^2}{3}\bigg(\frac{\varphi_{-3|0}}{\varphi_{-1|0}}\bigg)\sigma_{\lambda\alpha}\pi^{\lambda\alpha}
    - \frac{m^2}{3}\bigg(\frac{\phi_{-3|0}}{\phi_{-1|0}}\bigg)\theta\Pi
    + O(\epsilon^2)
\end{split}
\label{eq:Pi_2nd}
\end{equation}
Similarly, for $\pi^{\mu\nu}$, the second order evolution equation is
\begin{eqnarray}\label{eq:pi_second}
    \Delta_{\alpha \beta}^{\mu \nu}D\pi^{\alpha \beta} 
    &=&
    -\frac{\pi^{\mu\nu}}{\tau_R} 
    - \varphi_{-1|0}\sigma^{\mu\nu} 
    - \frac43\theta\pi^{\mu\nu}
    - \frac{m^2}{3}\theta\bigg(\frac{\varphi_{-3|0}}{\varphi_{-1|0}}\bigg)\pi^{\mu\nu}
    \nonumber\\ &&{} 
    - \frac{10}{7}\pi^{\lambda\langle\mu}\sigma_\lambda^{\nu\rangle}
    - \frac{4m^2}{7}\bigg(\frac{\varphi_{-3|0}}{\varphi_{-1|0}}\bigg) \pi^{\lambda\langle\mu}\sigma_\lambda^{\nu\rangle}
    \nonumber\\ &&{} 
    - \frac65\Pi\sigma^{\mu\nu}
    - \frac{2m^2}{5}\bigg(\frac{\phi_{-3|0}}{\phi_{-1|0}}\bigg)\Pi\sigma^{\mu\nu}
    \nonumber\\ && {}
    - 2\pi^{\lambda\langle\mu}\omega_\lambda^{\nu\rangle} + O(\epsilon^2)
\label{eq:pi_munu_2nd}
\end{eqnarray}
Note that these equations are hyperbolic, namely, involves the same number of temporal
and spatial derivatives. This fact does not automatically guarantee that the theory is
stable, but as long as $\tau_R > \eta/(\varepsilon + P)$, it is at least causal.

To go to the $O(\epsilon^2)$ order, one needs to examine $\rho_{-1}^\mu$
and $\rho_{-1}^{\mu_1\mu_2\mu_3}$ more closely. 
There is no need to consider $\rho_{-2}^{\mu_1\mu_2\mu_3\mu_4}$ any further since 
it can be expressed using $\pi^{\langle\mu_1\mu_2}\pi^{\mu_3\mu_4\rangle} = O(\epsilon^2)$.
But the first moment and the third moment
cannot be expressed solely in terms of $\Pi$ and $\pi^{\mu\nu}$
without involving their derivatives.
As such, if the $\epsilon$-expansion from section \ref{sec:Chapman_Enskog} is used, parabolic equations
will result. 
One way to remedy this problem is to promote the first moment $\rho_{-1}^{\mu}$ and
the third moment $\rho_{-1}^{\mu_1\mu_2\mu_3}$ to be dynamic variables.
Denoting $W^\mu = m^2\rho_{-1}^\mu$, its evolution equation can be obtained from
Eq.(\ref{eq:rank1_eq}) 
\begin{eqnarray}
\Delta_{\nu_1}^{\mu_1} D W^{\nu_1} 
& = &
-\frac{W^{\mu_1}}{\tau_R}
-m^2 F_{-2}^{\mu_1}
\nonumber\\ && {}
-\frac{2}{3} \theta W^{\mu_1}
-\frac{1}{5} \sigma_\lambda^{\mu_1} W^\lambda 
-\omega_\lambda^{\mu_1} W^\lambda
\nonumber \\ && {} 
+ \nabla^{\mu_1}\Pi 
+2 \Pi a^{\mu_1}
\nonumber \\ && {} 
-\frac{2}{3} m^4\theta \rho_{-3}^{\mu_1} 
-m^4\frac{4}{5} \sigma_\lambda^{\mu_1} \rho_{-3}^\lambda 
\nonumber \\ && {} 
-m^2\Delta_{\nu_1}^{\mu_1} \nabla_\lambda \rho_{-2}^{\lambda \nu_1}
+ m^2a_\alpha \rho_{-2}^{\alpha \mu_1} 
\nonumber\\ && {}
+ {m^4\over 3} \nabla^{\mu_1} \rho_{-2}
-\frac{m^4}{3} a^{\mu_1} \rho_{-2}
\nonumber\\ && {}
+2m^2 \sigma_{\lambda \alpha} \rho_{-3}^{\alpha \lambda \mu_1} 
\label{eq:W_eq}
\end{eqnarray}
Denoting $\xi^{\mu_1\mu_2\mu_3} = \rho_{-1}^{\mu_1\mu_2\mu_3}$, 
its evolution equation can be obtained from
Eq.(\ref{eq:rank3_eq}) 
\begin{eqnarray}
\Delta_{\nu_1 \nu_2 \nu_3}^{\mu_1 \mu_2 \mu_3} D \xi^{\nu_1 \nu_2 \nu_3} 
&=&
-\frac{\xi^{\mu_1 \mu_2 \mu_3}}{\tau_R} 
\nonumber\\ && {}
-\frac{18}{7} \pi^{\left\langle\mu_1 \mu_2\right.} a^{\left.\mu_3\right\rangle} 
-\frac{3}{7} \nabla^{\left\langle\mu_1\right.} \pi^{\left.\mu_2 \mu_3\right\rangle}
\nonumber\\ && {}
+\frac{3}{7} m^2 \nabla^{\left\langle\mu_1\right.} \rho_{-2}^{\left.\mu_2 \mu_3\right\rangle}
- \frac{3}{7} m^2 \rho_{-2}^{\left\langle\mu_1 \mu_2\right.} a^{\left.\mu_3\right\rangle}
\nonumber\\ && {}
-\frac{4}{3} \theta \xi^{\mu_1 \mu_2 \mu_3}
-\frac{15}{9} \sigma_\lambda^{\left\langle\mu_1\right.} \xi^{\left.\mu_2 \mu_3\right\rangle \lambda} 
-3 \omega_\lambda^{\left\langle\mu_1\right.} \xi^{\left.\mu_2 \mu_3\right\rangle \lambda}
\nonumber\\ && {}
+ \frac{18}{35} W^{\left\langle\mu_1\right.} \sigma^{\left.\mu_2 \mu_3\right\rangle}
+m^4 \frac{12}{35} \rho_{-3}^{\left\langle\mu_1\right.} \sigma^{\left.\mu_2 \mu_3\right\rangle} 
\nonumber\\ && {}
-m^2 \frac{2\theta}{3} \rho_{-3}^{\mu_1 \mu_2 \mu_3} 
-m^2 \frac{4}{3} \sigma_\lambda^{\left\langle\mu_1\right.}
\rho_{-3}^{\left.\mu_2 \mu_3\right\rangle \lambda} 
\nonumber\\ && {}
-\Delta_{\nu_1 \nu_2 \nu_3}^{\mu_1 \mu_2 \mu_3} \nabla_\lambda \rho_{-2}^{\lambda \nu_1 \nu_2 \nu_3}
+ a_\alpha \rho_{-2}^{\alpha \mu_1 \mu_2 \mu_3} 
\nonumber\\ && {}
+ O(\epsilon^3)
\label{eq:xi_eq}
\end{eqnarray}
We can use the $\epsilon$-expansions,
Eqs.(\ref{eq:rho_r_upto2}), (\ref{eq:rho_r_munu_upto2}), and
(\ref{eq:43rd}),
in place of $\rho_{-2}$, $\rho_{-2}^{\mu\nu}$, and $\rho_{-2}^{\mu_1\mu_2\mu_3\mu_4}$, respectively,
on the right-hand-sides of Eqs.(\ref{eq:W_eq}) and (\ref{eq:xi_eq}). These terms 
do not contain any derivatives.
We can also use the $\epsilon$-expansions, Eqs.(\ref{eq:rho_mu_r2}) and (\ref{eq:33rd}),
for $\rho^{\mu_1}_{-3|2}$
and
$\rho_{-3|2}^{\mu_1 \mu_2 \mu_3}$, respectively.
This replacement does involve derivatives, and since 
$\rho_{-3|2}^{\mu}$ and $\rho_{-3|2}^{\mu_1\mu_2\mu_3}$ above are accompanied by
either $\theta$ or $\sigma^{\mu\nu}$, that results in terms with two derivatives.
Fortunately, we can avoid having two derivatives by
associating the explicit factor of 
$\tau_R$ from Eqs.(\ref{eq:rho_mu_r2}) and (\ref{eq:33rd})
to the factors $\theta$ and $\sigma^{\mu\nu}$
to turn them into $\Pi$ and $\pi^{\mu\nu}$.
In this way, we have a closed set of equations for $\Pi, \pi^{\mu\nu}, W^\mu$ and
$\xi^{\mu_1\mu_2\mu_3}$ that involve no more than the first derivatives.
Furthermore, the relaxation time $\tau_R$ does not appears explicitly 
except for the collision integral term (the $1/\tau_R$ term).

\section{Third-Order Equations for $m=0$}\label{sec:n1}

The third-order hydrodynamic equations obtained in the previous subsections are
non-linear coupled differential equation of 20 degrees of freedom,
making them hard to analyze. For the sake of simplicity, from now on,
we take the massless limit.
In this imit,
the bulk pressure does not exist, $\Pi = 0$, and it is
consistent to set $W^\mu = 0$ as well.
As such, the dynamic degrees of freedom reduce to the energy density $\varepsilon$,
the flow vector ${\bf u}$,
the shear-stress tensor $\pi^{\mu\nu}$ and the third moment $\xi^{\mu_1\mu_2\mu_3}$.
In this limit, Eq.(\ref{shearstress}) reduces to
\begin{eqnarray}
\Delta_{\alpha \beta}^{\mu \nu} D \pi^{\alpha \beta} 
&=&
-\frac{\pi^{\mu \nu}}{\tau_R}-\varphi_{-1 \mid 0} \sigma^{\mu \nu}
-\Delta_{\alpha \beta}^{\mu \nu} \nabla_\lambda \xi^{\lambda \alpha \beta} 
\nonumber\\ && {}
-\frac{4}{3} \theta \pi^{\mu \nu}
+\sigma_{\lambda \alpha} \varsigma^{\alpha \lambda \mu \nu}
-\frac{10}{7} \pi^{\lambda\langle\mu} \sigma_\lambda^{\nu\rangle} \nonumber\\ && {}
-2 \pi^{\lambda\langle\mu} \omega_\lambda^{\nu\rangle}
\label{eq:pi_eq_m0}
\end{eqnarray}
where
\begin{equation}
\begin{aligned}
    \varsigma^{\alpha\beta\mu\nu}
    &= \rho_{-2}^{\alpha\beta\mu\nu}
    = -{8\over 7\varphi_{-1|0}} \pi^{\langle\alpha\beta}\pi^{\mu\nu\rangle} +
    O(\epsilon^3)
\end{aligned}
\label{eq:xi_and_varsigma}
\end{equation}
In the $m=0$ limit, Eq.(\ref{eq:xi_eq}) reduces to
\begin{eqnarray}
    \Delta_{\rho\alpha\beta}^{\lambda\mu\nu}D\xi^{\rho\alpha\beta}
    &=&
    -\frac{1}{\tau_R}\xi^{\lambda\mu\nu} 
    - \frac43\theta\xi^{\lambda\mu\nu} 
    - \frac53\xi^{\alpha\langle\lambda\mu}\sigma_\alpha^{\nu\rangle} 
    - 3\xi^{\alpha\langle\lambda\mu}\omega_\alpha^{\nu\rangle}
    \nonumber\\ && {}
    - \frac{18}{7}\pi^{\langle\lambda\mu}a^{\nu\rangle} 
    - \frac37 \nabla^{\langle\lambda}\pi^{\mu\nu\rangle}
    + a_\rho \varsigma^{\rho\lambda\mu\nu}
    - \Delta_{\rho\alpha\beta}^{\lambda\mu\nu}\nabla_\omega \varsigma^{\omega\rho\alpha\beta} 
    \nonumber\\ && {}
    + O(\epsilon^3)
    \label{eq:xi_eq_m0}
\end{eqnarray}
The dynamics variables are $\varepsilon, u^\mu, \pi^{\mu\nu},
\xi^{\lambda\mu\nu}$. The number of independent degrees of freedom is thus 16.

Eqs.(\ref{eq:pi_eq_m0}) and (\ref{eq:xi_eq_m0}) provide us
with the third-order dissipative equations for massless
particles without conservation of the net particle number. 
As far as terms linear in $\pi^{\mu\nu}$, $\xi^{\lambda\mu\nu}$, and $u^\mu$
are concerned,
these equations are equivalent to the stable third order theory postulated 
in Ref.~\cite{brito2022} with $\tau_\rho = \eta_\rho = \tau_\pi$ 
in their notation. Consequently, our 16 moment formuation is also linearly stable and causal.

What we would like to do further here is to analyze an alternative third order theory where 
$\varsigma^{\alpha\beta\mu\nu}$ is also promoted to be a dynamic variable.
Setting $r=-2$ and $m = 0$, Eq.(\ref{eq:rank4_eq}) becomes
\begin{eqnarray}
    \Delta_{\rho\lambda\omega\gamma}^{\alpha\beta\mu\nu}D\varsigma^{\rho\lambda\omega\gamma}
    &=&
    -\frac{1}{\tau_R}\varsigma^{\alpha\beta\mu\nu}
    - \frac{4}{3}\theta\varsigma^{\alpha\beta\mu\nu} 
    - \frac87\pi^{\langle\alpha\beta}\sigma^{\mu\nu\rangle}
    \nonumber\\ && {}
    - \frac{28}{9}\xi^{\langle\alpha\beta\mu}a^{\nu\rangle}
    - \frac49 \nabla^{\langle\alpha}\xi^{\beta\mu\nu\rangle}
    \nonumber\\ && {}
    -\frac{20}{11}\varsigma^{\lambda\langle\alpha\beta\mu}\sigma_\lambda^{\nu\rangle}
    - 4 \varsigma^{\lambda\langle\alpha\beta\mu}\omega_\lambda^{\nu\rangle}
    \nonumber\\ && {}
    + O(\epsilon^3)
\label{eq:varsigma_eq_m0}
\end{eqnarray}
Eqs.(\ref{eq:xi_eq_m0}) and Eq.(\ref{eq:varsigma_eq_m0}) are similar to, but not identical
to,
the equations for the 3rd and the 4th moments in Ref.~\cite{deBrito:2023tgb}.
This is because the 3rd and the 4th moments
used in Ref.~\cite{deBrito:2023tgb} are $\rho^{\mu_1\mu_2\mu_3}_0$ 
and $\rho_0^{\mu_1\mu_2\mu_3\mu_4}$ while ours are $\rho_{-1}^{\mu_1\mu_2\mu_3}$
and $\rho_{-2}^{\mu_1\mu_2\mu_3\mu_4}$ that naturally appear in the evolution equation of
$\pi^{\mu\nu}$.

One way of justifying the promotion of $\varsigma^{\mu_1\mu_2\mu_3\mu_4}$ to a dynamic variable is 
to note that both are $O(\epsilon^2)$ and in Eq.(\ref{eq:varsigma_eq_m0}),
$\Delta_{\rho\lambda\omega\gamma}^{\alpha\beta\mu\nu}D\varsigma^{\rho\lambda\omega\gamma}$
is linearly coupled to
$\nabla^{\langle\alpha}\xi^{\beta\mu\nu\rangle}$ while 
in Eq.(\ref{eq:xi_eq_m0}), 
$\Delta_{\rho\alpha\beta}^{\lambda\mu\nu}D\xi^{\rho\alpha\beta}$
is linearly coupled to 
$\Delta_{\rho\alpha\beta}^{\lambda\mu\nu}\nabla_\omega \varsigma^{\omega\rho\alpha\beta}$. 
Hence, a consistent linear analysis can be carried out that includes
both $\xi^{\mu_1\mu_2\mu_3}$ and $\varsigma^{\mu_1\mu_2\mu_3\mu_4}$.
This way of including $\varsigma^{\mu_1\mu_2\mu_3\mu_4}$ to close
the equations without incurring two derivatives, however, is possible only
when $m = 0$.
If $m\ne 0$, the right hand side of Eq.(\ref{eq:varsigma_eq_m0}) will contain
$\nabla^{\langle\mu_1}\rho^{\mu_2\mu_3\mu_4\rangle}_{-3}$ and
$a^{\langle\mu_1}\rho^{\mu_2\mu_3\mu_4\rangle}_{-3}$ resulting in two derivatives.
Even though we can argue that promoting $\varsigma^{\mu_1\mu_2\mu_3\mu_4}$ to a dynamic
variable is not strictly necessary, we feel that it is still beneficial to carry out a
linear analysis as these types of equations do appear elsewhere in literature (for instance
Ref.~\cite{deBrito:2023tgb}) without the full linear analysis.

In the next section, we carry out linear analysis of this extended 25-moment
theory.
Before we do so, let us consider the physical meaning of the third
moment $\xi^{\mu_1\mu_2\mu_3}$. We will not regard $\varsigma^{\mu_1\mu_2\mu_3\mu_4}$ as
a dynamic variable for this consideration.
Applying the thermodynamic identities 
$Ts = \varepsilon + P$ and $Tds = d\varepsilon$ to the local equilibrium part, the
energy conservation law, Eq.(\ref{eq:energy_conserv}), in the massless limit can be re-expressed as
\begin{equation}
\partial_\mu (su^\mu)
=
-\beta \pi^{\mu\nu}\sigma_{\mu\nu}
\label{eq:partials}
\end{equation}
where $s$ is the local equilibrium entropy density.
Within the first order approach,
the right hand side becomes non-negative 
upon using the first order constitutive equation, Eq.(\ref{eq:r_con3}),
affirming the second law of thermodynamics in this limit.
In our case,
upon using the full evolution equation for $\pi^{\mu\nu}$ (Eq.(\ref{eq:pi_eq_m0})) to replace $\sigma_{\mu\nu}$, 
Eq.(\ref{eq:partials}) can be re-arranged as
\begin{eqnarray}
\partial_\mu s_{\rm hyd}^\mu
\ampeq
{\beta\over \varphi_{-1|0}}
\bigg(
{1\over \tau_R}\pi_{\mu_1\mu_2}\pi^{\mu_1\mu_2}
+ {8\over 7}
\tau_R
\pi_{\langle\mu_1\mu_2}\sigma_{\lambda\alpha\rangle}
\pi^{\langle\alpha\lambda}\sigma^{\mu_1\mu_2\rangle}
\nlqq
- 
{5\over 2 I_{3,0}}\pi^{\mu_1\mu_2}\pi_{\mu_1\mu_2}
\pi^{\mu_3\mu_4}\sigma_{\mu_3\mu_4}
\nlqq
+{10\over 7}
\sigma_\lambda^{\langle\mu_2}\pi^{\mu_1\rangle\lambda} \pi_{\mu_1\mu_2}
\nlqq
-\xi^{\mu_1\mu_2\mu_3}
\left(
\nabla_{\langle\mu_1}\pi_{\mu_2\mu_3\rangle} + 6a_{\langle\mu_1}\pi_{\mu_2\mu_3\rangle}
\right)
\bigg)
+ O(\epsilon^4)
\label{eq:partialmusmu}
\end{eqnarray}
where
\begin{equation}
s_{\rm hyd}^\mu = 
\left(s - {\beta\over 2\varphi_{-1|0}}\pi_{\mu_1\mu_2}\pi^{\mu_1\mu_2}
\right)u^\mu
-
{\beta\over \varphi_{-1|0}}\pi_{\nu_1\nu_2}\xi^{\mu\nu_1\nu_2}
\label{eq:s_hyd}
\end{equation}
can be interpreted as the hydrodynamic non-equilibrium entropy current.
In deriving Eq.(\ref{eq:partialmusmu}), we used
Eqs.(\ref{eq:Dbeta}), (\ref{eq:I_rk}), and (\ref{eq:varphi_r0})
from Appendix \ref{app:Fintegrals}, and
the constitutive relationship
for the 4-th moment, Eq.(\ref{eq:43rd}).
The term in Eq.(\ref{eq:pi_eq_m0}) involving the 
vorticity tensor $\omega_{\mu_1\mu_2}$ 
does not contribute because of its anti-symmetric property.
Expressed this way, the meaning of $\xi^{\mu_1\mu_2\mu_3}$
is clear: It is a part of the dissipative entropy current.

In Eq.(\ref{eq:s_hyd}), the first term in the parenthesis indicates that
the non-equilibrium entropy density is {\em lower} than the equilibrium one, as it should be.
This $\pi_{\mu_1\mu_2}\pi^{\mu_1\mu_2}$ term appears in the original Israel-Stewart paper
\cite{IS:1979} and all subsequent second order and third order analyses.
The dissipative term is transverse to $u^\mu$ because of $\xi^{\mu\nu_1\nu_2}$.
Hence, the fact that one cannot assign definite sign to this term 
does not disturb the requirement that the non-equilibrium entropy to be lower
than the equilibrium one.

The second law of thermodynamics dictates that the entropy of a system must increase when
out of equilibrium. This is guaranteed if the right hand side of Eq.(\ref{eq:partialmusmu})
is non-negative.
On the right hand side of Eq.(\ref{eq:partialmusmu}), 
the first line is non-negative.
The second line is not guaranteed to be non-negative, but as $\pi^{\mu_3\mu_4}$ 
relaxes towards $-\tau_R\varphi_{-1|0}\sigma^{\mu_3\mu_4}$, it will become non-negative.
A similar argument applies to the last line which is the third order contribution.
As $\xi^{\mu_1\mu_2\mu_3}$ relaxes towards
$ 
-\tau_R {3\over 7}
\left( 
\nabla^{\langle\mu_1}\pi^{\mu_2\mu_3\rangle} + 6 a^{\langle\mu_1}\pi^{\mu_2\mu_3\rangle}
\right)
$
(e.g.~Eq.(\ref{eq:xi_eq})),
the last line in Eq.(\ref{eq:partialmusmu}) will become non-negative.
The third line
cannot be manipulated into a total derivative and/or a square even as $\pi^{\mu_1\mu_2}$
relaxes towards 
$-\tau_R\varphi_{-1|0}\sigma^{\mu_1\mu_2}$.
%
However, this may be an artifact of the way we defined the non-equilibrium entropy
\cite{Dore:2021xqq,Gavassino:2020ubn,Shokri:2020cxa}.

In Ref.~\cite{3rd:2015},
the entropy current was derived from the Chapman-Enskog expansion of $\delta f$.
Comparing the two expressions
one can see that they are almost the same except that
their entropy current 
contains the third order contribution
proportional to $(\pi_{\alpha}^{\gamma}\pi_{\gamma\beta}\pi^{\alpha\beta})u^\mu$. 
The entropy density found in Refs.~\cite{3rd:2010,Younus:2019rrt} also have
a similar term although
their entropy currents do not have our dissipative part.
The difference between our expression and the ones from
Refs.~\cite{3rd:2015,3rd:2010,Younus:2019rrt} 
mainly comes from the fact that they are using 
the Boltzmann's H-function definition of the entropy current whereas we 
are combining the energy conservation equation with thermodynamic identities to define the
entropy current following Israel and Stewart's work on the second order
hydrodynamics.
Unfortunately, it is not at all straightforward to make an exact correspondence because
expressing the H-function definition of
entropy (which involves $f\ln f$) as a linear combination of the energy-momentum moments
of $\delta f$ is highly non-trivial.

\section{Linear Stability and Causality Analysis
of the 25 moments}\label{sec:lin_stab_cau}

\subsection{Linearized Moment Equations}\label{sec:lin}

The previous section provided us with the third-order moment equations for massless
particles without conservation of net particle number.
The next step is to ensure
that these equations lead to stable and causal solutions. In general, analyzing the
stability and causality of non-linear partial differential equations is a challenging
task. In principle, one should carry out a full non-linear analysis as advocated in Ref.~\cite{Bemfica:2020xym}.
However, in this study we only perform the linear analysis of the 25-moment equations
as a first step towards establishing 
the stability and causality of our third order hydrodynamics.

Consider small fluctuations in the energy density $\varepsilon$, fluid 4-velocity
$u^\mu$, and shear-stress tensor $\pi^{\mu\nu}$:
\begin{equation}
        \varepsilon = \varepsilon_0 + \delta\varepsilon, \ \ u^\mu = u_0^\mu + \delta u^\mu,
        \ \ \pi^{\mu\nu} = \delta\pi^{\mu\nu}
\end{equation}
where $\varepsilon_0, u_0^\mu$ are constants.
Since $m=0$, the equation of state is simply $P = \varepsilon/3$.
Consider
the energy and momentum conservation laws Eqs.(\ref{eq:energy_conserv}) and
(\ref{eq:momentum_conserv}).
The linearized conservation
laws are straightforward to get:
\begin{equation}\label{lin_conserv}
\begin{aligned}
    & D_0 \delta \varepsilon + \frac43 \varepsilon_0 \nabla_{\mu,0} \delta u^\mu = 0\\
    & D_0 (\varepsilon_0 \delta u^\mu) + \frac14 \nabla_0^\mu \delta \varepsilon + \frac34
    \nabla_{\lambda, 0}\delta \pi^{\lambda \mu} = 0
\end{aligned}
\end{equation}
where we defined $\Delta_0^{\mu\nu} = g^{\mu\nu} + u_0^\mu u_0^\nu$ and $\nabla_0^{\mu}
= \Delta_0^{\mu\nu}\partial_\nu$. It is convenient to express the above equations in
Fourier space. We will use the following format of Fourier transform:
\begin{equation}
\begin{aligned}
& \widetilde{f}(k) = \int_{-\infty}^{\infty}d^4x \ e^{-ik_\mu x^\mu}f(x)\\
& f(x) = \int_{-\infty}^{\infty} \frac{d^4 k}{(2\pi)^4} e^{ik_\mu x^\mu}\widetilde{f}(k)
\end{aligned}
\end{equation}
Here, $k^\mu = (\omega, \textbf{k})$ is the wave 4-vector. Therefore, we can express
each Fourier component of the variables in the linearized equations as a plane wave
multiplied by a complex amplitude $\widetilde{\phi}$:
\begin{equation}\label{planewave}
    \phi = \widetilde{\phi}e^{ik_\mu x^\mu} =
    \widetilde{\phi}e^{i(\textbf{k}\cdot\textbf{x} - \omega t)}
\end{equation}
Note that since $g^{\mu\nu} = \hbox{diag}(-1, 1, 1, 1)$, we have $k_\mu x^\mu =
\textbf{k}\cdot\textbf{x} - \omega t$. Furthermore, we shall rewrite the linearized
equations in terms of the Lorentz-covariant variables defined below:
\begin{equation}
\begin{aligned}
    & \Omega \equiv u_0^\mu k_\mu \\
    & \kappa^\mu \equiv \Delta_0^{\mu\nu}k_\nu
\end{aligned}
\end{equation}
which correspond to $-\omega$ and ${\bf k}$ in the local rest frame of
the background system.
We also define the covariant wave number $\kappa$ as
\begin{equation}
    \kappa \equiv \sqrt{\kappa^\mu \kappa_\mu}
\end{equation}
In terms of the covariant variables, the linearized conservation laws
Eq.(\ref{lin_conserv}) can now be rewritten as
\begin{equation}\label{conserv_lin_fin}
\begin{aligned}
    & \Omega \delta \widetilde{\varepsilon} + \frac43 \varepsilon_0 \kappa_\mu \delta
    \widetilde{u}^\mu = 0 \\
    & \Omega \varepsilon_0 \delta \widetilde{u}^\mu + \frac14 \kappa^\mu \delta
    \widetilde{\varepsilon} + \frac34 \kappa_\alpha \delta \widetilde{\pi}^{\alpha\mu} = 0
\end{aligned}
\end{equation}

From now on, we will omit the tilde above the Fourier space variables.
All hydrodynamic
variables below are expressed in Fourier space. Furthermore, 
we scale $\Omega$ and $\kappa$ with the time scale $\tau_\eta = \eta/(\varepsilon_0 + P_0)$
so that they become dimensionless quantities following Refs.~\cite{brito2022, Denicol:2012cn}.
Here, $\eta = \tau_R \varphi_{-1|0}/2$ is the shear viscosity.

The next step is to linearize the $\pi^{\mu\nu}$ equation. To do this,
we drop all the higher-order terms in Eq.(\ref{eq:pi_eq_m0}) and keep only the terms
that are linear in $\delta\varepsilon$, $\delta u^\mu$, $\delta\pi^{\mu\nu}$, $\xi^{\mu_1\mu_2\mu_3}$,
and $\varsigma^{\mu_1\mu_2\mu_3\mu_4}$ to obtain the linearized $\pi^{\mu\nu}$ equation:
\begin{equation}
\begin{split}
    &\Delta_{\alpha\beta,0}^{\mu\nu}D_0\delta\pi^{\alpha\beta} 
    + \frac{1}{\tau_R}\delta\pi^{\mu\nu}
    + \varphi_{-1|0}\delta\sigma^{\mu\nu} +\Delta_{\alpha\beta,0}^{\mu\nu}\nabla_{\lambda,0
    }\xi^{\lambda\alpha\beta} = 0
\end{split}
\end{equation}
where $\delta\sigma^{\mu\nu} = \nabla^{\langle\mu}\delta u^{\nu\rangle}$.
Using (\ref{phi_10}) to express the coefficient
$\varphi_{-1|0}$ in terms of $\varepsilon_0$ leads us to the following linearized
$\pi^{\mu\nu}$ equation:
\begin{equation}\label{3rd_lin}
\begin{split}
    &\left(i\Omega + \frac{1}{\tau_R}\right) \delta\pi^{\mu\nu} +
    \frac{4i\varepsilon_0}{15} \bigg(\kappa^\mu \delta u^\nu + \kappa^\nu \delta u^\mu
    -\frac23 \kappa_\alpha \delta u^\alpha \Delta_0^{\mu\nu} \bigg) + i\kappa_\lambda
    \xi^{\lambda\mu\nu} = 0
\end{split}
\end{equation}
Similarly, the linearized equation for $\xi^{\lambda\mu\nu}$ is:
\begin{equation}
\begin{split}
    &\Delta_{\alpha\beta\gamma,0}^{\lambda\mu\nu}D_0\xi^{\alpha\beta\gamma}
    + \frac{1}{\tau_R}\xi^{\lambda\mu\nu} + \frac37
    \Delta_{\alpha\beta\gamma,0}^{\lambda\mu\nu}\nabla_0^\alpha \delta\pi^{\beta\gamma}
    + \Delta_{\alpha\beta\gamma,0}^{\lambda\mu\nu}\nabla_{\omega,0}
    \varsigma^{\omega\alpha\beta\gamma} = 0
\end{split}
\end{equation}
which becomes
\begin{equation}
\begin{aligned}
& \left(i\Omega + \frac{1}{\tau_R}\right)\xi^{\lambda\mu\nu} +
\frac{i}{7}\left(\kappa^\lambda \delta\pi^{\mu\nu} + \kappa^\mu \delta\pi^{\nu\lambda}
+ \kappa^\nu \delta\pi^{\mu\lambda}\right)\\
& - \frac{2i}{35}\left(\Delta_0^{\lambda\mu}\kappa^\omega \delta\pi_\omega^\nu +
\Delta_0^{\lambda\nu}\kappa^\omega \delta\pi_\omega^\mu + \Delta_0^{\mu\nu}\kappa^\omega
\delta\pi_\omega^\lambda\right)\\
& + i\kappa_\omega \varsigma^{\omega\lambda\mu\nu} = 0
\end{aligned}
\end{equation}
in the Fourier space after taking the derivatives $D_0$ and $\nabla_{\lambda,0}$. To
derive the above expression, we have used Eq.(\ref{eq:q_P_n}) from 
Appendix \ref{app:projectors} for $n=3$ to express
$\kappa^{\langle\lambda}\pi^{\mu\nu\rangle}$.
The linearized equation for $\varsigma^{\alpha\beta\mu\nu}$ is also straightforward
to obtain:
\begin{equation}
    \Delta_{\lambda\gamma\rho\theta,0}^{\alpha\beta\mu\nu}D_0
    \varsigma^{\lambda\gamma\rho\theta} +
    \frac{1}{\tau_R}\varsigma^{\alpha\beta\mu\nu} + \frac49
    \Delta_{\lambda\gamma\rho\theta,0}^{\alpha\beta\mu\nu} \nabla_0^\lambda
    \xi^{\gamma\rho\theta} = 0
\end{equation}
which becomes
\begin{equation}\label{sol54}
    \left(i\Omega + \frac{1}{\tau_R}\right) \varsigma^{\alpha\beta\mu\nu} +
    \frac{4i}{9}\Delta_{\lambda\gamma\rho\theta,0}^{\alpha\beta\mu\nu}\kappa^\lambda
    \xi^{\gamma\rho\theta} = 0
\end{equation}
in the Fourier space after taking the derivatives. Using 
Eq.(\ref{eq:q_P_n}) for $n=4$ from Appendix \ref{app:projectors}, one can show that
\begin{eqnarray}
\lefteqn{ \Delta_{\lambda\gamma\rho\theta,0}^{\alpha\beta\mu\nu}\kappa^\lambda
    \xi^{\gamma\rho\theta}
    = \frac14 \bigg(\kappa^\alpha \xi^{\beta\mu\nu} +
    \kappa^\beta \xi^{\alpha\mu\nu} + \kappa^\mu \xi^{\alpha\beta\nu} + \kappa^\nu
    \xi^{\alpha\mu\beta}\bigg) } &&
    \nonumber\\ && {}
    - \frac{1}{14}\Bigg(\Delta_0^{\beta\mu}\kappa_\lambda \xi^{\alpha\nu\lambda}
    + \Delta_0^{\beta\nu}\kappa_\lambda \xi^{\alpha\mu\lambda}
    + \Delta_0^{\mu\nu}\kappa_\lambda \xi^{\alpha\beta\lambda}+
    \Delta_0^{\alpha\mu}\kappa_\lambda \xi^{\beta\nu\lambda}
    + \Delta_0^{\alpha\nu}\kappa_\lambda \xi^{\beta\mu\lambda} +
    \Delta_0^{\alpha\beta}\kappa_\lambda \xi^{\mu\nu\lambda}\Bigg)
    \nonumber\\
\end{eqnarray}
Plugging this back into Eq.(\ref{sol54}) gives the complete linearized evolution
equation for $\varsigma^{\alpha\beta\mu\nu}$.

\subsection{Transverse Modes}
The linear stability and causality analysis presented in this work adheres to the
procedure outlined in de Brito \& Denicol's work \cite{Brito:2020, brito2022}. This
involves decomposing the linearized equations in Fourier space into longitudinal
(parallel to $\kappa^\mu$) and transverse (orthogonal to $\kappa^\mu$) components. This
method offers the advantage of decoupling the equations in the linear regime, allowing
them to be solved and analyzed independently and greatly simplifying the calculations
\cite{brito2022}. Due to the superposition principle of solutions to linear PDEs, this
procedure is equivalent to analyzing the complete 3-dimensional linearized equations
without decomposition.

It is beneficial to introduce a projector that is analogous to $\Delta^{\mu\nu}$
but with respect to $\kappa^\mu$:
\begin{equation}
    \Delta_\kappa^{\mu\nu} = g^{\mu\nu} - \frac{\kappa^\mu \kappa^\nu}{\kappa^2}
\end{equation}
where $\kappa^2$ is introduced to ensure normalization. Then, any 4-vector $A^\mu$
can be decomposed into a linear combination of the longitudinal and transverse parts:
\begin{equation}
    A^\mu = A_{||}\frac{\kappa^\mu}{\kappa} + A_\perp^\mu
\end{equation}
where $A_{||} = \kappa_\mu A^\mu / \kappa$ and $A_\perp^\mu =
\Delta_\kappa^{\mu\nu}A_\nu$. Similarly, a rank-2 tensor $A^{\mu\nu}$ can also be
decomposed as
\begin{equation}
    A^{\mu\nu} = A_{||}\frac{\kappa^\mu \kappa^\nu}{\kappa^2} + \frac13
    A_{\perp}\Delta_\kappa^{\mu\nu} + A_\perp^\mu \frac{\kappa^\nu}{\kappa} + +
    A_\perp^\nu \frac{\kappa^\mu}{\kappa} + A_\perp^{\mu\nu}
\end{equation}
where $A_{||} = \kappa_\mu \kappa_\nu A^{\mu\nu}/\kappa^2$, $A_{\perp}
= \Delta_\kappa^{\mu\nu}A_{\mu\nu}$, $A_\perp^\mu = \kappa^\lambda
\Delta_\kappa^{\mu\nu} A_{\lambda\nu}/\kappa$, and $A_\perp^{\mu\nu} =
\Delta_\kappa^{\mu\nu,\alpha\beta}A_{\alpha\beta}$. Here, we defined the rank-2
$\kappa$-projector to be
\begin{equation}
    \Delta_\kappa^{\mu\nu,\alpha\beta} =
    \frac12\bigg(\Delta_\kappa^{\mu\alpha}\Delta_\kappa^{\nu\beta}
    + \Delta_\kappa^{\mu\beta}\Delta_\kappa^{\nu\alpha} -
    \frac23\Delta_\kappa^{\mu\nu}\Delta_\kappa^{\alpha\beta}\bigg)
\end{equation}
In this section, we will analyze the linear stability and causality of the transverse
components of third-order regularized hydrodynamics for $m=0$. We will discuss two cases:
in the first, the wave vector $\textbf{k}$ is parallel to the background fluid velocity
$\textbf{v}$, while in the second, the wave vector is orthogonal to $\textbf{v}$.

\subsubsection{Case 1: $\textbf{k}$ is parallel to $\textbf{v}$}
\begin{figure*}
    \centering
    \includegraphics[width = 1.\linewidth]{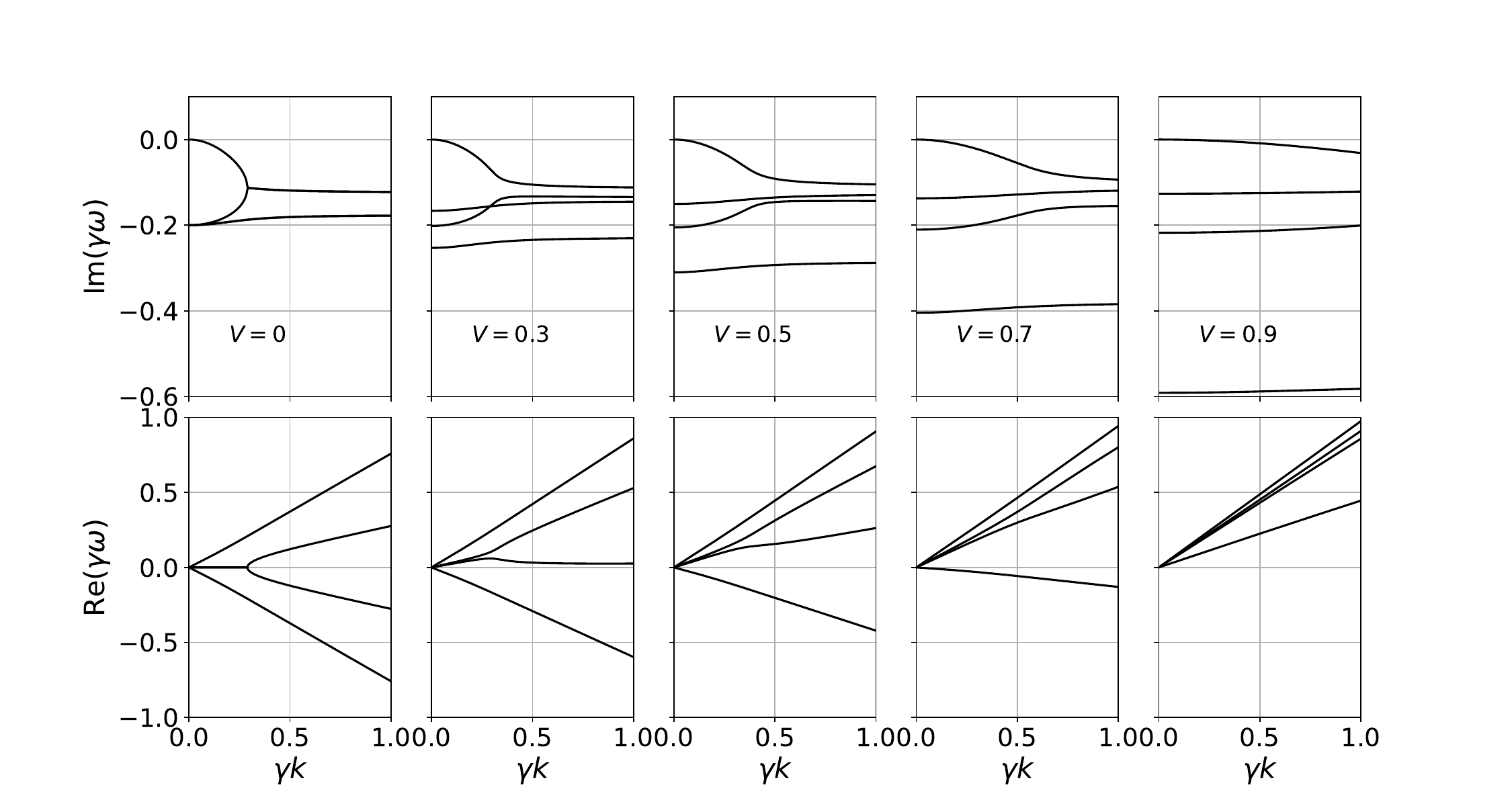}
    \caption{Real and Imaginary parts of the transverse modes of the massless third-order
    hydrodynamics without conservation of net particle number, in the case of fluid
    velocity vector being parallel to the wave vector. The relaxation time is chosen
    to be $\tau_R = 5$.}
    \label{fig:3rd_trans_stab}
\end{figure*}

For simplicity and without loss of generality, we will assume that $\textbf{k}$ and
$\textbf{v}$ are both in the $x$-axis:
\begin{equation}\label{para1}
\begin{aligned}
    & u_0^\mu = \gamma(1, v, 0, 0)\\
    & k^\mu = (\omega, k, 0, 0)
\end{aligned}
\end{equation}
It immediately follows that
\begin{equation}\label{para2}
\begin{aligned}
    & \Omega = \gamma(vk-\omega)\\
    & \kappa^2 = \gamma^2 (k-v\omega)^2
\end{aligned}
\end{equation}
Note that the first equation in Eq.(\ref{conserv_lin_fin}), which corresponds
to the energy conservation law, is a scalar equation. Thus it is purely longitudinal
and does not contribute to the transverse analysis. The transverse component of the
momentum conservation law and the moment equations can be easily obtained by applying
the projector $\Delta_\kappa^{\mu\nu}$ and $\kappa^\mu$. Doing so gives us
\begin{equation}
\begin{aligned}
    & \Omega \varepsilon_0 \delta u_\perp^\mu + \frac34 \kappa \delta \pi_\perp^\mu = 0\\
    & \left(i\Omega + \frac{1}{\tau_R}\right)\delta\pi_{\perp}^\mu +
    \frac{4}{15}i\kappa\varepsilon_0  \delta u_{\perp}^\mu + i\kappa \xi_\perp^\mu = 0 \\
    & \left(i\Omega + \frac{1}{\tau_R}\right)\xi_\perp^\mu +
    \frac{8}{35}i\kappa\delta\pi_\perp^\mu + i\kappa\varsigma_\perp^\mu = 0\\
    & \left(i\Omega + \frac{1}{\tau_R}\right)\varsigma_\perp^\mu +
    \frac{5}{21}i\kappa\xi_\perp^\mu = 0
\end{aligned}
\end{equation}
where we defined $\xi_\perp^\mu = \kappa_\alpha \kappa_\lambda \Delta_{\nu,\kappa}^\mu
\xi^{\alpha\lambda\nu}/\kappa^2$ and $\varsigma_\perp^\mu = \kappa_\alpha \kappa_\beta
\kappa_\lambda \Delta_{\nu,\kappa}^\mu \varsigma^{\alpha\beta\lambda\nu}/\kappa^3$. This
can be written in the following matrix form:
\begin{equation}
\begin{pmatrix}
\Omega & \frac34 \kappa & 0 & 0\\
\frac{4}{15}i\kappa & i\Omega+\frac{1}{\tau_R} & i\kappa & 0 \\
0 & \frac{8}{35}i\kappa & i\Omega + \frac{1}{\tau_R} & i\kappa\\
0 & 0 & \frac{5}{21}i\kappa & i\Omega + \frac{1}{\tau_R}
\end{pmatrix}
\begin{pmatrix}
\varepsilon_0 \delta u_\perp^\mu \\
\delta \pi_\perp^\mu\\
\xi_\perp^\mu \\
\varsigma_\perp^\mu
\end{pmatrix} = 0
\label{eq:trans_vec_eq}
\end{equation}
We require that the determinant of the $4 \times 4$ matrix be zero to obtain non-trivial
solutions, the resulting equation is the dispersion relation. However, we should note
that the dispersion relation is extremely complicated, even displaying the leading-order
terms is not feasible. Therefore, we will only present the numerical solutions to the
dispersion relation shown in Fig.~\ref{fig:3rd_trans_stab}, assuming $\tau_R = 5$
in the unit of $\tau_\eta$ \cite{brito2022, Denicol:2012cn}.
This particular value for the shear relaxation time
$\tau_R$ is calculated from the Boltzmann equation in the ultra-relativistic limit,
using the 14 moments approximation.
However, since the matrix is linear in $1/\tau_R$, $\Omega$ and $\kappa$, the value of
$\tau_R$ does not really matter in the current and the subsequent linear analysis. 
We chosen value for $\tau_R$ is just to facilitate the comparison between our results and
those in Refs.~\cite{brito2022, Denicol:2012cn} by having a common scale.

To determine whether these solutions are linearly stable, we first take a look at the
plane waves formula (Eq.(\ref{planewave})):
\begin{equation}\label{damp1}
\begin{aligned}
\phi &\sim e^{i(kx-\omega t)} = e^{ikx} e^{-i\omega_r t} e^{\omega_i t}
\end{aligned}
\end{equation}
where $\omega = \omega_r + i \omega_i$ is complex. Note that the first two exponential
terms are simply oscillating waves, therefore only the third term contributes to the
damping, and thus, stability. To ensure exponential suppression of Eq.(\ref{damp1})
for $t \geq 0$, it is necessary that $\omega_i$ be less than or equal to zero. Thus,
in general, stability requires
\begin{equation}\label{req1}
    \omega_i \leq 0
\end{equation}
for all $t \geq 0$. 
The determinant of the matrix in Eq.(\ref{eq:trans_vec_eq}) 
results in a 4-th order polynomial in $\omega$, thus we should expect to obtain four modes. 
Indeed,
Fig.~\ref{fig:3rd_trans_stab} shows four distinct curves, two of which have
the same imaginary parts for static fluids, i.e. $v=0$. As one can easily see,
all the modes have non-positive imaginary parts for small $k$.
We have also ascertained that the imaginary parts of all 4 modes 
become non-positive constants for large $k$.

\begin{figure}
    \centering
    \includegraphics[width = 1.0\linewidth]{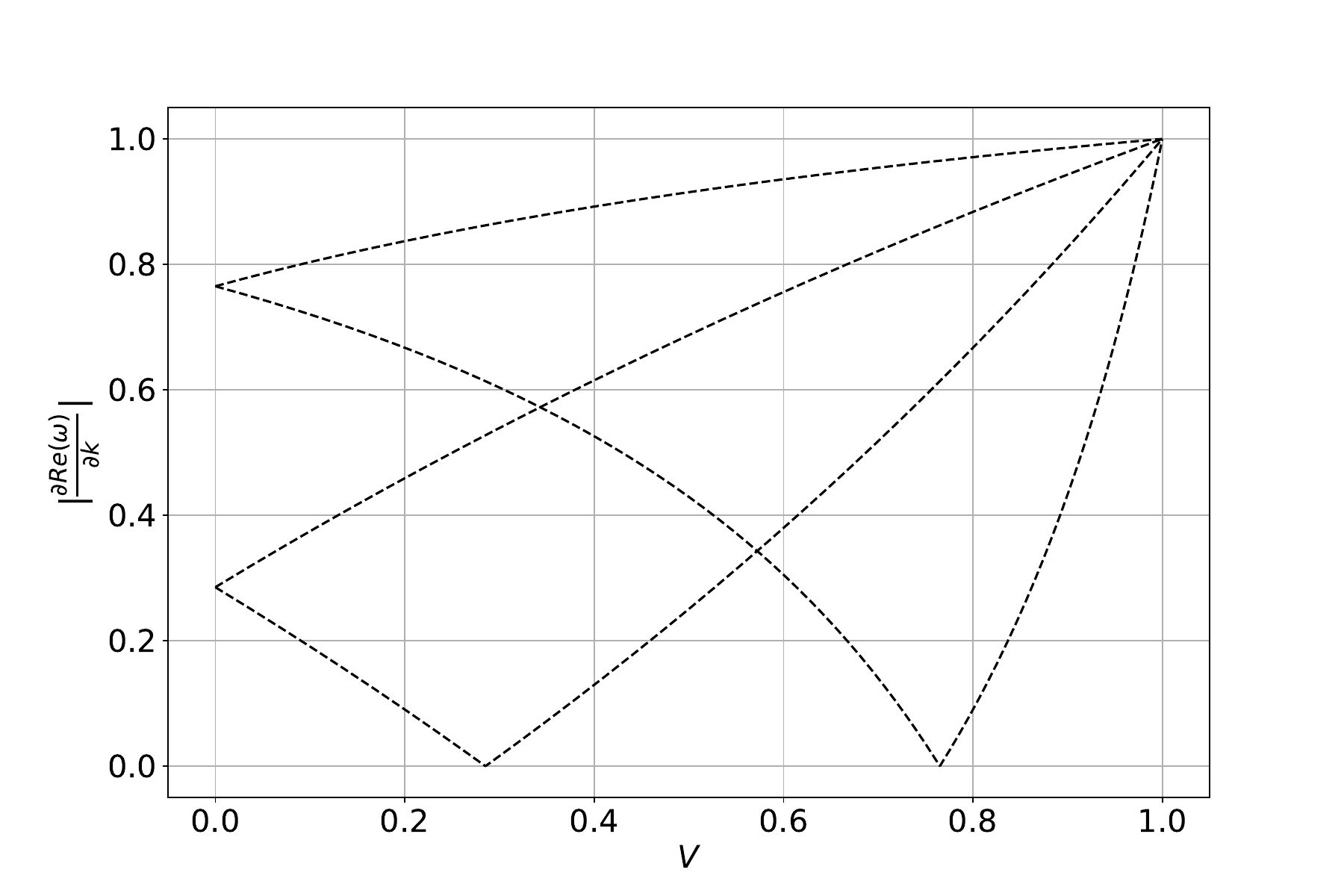}
    \caption{Magnitude of the group velocity for the transverse modes of the massless
    third-order hydrodynamics without conservation of net particle number, as a
    function of the fluid velocity $v$ in the large $k$ limit and with $\tau_R=5$,
    in the case of the fluid velocity vector is parallel to the wave vector.}
    \label{fig:3rd_trans_caus}
\end{figure}
For the causality analysis, 
we plot the asymptotic behaviour of the group velocity of the 4 modes in 
Figure \ref{fig:3rd_trans_caus}. In the large $k$ limit,
the magnitude of the group velocity remains 
subluminal for all values of the fluid velocity $v$.
Thus, the linear theory is causal.

\subsubsection{Case 2: $\textbf{k}$ is orthogonal to $\textbf{v}$}
We will now discuss the second case in which the wave vector is orthogonal to the
fluid velocity vector. Without loss of generality, we will assume that \textbf{V}
is still in the $x$-axis, but $\textbf{k}$ is now in the $y$-axis:
\begin{equation}\label{ortho1}
\begin{aligned}
    & u_0^\mu = \gamma(1, v, 0, 0)\\
    & k^\mu = (\omega, 0, k, 0)
\end{aligned}
\end{equation}
It follows that
\begin{equation}\label{ortho2}
\begin{aligned}
    & \Omega = -\gamma\omega\\
    & \kappa^2 = \gamma^2 v^2 \omega^2 + k^2
\end{aligned}
\end{equation}

\begin{figure}
    \centering
    \includegraphics[width = 1.0\linewidth]{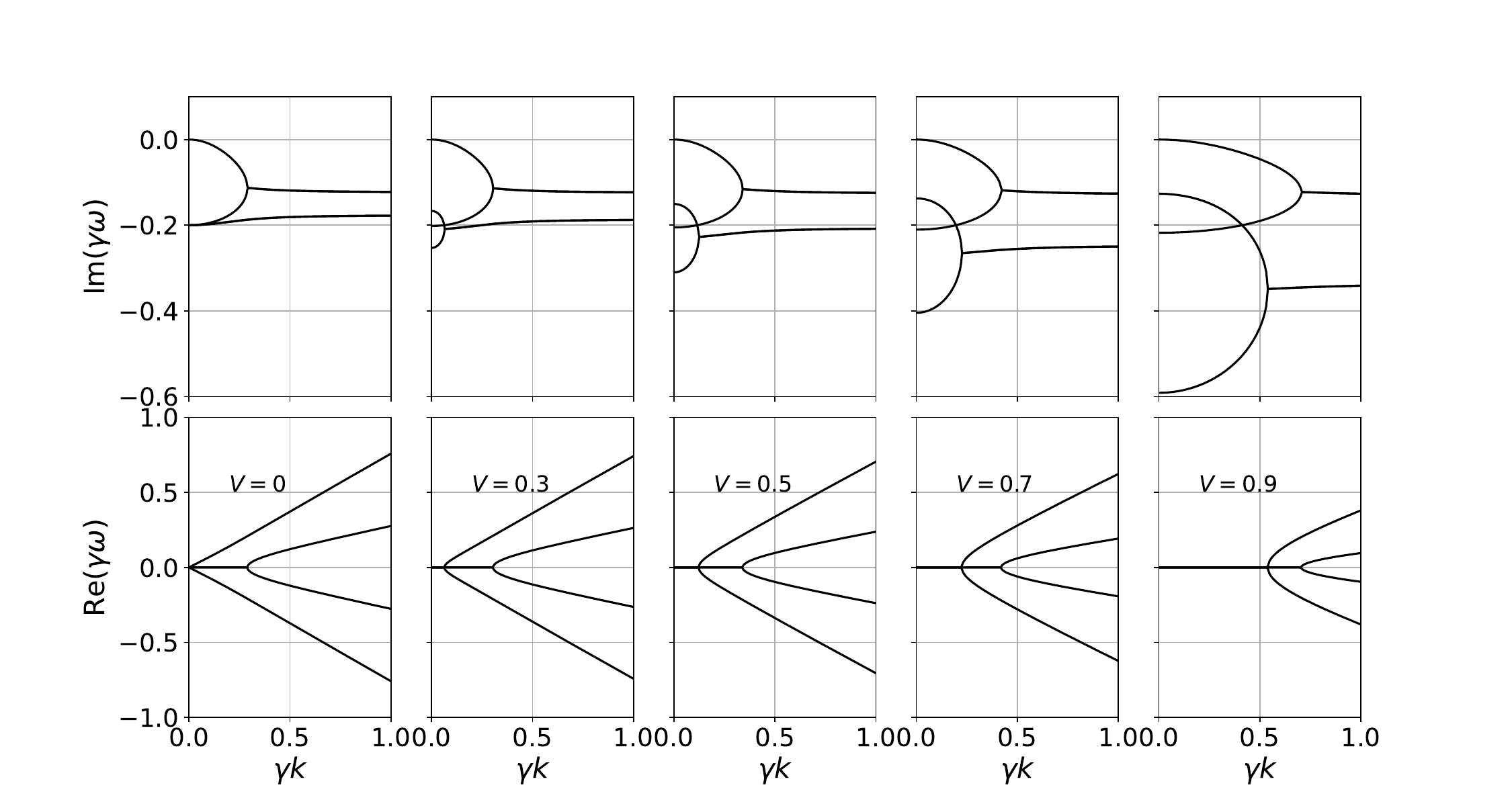}
    \caption{Real and Imaginary parts of the transverse modes of the massless third-order
    hydrodynamics without conservation of net particle number, in the case of fluid
    velocity vector being orthogonal to the wave vector and with $\tau_R = 5$.}
    \label{fig:3rd_trans_stab_yk}
\end{figure}

It is then straightforward to obtain the solutions for this case by substituting
Eq.(\ref{ortho2}) into the dispersion relation and then solving it numerically. The
results are shown in Fig.~\ref{fig:3rd_trans_stab_yk}. From the figure, we can
again see that all the modes are linearly stable as their imaginary parts are always
non-positive for small $k$, regardless of the background fluid velocity. As before,
we can further extend the linear stability of the modes to all $k \geq 0$ from the
asymptotic behavior of the modes which asymptote to contant negative values. 

\begin{figure}
    \centering
    \includegraphics[width = 1.0\linewidth]{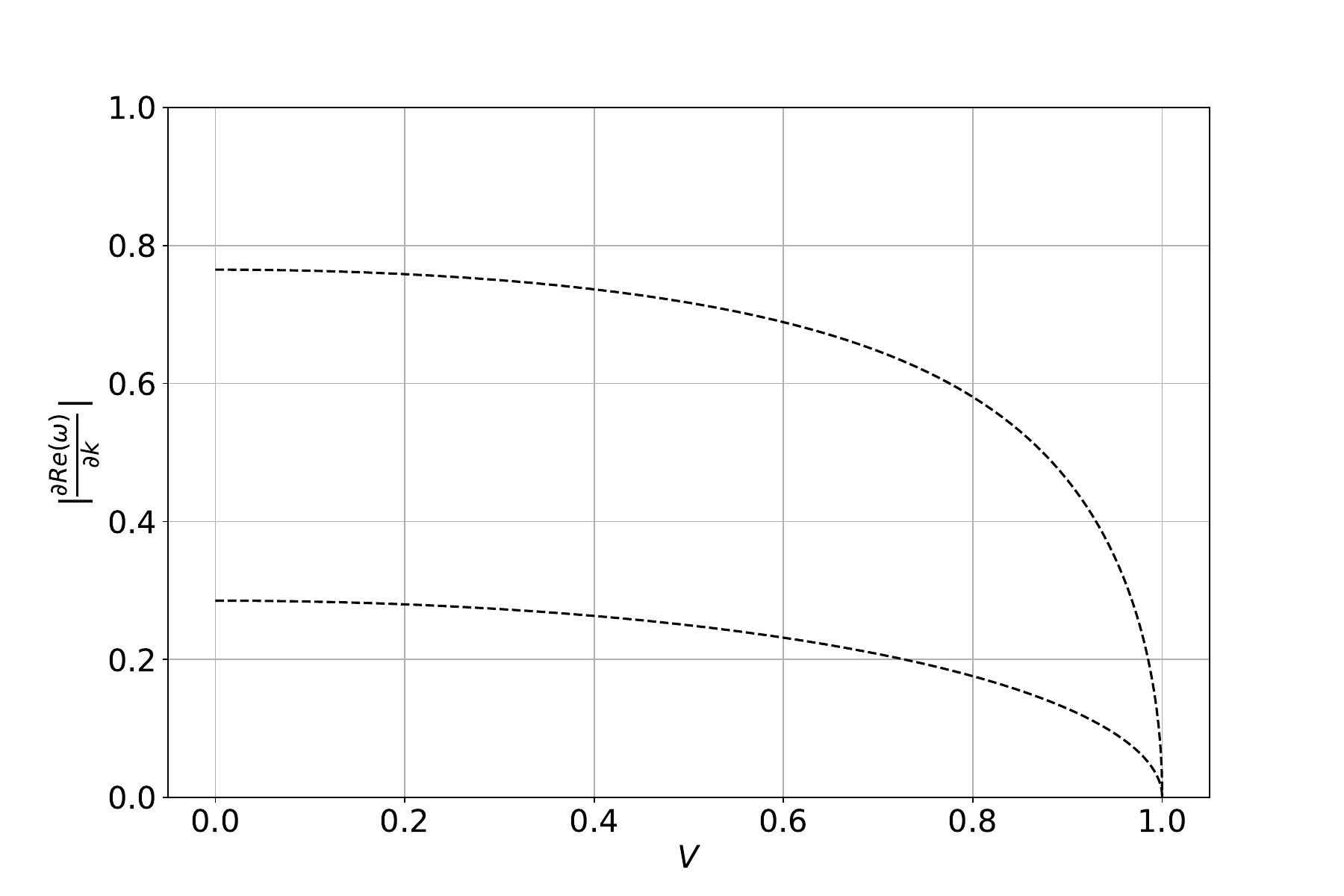}
    \caption{Magnitude of the group velocity for the transverse modes of the massless
    third-order hydrodynamics without conservation of net particle number, as a
    function of the fluid velocity $v$ in the large $k$ limit and with $\tau_R=5$,
    in the case of fluid velocity vector being orthogonal to the wave vector.}
    \label{fig:3rd_trans_caus_yk}
\end{figure}

Fig.~\ref{fig:3rd_trans_caus_yk} shows asymptotic group velocity
as a function of $v$. Note that there are only two curves for four solutions. This is
because the group velocities for each pair of solutions are only off by a sign. Since
the $y$-axis is the absolute value of the group velocity, both solutions coincide in
this case. Also, note that both curves approach zero when the fluid velocity reaches
the speed of light. This is expected since the plane wave propagates in the orthogonal
direction with respect to the fluid flow. As the fluid moves faster and faster, the
wave is eventually ``dragged'' by the fluid flow under the effect of shear viscosity
and moves in the fluid flow direction eventually, resulting in zero group velocity in
the orthogonal direction.

\subsection{Longitudinal Modes}
\subsubsection{Case 1: $\textbf{k}$ is parallel to $\textbf{v}$}
\begin{figure}[h]
    \centering
    \includegraphics[width = 1.0\linewidth]{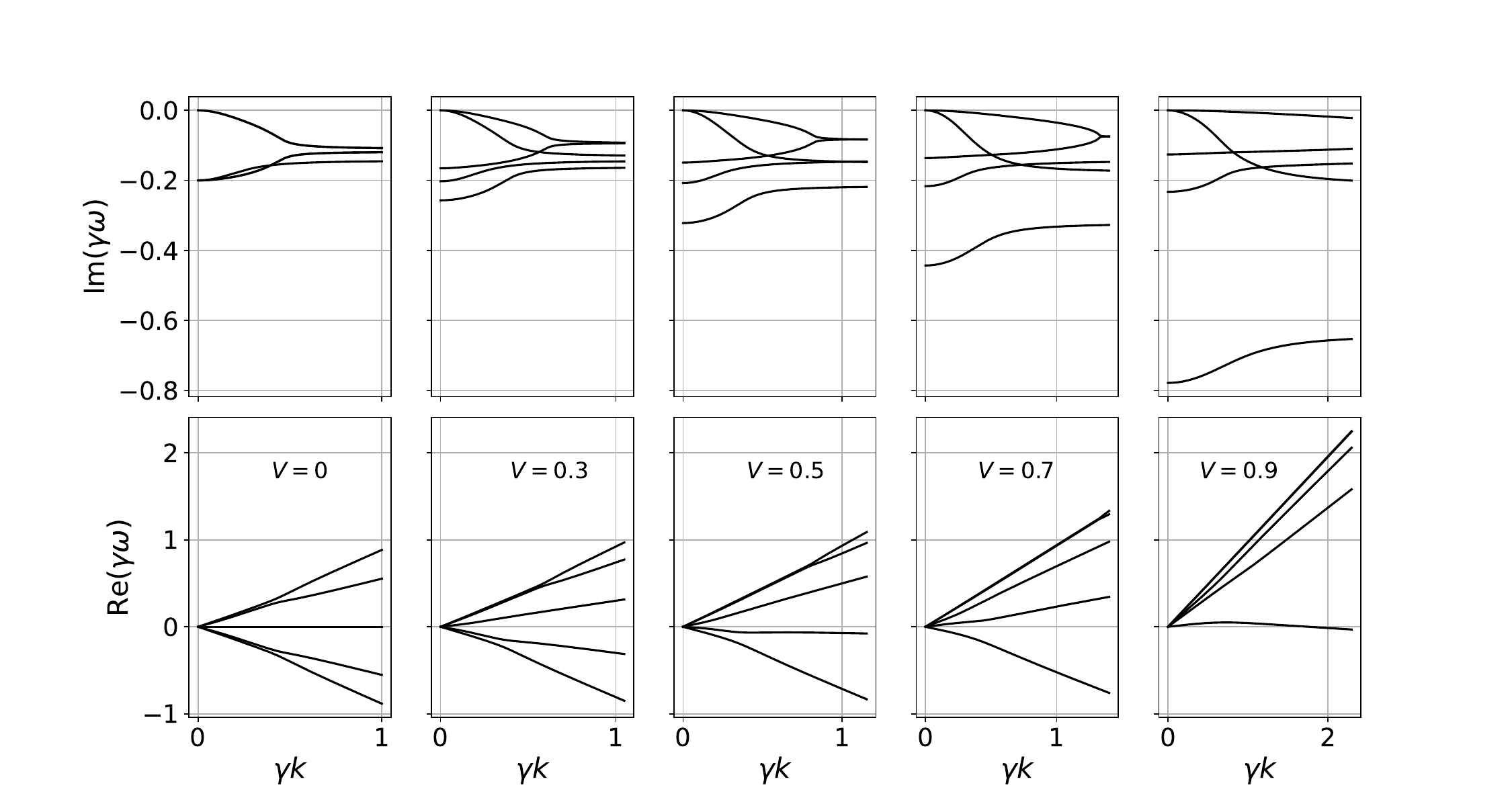}
    \caption{Real and Imaginary parts of the longitudinal modes of the massless
    third-order hydrodynamics without conservation of net particle number, in the case
    of fluid velocity vector being parallel to the wave vector and with $\tau_R = 5$.}
    \label{fig:3rd_long_stab}
\end{figure}

Similar to the second-order case, the first step is to obtain the longitudinal components
of the conservation laws and the equations for $\pi^{\mu\nu}$, $\xi^{\lambda\mu\nu}$,
and $\varsigma^{\alpha\beta\mu\nu}$. Applying $\kappa^\mu \kappa^\nu$ and $\kappa^\mu$
to the corresponding equations, we get
\begin{equation}
\begin{aligned}
    & \Omega \delta\varepsilon + \frac43 \varepsilon_0 \kappa \delta u_{||} = 0\\
    & \Omega \varepsilon_0 \delta u_{||} + \frac14 \kappa \delta \varepsilon + \frac34 \kappa
    \delta \pi_{||} = 0\\
    &\left(i\Omega + \frac{1}{\tau_R}\right)\delta\pi_{||} + \frac{16}{45}i\varepsilon_0
    \kappa \delta u_{||} + i\kappa \xi_{||} = 0\\
    & \left(i\Omega + \frac{1}{\tau_R}\right)\xi_{||} + \frac{9}{35}i\kappa
    \delta\pi_{||} + i\kappa \varsigma_{||} = 0 \\
    & \left(i\Omega + \frac{1}{\tau_R}\right)\varsigma_{||} + \frac{16}{63}i\kappa
    \xi_{||} = 0
\end{aligned}
\end{equation}
where we defined $\xi_{||} = \kappa_\alpha \kappa_\beta \kappa_\lambda
\xi^{\alpha\beta\lambda}/\kappa^3$ and $\varsigma_{||} = \kappa_\alpha \kappa_\beta
\kappa_\mu \kappa_\nu \varsigma^{\alpha\beta\mu\nu}/\kappa^4$.
Note that we have included the purely-longitudinal energy conservation law in this
system of equations. Written in the matrix form, this is equivalent to
\begin{equation}
\begin{pmatrix}
\Omega & \frac43\kappa & 0 & 0 & 0\\
\frac{\kappa}{4} & \Omega & \frac34 \kappa & 0 & 0 \\
0 & \frac{16}{45}i\kappa & i\Omega + \frac{1}{\tau_R} & i\kappa & 0\\
0 & 0 & \frac{9}{35}i\kappa & i\Omega + \frac{1}{\tau_R} & i\kappa \\
0 & 0 & 0 & \frac{16}{63}i\kappa & i\Omega + \frac{1}{\tau_R}
\end{pmatrix}
\begin{pmatrix}
\delta\varepsilon\\
\varepsilon_0 \delta u_{||} \\
\delta \pi_{||}\\
\xi_{||} \\
\varsigma_{||}
\end{pmatrix} = 0
\end{equation}
Since $\Omega$ is of fifth-order in the determinant, we should expect to obtain five
modes. Indeed, Fig.~\ref{fig:3rd_long_stab} shows that all five solutions are linearly
stable since for small $k$, their imaginary parts are all non-positive for various background fluid
velocities. Again, one can numerically show that all 5 modes asymptote to non-positive
contants.

\begin{figure}[h]
    \centering
    \includegraphics[width = 1.0\linewidth]{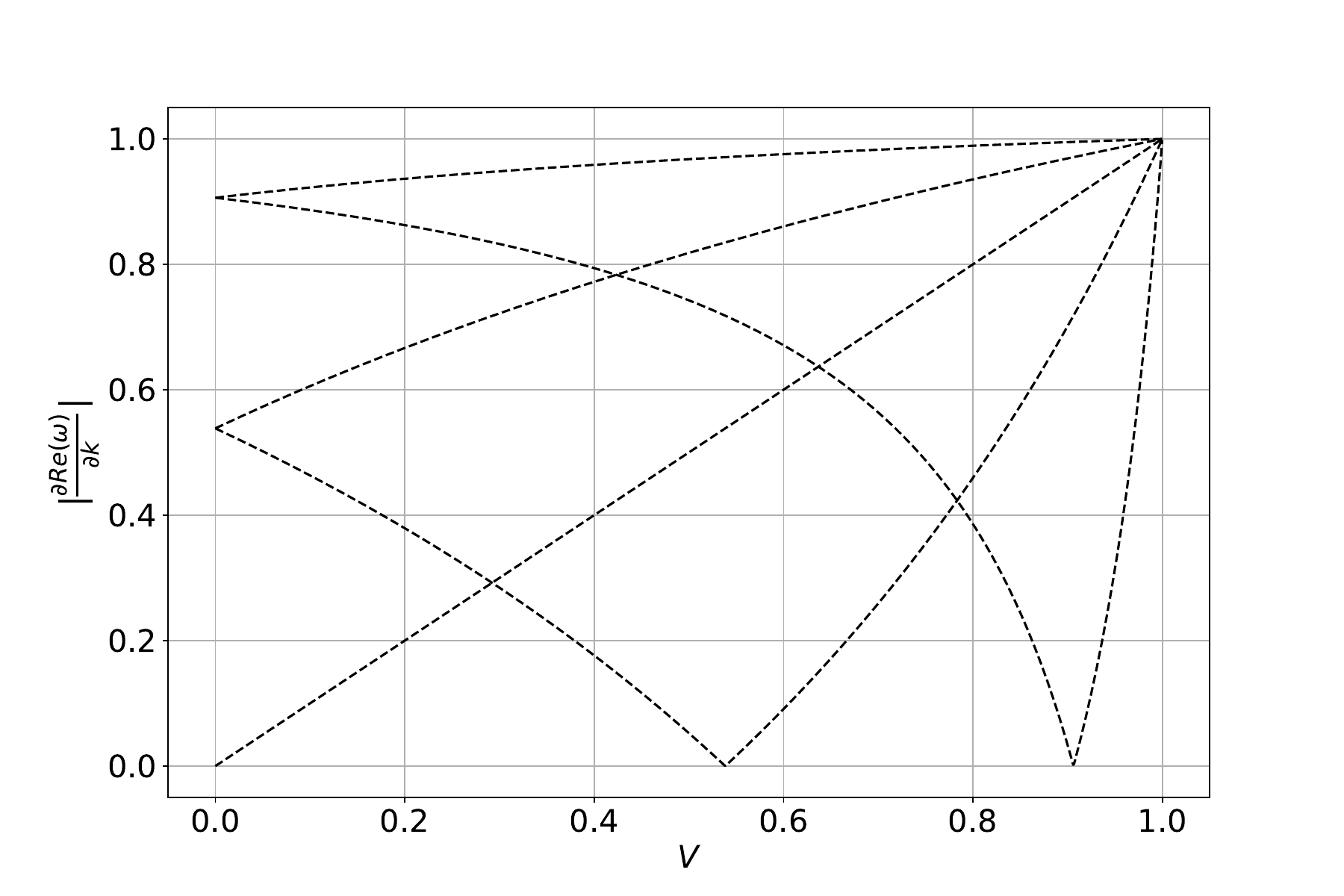}
    \caption{Magnitude of the group velocity for the longitudinal modes of the
    massless third-order hydrodynamics without conservation of net particle number,
    as a function of the fluid velocity $v$ in the large $k$ limit and with $\tau_R=5$,
    in the case of fluid velocity vector being parallel to the wave vector.}
    \label{fig:3rd_long_caus}
\end{figure}

In Fig.~\ref{fig:3rd_long_caus}, we show asymptotic group velocities of 5 modes
as a function of $v$.
One can see that all solutions are linearly causal since the magnitude of the group
velocity is less than 1 for all of them, in the large-$k$ limit. Also, note that the
straight diagonal line in the figure corresponds to a stationary mode in the fluid rest
frame since its group velocity is simply the fluid flow velocity.

\subsubsection{Case 2: $\textbf{k}$ is orthogonal to $\textbf{v}$}
\begin{figure}[h]
    \centering
    \includegraphics[width = 1.0\linewidth]{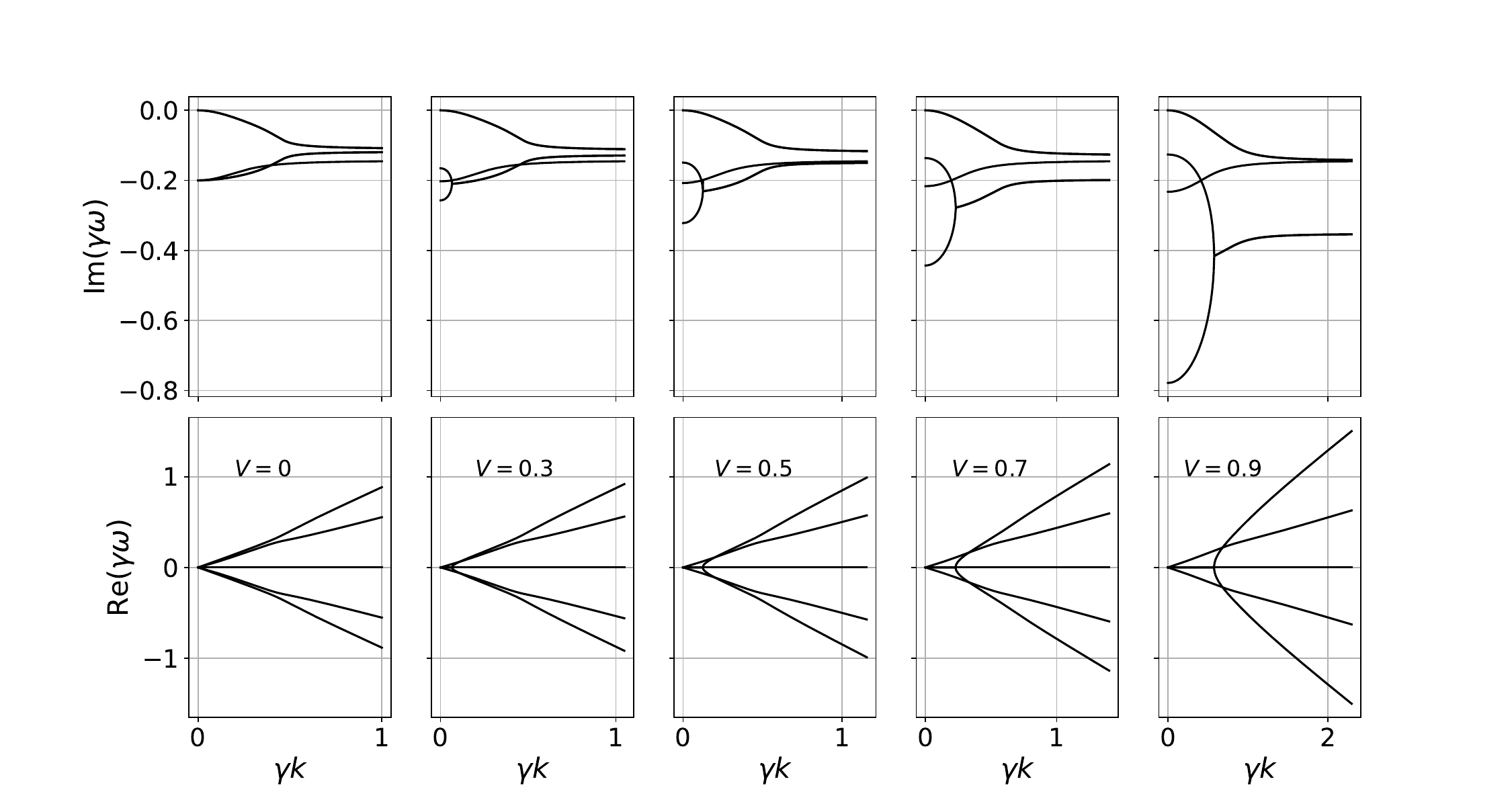}
    \caption{Real and Imaginary parts of the longitudinal modes of the massless
    third-order hydrodynamics without conservation of net particle number, in the case
    of fluid velocity vector being orthogonal to the wave vector and with $\tau_R = 5$.}
    \label{fig:3rd_long_stab_yk}
\end{figure}

As before, we insert Eq.(\ref{ortho2}) into the dispersion relation and solve
numerically for the solutions. Fig. (\ref{fig:3rd_long_stab_yk}) shows the result.
Note that two out of the five solutions have the same imaginary parts, and we can
see that all solutions are linearly stable since they all have non-positive imaginary
parts for small $k$. Again, we checked that all modes asymptote to non-positive constants
in the large $k$ limit.

\begin{figure}[h]
    \centering
    \includegraphics[width = 1.0\linewidth]{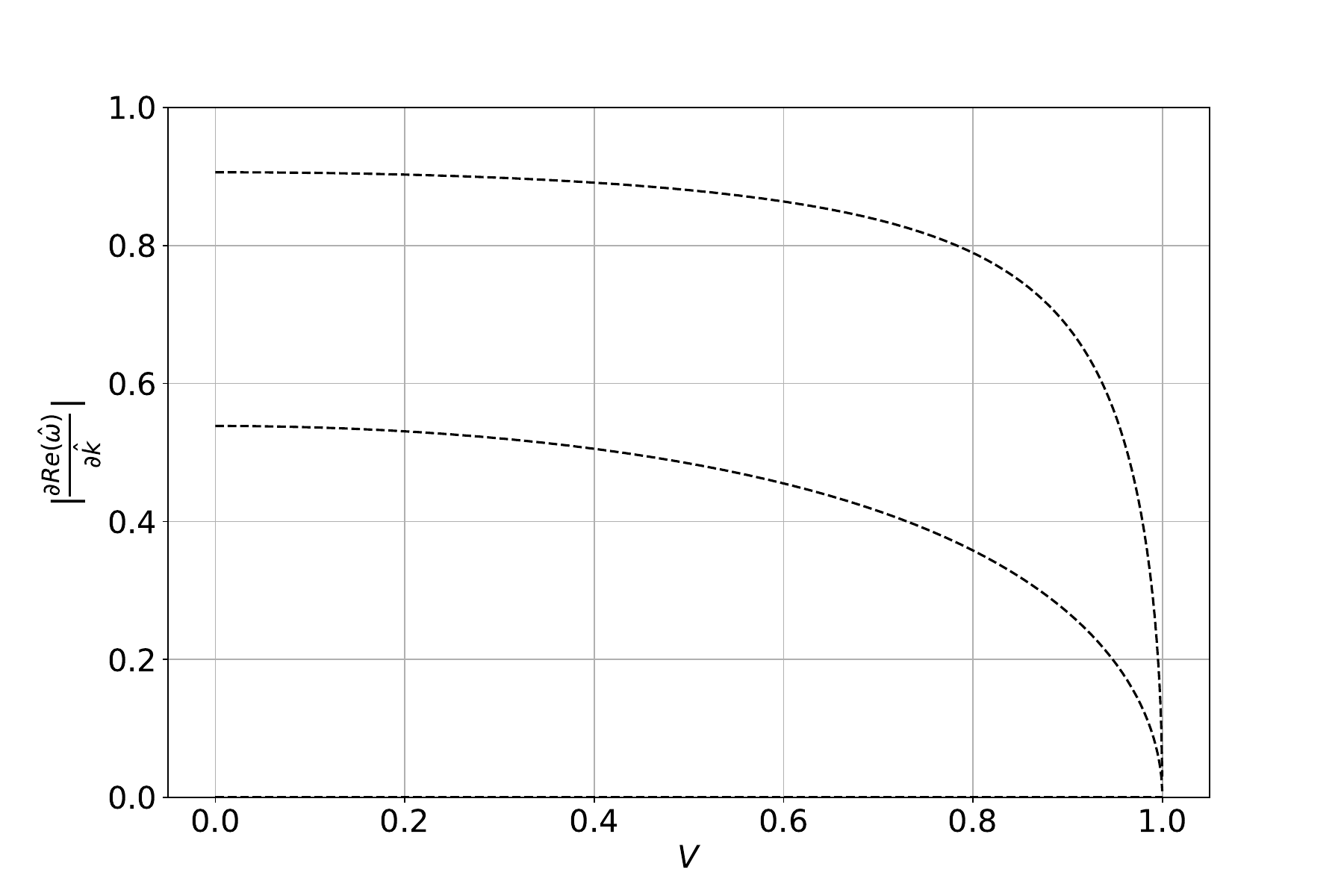}
    \caption{Magnitude of the group velocity for the longitudinal modes of the massless
    third-order hydrodynamics without conservation of net particle number, as a
    function of the fluid velocity $v$ in the large $k$ limit and with $\tau_R=5$,
    in the case of fluid velocity vector being orthogonal to the wave vector. Notice
    that there is a stationary mode with zero group velocity along the direction of
    wave's propagation.}
    \label{fig:3rd_long_caus_yk}
\end{figure}

To verify the causality of these solutions, we repeat the process from the previous
sections.  The group velocities of the solutions are shown
in Fig.~\ref{fig:3rd_long_caus_yk} as a function of the fluid flow velocity.
Note that there are three curves in this figure, one of them lies on the $x$-axis and
corresponds to the stationary mode with zero group velocity.

\section{Discussions and Conclusions}\label{sec:discussion}

The main results of this work are the derivation of the evolution equation for the general
energy-momentum moment of the phase space density function, introduction of the
regularized hydrodynamics, and the derivation and the analysis of the third-order
hydrodynamics.
As far as we can find out, this is the first time that the derivation of the evolution equation for 
a general energy-momentum moment has appeared in literature.

Our derivation of hydrodynamic equations from the general moment equations follows closely
the derivation of the regularized hydrodynamics by Struchtrup and Torrilhon in which
the Chapman-Enskog-like expansion is applied to the moments, not to the density function,
except for hydrodynamic variables.
In this way, we avoided the inherent ambiguity in the method of moments \cite{Denicol:2012es}
as well as possible acausality in the Chapman-Enskog method 
\cite{Jaiswal:20132,brito2022}. The third-order hydrodynamics 
unambigously derived this way includes additional rank-1 moment and rank-3 moment
as dynamic variable.

In recent literature, other versions of third-order theories appeared. 
The versions most closely related to ours are those from 
Refs.~\cite{brito2022,deBrito:2023tgb}.
The authors of Ref.~\cite{brito2022} proposed a third order theory
based on Ref.~\cite{Jaiswal:20132}
in which they promoted the gradient of $\pi^{\mu\nu}$
to a new hydrodynamic variable
\begin{equation}
    \nabla^{\langle\alpha}\pi^{\mu\nu\rangle} \rightarrow \rho^{\alpha\mu\nu}
\end{equation}
to eliminate the second-order gradients in the evolution equation of $\pi^{\mu\nu}$. This
is analogous to $\xi^{\lambda\mu\nu}$ we defined in Eqs.(\ref{eq:xi_and_varsigma}) and
(\ref{eq:xi_eq_m0}), but it was done in a heuristic way.
This situation was remedied
by the same authors in Ref.~\cite{deBrito:2023tgb}
where they derived the equations for the 3rd and the 4th moments 
using $\rho_{0}^{\mu_1\mu_2\mu_3}$ and $\rho_0^{\mu_1\mu_2\mu_3\mu_4}$,
while we use $\rho_{-1}^{\mu_1\mu_2\mu_3}$
and $\rho_{-2}^{\mu_1\mu_2\mu_3\mu_4}$. In their approach, all
$\rho_{r}^{\mu_1\cdots\mu_n}$ up to $n=4$ are proportional to $\rho_0^{\mu_1\cdots\mu_n}$
while ours clearly differ. Nevertheless, linear analysis should yield similar results.

To further analyze the properties of
this theory and for simplicity, we assume the particles to be massless.
A series of linear stability and causality analysis was then performed,
and we showed that all the modes of the massless third-order theory
are linearly stable and causal.

The hybrid method advocated in this work may be extended to higher orders.
However, given that the Chapman-Enskog expansion is asymptotic in nature \cite{PhysRevLett.56.1571},
and the fact that we will need to promote higher and higher order moments to be dynamic,
this path may not be a profitable one to study the effect of higher order moments.
Instead, one may consider resummation approaches such as the generalized hydrodynamics
formulated by B.C.~Eu \cite{eu1986modified,EU199265}.
Other ways to extend our method include 
applying it to systems with multiple species and multiple conserved charges, to 
spin hydrodynamics 
\cite{Vasak:1987um,Gao:2012ix,Hidaka:2016yjf,Florkowski:2017dyn,Shi:2020htn,Li:2020eon,Peng:2021ago,Hu:2021pwh,Sarwar:2022yzs}, and to the general-frame
theories with off-shell transport parameters \cite{Shokri:2020cxa,Kovtun:2019hdm}.
We will leave these as possible venues for further investigations.



\begin{acknowledgments}
We acknowledge the support of the Natural Sciences and Engineering Research Council of Canada (NSERC), 
[funding reference number SAPIN-2020-00048 
and SAPIN-2018-00024].

Cette recherche a
\'et\'e financ\'ee
par le Conseil de recherches en sciences naturelles et en 
g\'enie
du Canada (CRSNG), 
[num\'ero de r\'ef\'erence SAPIN-2020-00048 et
SAPIN-2020-00048].

%
%
S.J.~is grateful to H.~Struchtrup, M.~Kr\"oger, G.~Denicol and A.~Jaiswal for fruitful
discussions.

S.J.~also acknowledges insightful discussions with late B.C.~Eu whose work on non-equilibrium
statistical mechanics and generalized hydrodynamics inspired this paper.
\end{acknowledgments}

\newpage
\appendix

\section{On Projectors}
\label{app:projectors}

The definition of the rank $n$ projector is a tensor of rank $(n,n)$ that 
selects the symmetric and traceless part of a tensor or rank $(m,n)$ or
rank $(n,m)$.
The basic building block is 
the spatial metric tensor for a fluid cell moving with
the flow velocity $u^\mu$:
\begin{equation}
\Delta^{\mu}_{\nu} = g^{\mu}_{\nu} + u^\mu u_\nu
\end{equation}
which is the rank 1 projector. 
When applied to a 4-momentum, it gives
\begin{eqnarray}
p^\ave{\mu} \ampeq \Delta^{\mu}_{\nu}p^\nu
\non
\ampeq
p^\mu - (\calE_p) u^\mu
\end{eqnarray}
where $\calE_p = -u_\nu p^\nu$ is the time-component of the 4-vector in the fluid-cell
rest frame. From here on the angular bracke around indices indicate the symmetric and
traceless part of the tensor. 
For $n=2$
\begin{equation}
\Delta_{\nu_1\nu_2}^{\mu_1\mu_2}
= 
{1\over 2}
\left(
\Delta^{\mu_1}_{\nu_1} \Delta^{\mu_2}_{\nu_2}
+
\Delta^{\mu_1}_{\nu_2} \Delta^{\mu_2}_{\nu_1}
-
{2\over 3}\Delta_{\nu_1\nu_2}\Delta^{\mu_1\mu_2}
\right)
\end{equation}
and for $n=3$
\begin{eqnarray}
\Delta_{\nu_1\nu_2\nu_3}^{\mu_1\mu_2\mu_3}
\ampeq
{1\over 6}
\bigg[
\Delta_{\nu_1}^{\mu_1}\Delta_{\nu_2}^{\mu_2}\Delta_{\nu_3}^{\mu_3}
+ \Delta_{\nu_1}^{\mu_1}\Delta_{\nu_3}^{\mu_2}\Delta_{\nu_2}^{\mu_3}
+ \Delta_{\nu_2}^{\mu_1}\Delta_{\nu_1}^{\mu_2}\Delta_{\nu_3}^{\mu_3}
\nonumber\\ && {}\qquad
+ \Delta_{\nu_2}^{\mu_1}\Delta_{\nu_3}^{\mu_2}\Delta_{\nu_1}^{\mu_3}
+ \Delta_{\nu_3}^{\mu_1}\Delta_{\nu_1}^{\mu_2}\Delta_{\nu_2}^{\mu_3}
+ \Delta_{\nu_3}^{\mu_1}\Delta_{\nu_2}^{\mu_2}\Delta_{\nu_1}^{\mu_3}
\bigg]
\nl
- {1\over 15}
\bigg[
\Delta^{\mu_1\mu_2}
\left( \Delta_{\nu_1\nu_2}\Delta^{\mu_3}_{\nu_3}
+\Delta_{\nu_2\nu_3}\Delta^{\mu_3}_{\nu_1}
+\Delta_{\nu_3\nu_1}\Delta^{\mu_3}_{\nu_2} \right)
\nl\qquad
+
\Delta^{\mu_2\mu_3}
\left( \Delta_{\nu_1\nu_2}\Delta^{\mu_1}_{\nu_3}
+\Delta_{\nu_2\nu_3}\Delta^{\mu_1}_{\nu_1}
+\Delta_{\nu_3\nu_1}\Delta^{\mu_1}_{\nu_2} \right)
\nl\qquad
+
\Delta^{\mu_3\mu_1}
\left( \Delta_{\nu_1\nu_2}\Delta^{\mu_2}_{\nu_3}
+\Delta_{\nu_2\nu_3}\Delta^{\mu_2}_{\nu_1}
+\Delta_{\nu_3\nu_1}\Delta^{\mu_2}_{\nu_2} \right)
\bigg]
\end{eqnarray}

The above projectors are constructed in such a way that
they are symmetric and traceless in both $(\mu_1,\cdots,\mu_n)$
and $(\nu_1,\cdots,\nu_n)$.
For the sake of projecting $T^{\nu_1\cdots\nu_n}$ to $T^{\langle\mu_1\cdots\mu_n\rangle}$,
this is actually not necessary.
It turned out that we just need to make sure that the superscripted indices are symmetric and traceless.
In that case, the following recursive construction
works just as well as a projector \cite{kroeger}
\begin{eqnarray}
\tilde\Delta^{\mu_1\cdots\mu_n}_{\nu_1\cdots\nu_n}
\ampeq
{1\over n} \sum_{i=1}^n \Delta^{\mu_i}_{\nu_1}
\tilde\Delta^{\mu_1\cdots\mu_{i-1}\mu_{i+1}\cdots\mu_n}_{\nu_2\cdots\nu_n}
\nl
-
{2\over n(2n-1)}
\sum_{i=1}^n\sum_{j=i+1}^n \Delta^{\mu_i\mu_j} \Delta_{\nu_1\alpha}
\tilde\Delta^{\alpha\mu_1\cdots\mu_{i-1}\mu_{i+1}\cdots\mu_{j-1}\mu_{j+1}\cdots\mu_n}_{\nu_2\cdots\nu_n}
\label{eq:proj_kroeger}
\end{eqnarray}
This is explicitly constructed so that it is symmetric and traceless in
$(\mu_1,\cdots,\mu_n)$, but not necessarily in $(\nu_1,\cdots,\nu_n)$. 
We do have $\Delta^{\mu_1\mu_2}_{\nu_1\nu_2} = \tilde\Delta^{\mu_1\mu_2}_{\nu_1\nu_2}$,
but for $n > 2$, $\tilde\Delta^{\mu_1\cdots\mu_n}_{\nu_1\cdots\nu_n}$ 
is neither symmetric nor traceless in $(\nu_1,\cdots,\nu_n)$.
As an example, applying this projector to $q^{\nu_1}P^{\nu_2\cdots\nu_n}$ yields 
\begin{eqnarray}
q^{\langle\mu_1}P^{\mu_2\cdots\mu_n\rangle}
\ampeq
\tilde\Delta^{\mu_1\cdots\mu_n}_{\nu_1\cdots\nu_n}
q^{\nu_1}P^{\nu_2\cdots\nu_n}
\non
\ampeq
{1\over n} \sum_{i=1}^n q^{\langle\mu_i\rangle}
P^{\langle\mu_1\cdots\mu_{i-1}\mu_{i+1}\cdots\mu_n\rangle}
\nl
-
{2\over n(2n-1)}
\sum_{i=1}^n\sum_{j=i+1}^n \Delta^{\mu_i\mu_j} q_{\langle\alpha\rangle}
P^{\langle\alpha\mu_1\cdots\mu_{i-1}\mu_{i+1}\cdots\mu_{j-1}\mu_{j+1}\cdots\mu_n\rangle}
\label{eq:q_P_n}
\end{eqnarray}
where $q^{\nu_1}$ is an arbitrary 4-vector and
$P^{\nu_2\cdots\nu_n}$ is an arbitrary rank-$(n{-}1)$ tensor. 
Eq.(\ref{eq:p1_pn}) is a particular example of this identity.

The full rank-$n$ projector that are symmetric and tracelss in both sets of indices
can be recursively built by 
averaging Eq.(\ref{eq:proj_kroeger}) over $n$ different choices of $\nu_k$ that can be
isolated
\begin{eqnarray}
\Delta^{\mu_1\cdots\mu_n}_{\nu_1\cdots\nu_n}
\ampeq
{1\over n^2}
\sum_{i=1}^n\sum_{k=1}^n
\Delta^{\mu_i}_{\nu_k}
\Delta^{\mu_1\cdots\mu_{i-1}\mu_{i+1}\cdots\mu_n}_{\nu_1\cdots\nu_{k-1}\nu_{k+1}\cdots\nu_n}
\nl
-{2\over n^2(2n-1)}
\sum_{i=1}^n
\sum_{j=i+1}^n
\Delta^{\mu_i\mu_j}
\sum_{k=1}^n
\Delta_{\nu_k\alpha}
\Delta^{\alpha\mu_1\cdots\mu_{i-1}\mu_{i+1}\cdots\mu_{j-1}\mu_{j+1}\cdots\mu_n}_{\nu_1\cdots
\nu_{k-1}\nu_{k+1}\cdots\nu_n}
\label{eq:proj_recursion}
\end{eqnarray}
The right hand side is explicity constructed in such a way that 
it is symmetric and traceless in $(\mu_1,\cdots,\mu_n)$.
It looks only symmetric in $(\nu_1,\cdots,\nu_n)$, but
it would be also traceless provided that the following identities holds
\begin{eqnarray}
\Delta^{\alpha\mu_2\cdots\mu_{n-1}}_{\alpha\nu_2\cdots\nu_{n-1}}
=
{(2n-1)\over (2n-3)}
\Delta^{\mu_2\cdots\mu_{n-1}}_{\nu_2\cdots\nu_{n-1}}
\label{eq:contraction}
\end{eqnarray}
\begin{eqnarray}
\sum_{i=1}^n
\Delta^{\mu_i\alpha}
\Delta^{\mu_1\cdots\mu_{i-1}\mu_{i+1}\cdots\mu_n}_{
\alpha\nu_3\cdots\nu_{n}}
\ampeq
{2\over (2n-3)}
\sum_{i=1}^n
\sum_{j=i+1}^n
\Delta^{\mu_i\mu_j}
\Delta^{\mu_1\cdots\mu_{i-1}\mu_{i+1}\cdots\mu_{j-1}\mu_{j+1}\cdots\mu_n}_{
\nu_3\cdots\nu_{n}}
\label{eq:Key_Id}
\end{eqnarray}
and
\begin{eqnarray}
\Delta^{\mu_2\cdots\mu_{n}}_{\nu_2\cdots\nu_{n}}
\ampeq
{(2n-1)\over n(n-1)}
\Bigg(
{1\over (2n-3)}
\sum_{i=2}^{n}\sum_{k=2}^{n}
\Delta^{\mu_i}_{\nu_k}
\Delta^{\mu_2\cdots\mu_{i-1}\mu_{i+1}\cdots\mu_{n}}_{
\nu_2\cdots\nu_{k-1}\nu_{k+1}\cdots\nu_{n}}
\nlqqqq
-{1\over (2n-1)}
\sum_{j=2}^{n}
\sum_{k=2}^{n}
\Delta^{\mu_j\beta}
\Delta_{\nu_k\alpha}
\Delta^{\alpha\mu_2\cdots\mu_{j-1}\mu_{j+1}\cdots\mu_{n}}_{
\beta\nu_2\cdots \nu_{k-1}\nu_{k+1}\cdots\nu_{n}}
\Bigg)
\label{eq:proj_id_2}
\end{eqnarray}
These identities can be proven by using the following mathematical induction strategy:
\begin{enumerate}
\setlength{\itemsep}{0pt}
 \item Show that Eqs.(\ref{eq:proj_recursion}), (\ref{eq:contraction}), (\ref{eq:Key_Id}), 
 and (\ref{eq:proj_id_2}) are valid for $n=2$.
 \item Assume that Eqs.(\ref{eq:contraction}), (\ref{eq:Key_Id}), 
 and (\ref{eq:proj_id_2}) are valid for an arbitrary $n$.

 \item Show that the projector recursion relationship, Eq.(\ref{eq:proj_recursion}), is valid
 for this $n$.

 \item Using Eqs.(\ref{eq:proj_recursion} -- \ref{eq:proj_id_2}) for $n$,
 show that Eqs.(\ref{eq:contraction} -- \ref{eq:proj_id_2}) are valid for $n+1$.
\end{enumerate}
Due to the symmetry between $\mu$'s and $\nu$'s, the following is equivalent to
Eq.(\ref{eq:Key_Id})
\begin{eqnarray}
\sum_{i=1}^n
\Delta_{\nu_i\alpha}
\Delta_{\nu_1\cdots\nu_{i-1}\nu_{i+1}\cdots\nu_n}^{
\alpha\mu_3\cdots\nu_{n}}
\ampeq
{2\over (2n-3)}
\sum_{i=1}^n
\sum_{j=i+1}^n
\Delta_{\nu_i\nu_j}
\Delta_{\nu_1\cdots\nu_{i-1}\nu_{i+1}\cdots\nu_{j-1}\nu_{j+1}\cdots\nu_n}^{
\mu_3\cdots\mu_{n}}
\label{eq:Key_Id2}
\end{eqnarray}
By combining Eq.(\ref{eq:proj_recursion}) and Eq.(\ref{eq:Key_Id2}),
we can have a recursion relationship which is explicitly symmetric under $\mu_i
\leftrightarrow \nu_i$ swapping:
\begin{eqnarray}
\lefteqn{
\Delta^{\mu_1\cdots\mu_n}_{\nu_1\cdots\nu_n}
} &&
\non
\ampeq
{1\over n^2}
\sum_{i=1}^n\sum_{k=1}^n
\Delta^{\mu_i}_{\nu_k}
\Delta^{\mu_1\cdots\mu_{i-1}\mu_{i+1}\cdots\mu_n}_{\nu_1\cdots\nu_{k-1}\nu_{k+1}\cdots\nu_n}
\nl
-{4\over n^2(2n-1)(2n-3)}
\sum_{l=1}^n
\sum_{m=l+1}^n
\sum_{i=1}^n
\sum_{j=i+1}^n
\Delta_{\nu_l\nu_m}
\Delta^{\mu_i\mu_j}
\Delta^{\mu_1\cdots\mu_{i-1}\mu_{i+1}\cdots\mu_{j-1}\mu_{j+1}\cdots\mu_n}_{\nu_1\cdots\nu_{l-1}
\nu_{l+1}\cdots\nu_{m-1}\nu_{m+1}\cdots\nu_n}
\non
\end{eqnarray}

\section{Irreducible Polynomials}
\label{app:polynomials}

In the rest-frame of the fluid cell,
the irreducible tensors of rank $n$ is defined as
the symmetric and traceless combinations of the $n$ factors of $p^{m}$
where $m = 1,2,3$ is the spatial index.
For instance, the rank-1 tensor is just $p^m$ and the rank-2 tensor is
\begin{equation}
p^{\langle m_1} p^{m_2\rangle}
=
p^{m_1} p^{m_2} - {\Delta^{m_1 m_2}\over 3}p^2
\end{equation}
where $\Delta^{m_1 m_2} = \delta^{m_1 m_2}$ is the spatial metric tensor in the rest frame
and $p^2 = p_{m_1}p_{m_2}\Delta^{m_1 m_2}$. Here the angular bracket over indices indicate
the symmetric and traceless part.
For $n=3$,
\begin{eqnarray}
p^{\langle m_1}p^{m_2}p^{m_3\rangle}
\ampeq
p^{m_1} p^{m_2} p^{m_3}
-{p^2\over 5} \left(\Delta^{m_1 m_2} p^{m_3} 
+ \Delta^{m_1 m_3} p^{m_2} + \Delta^{m_2 m_3} p^{m_1} \right)
\end{eqnarray}

Higher rank irreducible tensors can be built using lower rank ones
by using the following recursion relationship
\begin{eqnarray}
\lefteqn{
{p^{\langle  m_1} p^{ m_2} \cdots p^{ m_n \rangle}}
} &&
\non
\ampeq
{1\over n}\sum_{i=1}^n p^{ m_i} p^{\langle  m_1} \cdots p^{ m_{i-1}}p^{ m_{i+1}}
\cdots p^{ m_n\rangle}
\nl
-
{2\over n(2n-1)}
\sum_{i=1}^n\sum_{j=i+1}^n
\Delta^{ m_i m_j} 
p_a p^{\langle a} p^{ m_1} \cdots p^{ m_{i-1}}p^{ m_{i+1}}\cdots
p^{ m_{j-1}}p^{ m_{j+1}}\cdots p^{ m_n\rangle}
\label{eq:p1_pn}
\end{eqnarray}
which comes from applying Eq.(\ref{eq:q_P_n}) to $p^{k_1}p^{\langle k_2}\cdots
p^{k_n\rangle}$.

When the fluid-cell has a non-zero flow velocity $u^\mu$, 
then the spatial metric tensor is
\begin{eqnarray}
\Delta^{\mu\nu} = g^{\mu\nu} + u^\mu u^\nu
\end{eqnarray}
and the spatial part of a 4-momentum is 
\begin{eqnarray}
p^\ave{\mu} \ampeq \Delta^{\mu}_{\nu} p^\nu
\non
\ampeq
p^\mu - u^\mu \calE_p
\end{eqnarray}
where $\calE_p = -p_\mu u^\mu$ is the time component of the 4-momentum in the fluid-cell
rest frame. All results in this sections can be generalized to the non-zero fluid
velocity case by changing $m_i \to \ave{\mu_i}$ and $p^2 \to (\calE_p^2 - m^2)$
where $m^2 = -p^\mu p_\mu$.

The orthogonality condition for the momentum polynomial is
\cite{Denicol:2012cn,degroot}
\begin{eqnarray}
\int {d^3p\over (2\pi)^3 p^0}\,
F({\cal E}_p)
p^{\langle\mu_1}\cdots p^{\mu_n\rangle}
p_{\langle\nu_1}\cdots p_{\nu_m\rangle}
=
{n!\over (2n+1)!!}\delta_{mn} \Delta^{\mu_1\cdots\mu_n}_{\nu_1\cdots\nu_n}
\int {d^3p\over (2\pi)^3 p^0}\,
F({\cal E}_p)\left({\cal E}_p^2 - m^2\right)^n
\non
\label{eq:p_poly_ortho}
\end{eqnarray}

In deriving the evolution equation for a general energy-momentum moment, the following
identity is frequently needed:
\begin{eqnarray}
p^\ave{\lambda} p^{\langle\mu_1}\cdots p^{\mu_n\rangle}
=
p^{\langle\lambda}p^{\mu_1}\cdots p^{\mu_n\rangle}
+
{n\over 2n+1}(\calE_p^2 - m^2)p^{\langle\mu_1}p^{\mu_2}\cdots p^{\mu_{n-1}}\Delta^{\mu_n\rangle\lambda}
\label{eq:irreducible_recursion}
\end{eqnarray}
To prove this, first we go to the rest frame where $u^\mu = (1,0,0,0)$.
In that case, 
\begin{eqnarray}
p^\ave{\mu} \to p^m
\end{eqnarray}
where $m = 1, 2, 3$ are the spatial component of a momentum and
\begin{eqnarray}
\calE_p^2 - m^2 \to p^2
\end{eqnarray}
where $p^2 = p_i p^i$.

The identity to prove is then
\begin{eqnarray}
p^{\langle l}p^{m_1}\cdots p^{m_n\rangle}
=
p^l p^{\langle m_1}\cdots p^{m_n\rangle}
-
{n\over 2n+1}p^2 p^{\langle m_1}p^{m_2}\cdots p^{m_{n-1}}\Delta^{m_n\rangle l}
\label{eq:irreducible_recursion2}
\end{eqnarray}
Our starting point is the fact that
these polynomials can be obtained 
from
\begin{eqnarray}
\partial_{m_n}\cdots\partial_{m_2}\partial_{m_1}{1\over p}
\ampeq
(-1)^n (2n-1)!! {p^{\langle m_1}p^{m_2}\cdots p^{m_n\rangle} \over p^{2n+1}}
\label{eq:Waldmann_formula}
\end{eqnarray}
where $\partial_{m} = \partial/\partial p^m$.
This expression is explicitly symmetric since derivatives commute. It is also traceless since
\begin{eqnarray}
\nabla_p^2 {1\over p} \propto \delta(p) 
\end{eqnarray}
The nomalization constant is chosen so that 
the coefficient of $p^{m_1}\cdots p^{m_n}$ in 
$p^{\langle m_1}\cdots p^{m_n\rangle}$ is
1.

We can get the following recursion relation by considering
the product rule of taking one more derivative of Eq.(\ref{eq:Waldmann_formula}).
\begin{eqnarray}
\lefteqn{
(-1)^{n+1}(2n+1)!! 
{p^{\langle m_1}p^{m_2}\cdots p^{m_n}p^{m_{n+1}\rangle} \over p^{2n+3}}
} &&
\non
\ampeq
\partial_{m_{n+1}}\partial_{m_n}\cdots\partial_{m_2}\partial_{m_1}{1\over p}
\non
\ampeq
(-1)^n (2n-1)!! 
\left(
(-1)(2n+1)
{p^{m_{n+1}}p^{\langle m_1}p^{m_2}\cdots p^{m_n\rangle} \over p^{2n+3}}
+ 
{\partial_{m_{n+1}}(p^{\langle m_1}p^{m_2}\cdots p^{m_n\rangle}) \over p^{2n+1}}
\right)
\non
\end{eqnarray}
which yields
\begin{eqnarray}
p^{\langle m_1}p^{m_2}\cdots p^{m_n}p^{m_{n+1}\rangle} 
=
p^{m_{n+1}}p^{\langle m_1}p^{m_2}\cdots p^{m_n\rangle} 
-
{p^2 \over 2n+1}
\partial_{m_{n+1}}p^{\langle m_1}p^{m_2}\cdots p^{m_n\rangle}
\end{eqnarray}
The identity Eq.(\ref{eq:irreducible_recursion2}) is proven if we can show
\begin{eqnarray}
\partial_{m_{n+1}}p^{\langle m_1}p^{m_2}\cdots p^{m_n\rangle}
=
n\, p^{\langle m_1}p^{m_2}\cdots \Delta^{m_n\rangle m_{n+1}}
\label{eq:derivative_pDelta}
\end{eqnarray}
To start mathematical induction, consider $n=2$:
\begin{eqnarray}
\partial_{m_3}p^{\langle m_1}p^{m_2\rangle}
\ampeq
\partial_{m_3}
\left(
p^{m_1}p^{m_2}
-
{\Delta^{m_1 m_2}\over 3} p^2
\right)
\non
\ampeq
\Delta^{m_1 m_3}p^{m_2}
+
\Delta^{m_2 m_3}p^{m_1}
-
2{\Delta^{m_1 m_2}\over 3} p^{m_3}
\non
\ampeq
2
\left(
{1\over 2}
\left(
\Delta^{m_1 m_3}p^{m_2}
+
\Delta^{m_2 m_3}p^{m_1}
\right)
-
{\Delta^{m_1 m_2}\over 3} p^{m_3}
\right)
\non
\ampeq
2 p^{\langle m_1}\Delta^{m_2\rangle m_3}
\end{eqnarray}
which gives the correct expression.

To prove Eq.(\ref{eq:derivative_pDelta}) for general $n$,
we need some identities first.
The right hand side of the following expression
\begin{eqnarray}
\lefteqn{
p^{\langle m_1}p^{m_2}\cdots p^{m_{n-1}} \Delta^{m_n\rangle m_{n+1}}
} &&
\non
\ampeq
{1\over n}\sum_{i=1}^n p^{m_i} p^{\langle m_1} p^{m_2} \cdots
p^{m_{i-1}}p^{m_{i+1}}\cdots
\Delta^{m_n\rangle m_{n+1}}
\nl
-{2\over n(2n-1)}
\sum_{i=1}^n \sum_{j+1}^n \Delta^{m_i m_j}
p_a p^{\langle a}p^{m_1} p^{m_2}\cdots 
p^{m_{i-1}}p^{m_{i+1}}\cdots 
\Delta^{m_n\rangle m_{n+1}}
\label{eq:pD_Id_1}
\end{eqnarray}
is explicitly
constructed in such a way that it is symmetric and traceless 
in $(m_1,\cdots,m_n)$.
The tensor 
$p^{\langle m_1}p^{m_2}\cdots p^{m_{n-1}} \Delta^{m_n\rangle m_{n+1}}$ can be also
expressed as
\begin{eqnarray}
\lefteqn{
p^{\langle m_1}p^{m_2}\cdots p^{m_{n-1}}\Delta^{m_{n}\rangle m_{n+1}}
} &&
\non
\ampeq
{1\over n}
\sum_{i=1}^{n}
\Delta^{m_{n+1}m_i}p^{\langle m_1}\cdots p^{m_{i-1}}p^{m_{i+1}}\cdots
p^{m_{n}\rangle}
\nl
-
{2\over n(2n-1)}
\sum_{i=1}^{n}\sum_{j=i+1}^{n}
\Delta^{m_i m_j} p^{\langle m_1}\cdots p^{m_{i-1}}p^{m_{i+1}}\cdots
p^{m_{j-1}}p^{m_{j+1}}\cdots p^{m_{n}}p^{m_{n+1}\rangle}
\label{eq:pD_Id_2}
\end{eqnarray}
Again the RHS is explicitly constructed so that it is symmetric and traceless 
in $(m_1,\cdots,m_n)$.

To prove Eq.(\ref{eq:derivative_pDelta}), assume that it works for $n-1$.
We then take another derivative of Eq.(\ref{eq:p1_pn}):
\begin{eqnarray}
\lefteqn{
{\partial\over \partial p^{m_{n+1}}}
{p^{\langle  m_1} p^{ m_2} \cdots p^{ m_n \rangle}}
} &&
\non
\ampeq
{1\over n}\sum_{i=1}^n 
{\partial\over \partial p^{m_{n+1}}}
\left(p^{ m_i} p^{\langle  m_1} \cdots p^{ m_{i-1}}p^{ m_{i+1}}
\cdots p^{ m_n\rangle}
\right)
\nl
-
{2\over n(2n-1)}
\sum_{i=1}^n\sum_{j=i+1}^n
\Delta^{ m_i m_j} 
{\partial\over \partial p^{m_{n+1}}}
\left(
p_a p^{\langle a} p^{ m_1} \cdots p^{ m_{i-1}}p^{ m_{i+1}}\cdots
p^{ m_{j-1}}p^{ m_{j+1}}\cdots p^{ m_n\rangle}
\right)
\non
\end{eqnarray}
One can then show Eq.(\ref{eq:derivative_pDelta}) can be reproduced with for $n+1$
using the identities Eq.(\ref{eq:pD_Id_1}) and Eq.(\ref{eq:pD_Id_2}).

This proves 
\begin{eqnarray}
p^{\langle l}p^{m_1}\cdots p^{m_n\rangle}
=
p^l p^{\langle m_1}\cdots p^{m_n\rangle}
-
{n\over 2n+1}p^2 p^{\langle m_1}p^{m_2}\cdots p^{m_{n-1}}\Delta^{m_n\rangle l}
\end{eqnarray}
which can be found in Ref.~\cite{struchtrup_phd}.
In a moving frame, this becomes
\begin{eqnarray}
p^\ave{\lambda} p^{\langle\mu_1}\cdots p^{\mu_n\rangle}
=
p^{\langle\lambda}p^{\mu_1}\cdots p^{\mu_n\rangle}
+
{n\over 2n+1}(\calE_p^2 - m^2)p^{\langle\mu_1}p^{\mu_2}\cdots p^{\mu_{n-1}}\Delta^{\mu_n\rangle\lambda}
\label{eq:Waldmann_recursion}
\end{eqnarray}

One can also show
\begin{eqnarray}
\lefteqn{
p^\ave{\alpha} p^\ave{\lambda}
p^{\langle\mu_1}\cdots p^{\mu_{n-1}}p^{\mu_n\rangle}
}&&
\non
& = &
p^{\langle\alpha}p^{\lambda} p^{\mu_1}\cdots p^{\mu_{n-1}}p^{\mu_n\rangle}
\nl
+ 
{1\over (2n+3)} (\calE_p^2-m^2)
\sum_{i=1}^n 
\Delta^{\mu_i\alpha}
p^{\langle\lambda}p^{\mu_1}\cdots p^{\mu_{i-1}} p^{\mu_{i+1}}\cdots p^{\mu_n\rangle}
\nl
+
{1\over (2n+3)} (\calE_p^2-m^2)
\sum_{i=1}^n \Delta^{\mu_i\lambda}
p^{\langle\alpha} 
p^{\mu_1}\cdots p^{\mu_{i-1}} p^{\mu_{i+1}}\cdots p^{\mu_n\rangle}
\nl
-{4\over (2n+3)(2n-1)} (\calE_p^2-m^2)
\sum_{i<j}^n \Delta^{\mu_i\mu_j}
p^{\langle\alpha} 
p^{\lambda}p^{\mu_1}
\cdots p^{\mu_{i-1}} p^{\mu_{i+1}}
\cdots p^{\mu_{j-1}} p^{\mu_{j+1}}
\cdots p^{\mu_n\rangle}
\nl
+ 
{1\over (2n+3)}
(\calE_p^2-m^2)
\left(
\Delta^{\lambda\alpha} p^{\langle\mu_1}\cdots p^{\mu_n\rangle}
\right)
\nl
+
{n(n-1) \over (2n+1)(2n-1)}
(\calE_p^2-m^2)^2
\left(
p^{\langle\mu_1}\cdots p^{\mu_{i-1}} p^{\mu_{i+1}}\cdots 
\Delta^{\mu_{n-1}}_{\alpha'} 
\Delta^{\mu_n\rangle}_{\lambda'} 
\right)
\Delta^{\alpha\alpha'}\Delta^{\lambda\lambda'}
\non
\label{eq:alpha_lambda_n}
\end{eqnarray}
by using
\begin{eqnarray}
\lefteqn{
p^\ave{\lambda}
p^{\langle\mu_1}\cdots p^{\mu_{n-1}}\Delta^{\mu_n\rangle\alpha}
}&&
\non
& = &
{1\over n}
\sum_{i=1}^n 
\Delta^{\mu_i\alpha}
p^{\langle\lambda}
p^{\mu_1}\cdots p^{\mu_{i-1}} p^{\mu_{i+1}}\cdots p^{\mu_n\rangle}
\nl
-
{2\over n(2n-1)} 
\sum_{i<j}^n \Delta^{\mu_i\mu_j}
p^{\langle\lambda}
p^{\alpha}p^{\mu_1}
\cdots p^{\mu_{i-1}} p^{\mu_{i+1}}
\cdots p^{\mu_{j-1}} p^{\mu_{j+1}}
\cdots p^{\mu_n\rangle}
\nl
+
(\calE_p^2-m^2) {n-1\over (2n-1)}
p^{\langle\mu_1}\cdots
p^{\mu_{n-2}}\Delta^{\mu_{n-1}}_{\lambda'}\Delta^{\mu_{n}\rangle}_{\alpha'}
\Delta^{\lambda\lambda'}\Delta^{\alpha\alpha'}
\end{eqnarray}

\section{A useful mathetical identity}\label{app:Dpn}
Consider the following rank-$n$ tensor:
\begin{equation}
\begin{split}
    A^{\mu_1...\mu_n} =
    \Delta_{\nu_1...\nu_n}^{\mu_1...\mu_n}Dp^{\langle\nu_1...}p^{\nu_n\rangle}
\end{split}
\end{equation}
Following Eqs.(C.8) and (C.9) in \cite{struchtrup_phd}, for any symmetric tensor $\Pi$
we have:
\begin{equation}\label{a1}
\begin{split}
    \Pi_{\langle i_1...i_n\rangle} &= \Pi_{i_1...i_n}\\
    & + \alpha_{n1}(\Delta_{i_1 i_2}\Pi_{i_3...i_nkk} + permutation)\\
    & + \alpha_{n2}(\Delta_{i_1 i_2}\Delta_{i_3 i_4}\Pi_{i_5...i_nkkll} + permutation)\\
    & + ...
\end{split}
\end{equation}
where
\begin{equation}
    \alpha_{nk} = \frac{(-1)^k}{\Pi_{j=0}^{k-1}(2n-2j-1)}
\end{equation}
Now, if we let
\begin{equation}
    \Pi_{i_1...i_n} = p_{\langle i_1\rangle}...p_{\langle i_n\rangle}
\end{equation}
then all terms in Eq.(\ref{a1}) except the first one vanish under
$\Delta_{j_1...j_n}^{i_1...i_n} D$
since
\begin{equation}
D(\Delta_{i_k i_l}F) = (D\Delta_{i_k i_l})F + \Delta_{i_k i_l}DF 
= (a_{i_k} u_{i_l} + u_{i_k} a_{i_l})F + \Delta_{i_k i_l}DF
\end{equation}
This expression vanishes when the projector is applied
due to the presence of $u_{i_k}$, $u_{i_l}$, or $\Delta_{i_l i_k}$.
Consequently, we arrive at the following useful identity:
\begin{equation}
\begin{split}
    \Delta_{\nu_1...\nu_n}^{\mu_1...\mu_n}Dp^{\langle\nu_1...}p^{\nu_n\rangle} =
    \Delta_{\nu_1...\nu_n}^{\mu_1...\mu_n}Dp^{\langle\nu_1\rangle}...p^{\langle\nu_n\rangle}
\end{split}
\end{equation}
Note that
\begin{equation}
\begin{split}
    Dp^{\langle\mu\rangle} &= D\Delta^{\mu\nu}p_\nu \\
    & = D(g^{\mu\nu}+u^\mu u^\nu)p_\nu \\
    & = (u^\mu D u^\nu + u^\nu D u^\mu) p_\nu \\
    & = u^\mu p_\nu a^\nu -{\cal E}_p a^\mu \\
    & = u^\mu (p^{\langle\nu\rangle} + {\cal E}_p u^\nu) a_\nu - {\cal E}_p a^\mu \\
    & = u^\mu p^{\langle\nu\rangle}a_\nu - {\cal E}_p a^\mu
\end{split}
\end{equation}
where the term with $u^\mu$ vanishes when being projected. With some simple algebraic
manipulations, we get
\begin{equation}\label{id1}
    \Delta_{\nu_1...\nu_n}^{\mu_1...\mu_n}Dp^{\langle\nu_1...}p^{\nu_n\rangle}
    = -n{\cal E}_p p^{\langle\mu_1...}p^{\mu_{n-1}}a^{\mu_n\rangle}
\end{equation}
Similarly, one can also argue for the same reasons:
\begin{equation}
\begin{split}
\Delta_{\nu_1...\nu_n}^{\mu_1...\mu_n}\nabla_\lambda
(p^{\langle\nu_1...}p^{\nu_n\rangle}) =
\Delta_{\nu_1...\nu_n}^{\mu_1...\mu_n}\nabla_\lambda
(p^{\langle\nu_1\rangle}...p^{\langle\nu_n\rangle})
\end{split}
\end{equation}
and
\begin{equation}
\begin{split}
    \nabla_\lambda p^{\langle\nu\rangle} &= \nabla_\lambda(p^\nu - {\cal E}_p u^\nu)\\
    & = -u^\nu \nabla_\lambda {\cal E}_p - {\cal E}_p(\nabla_\lambda u^\nu)
\end{split}
\end{equation}
once again, the first term vanishes when being projected. After some manipulations,
we get:
\begin{equation}\label{id3}
\begin{split}
    \Delta_{\nu_1...\nu_n}^{\mu_1...\mu_n}\nabla_\lambda
    (p^{\langle\nu_1...}p^{\nu_n\rangle}) = -n{\cal E}_p
    p^{\langle\mu_1}...p^{\mu_{n-1}}\nabla_\lambda u^{\nu\rangle}
\end{split}
\end{equation}

\section{Derivation of the General Moment Equation}\label{app:derivation}

The starting point is the general rank-$n$ energy-momentum moments of $\delta f$:
\begin{equation}
    \rho_r^{\mu_1...\mu_n} = \intvar \delta f {\cal E}_p^r p^{\langle
    \mu_1}p^{\mu_2}...p^{\mu_n\rangle}
\end{equation}
Taking the comoving derivative $D = u^\mu \partial_\mu$, which corresponds to
the time derivative in the fluid rest-frame, and then projecting onto the transverse
space, we get
\begin{equation}
\begin{split}
     \proj D \rho_r^{\nu_1...\nu_n} &= \proj \intvar (D\delta f){\cal E}_p^r p^{\langle
     \nu_1}p^{\nu_2}...p^{\nu_n\rangle}\\
     &\quad + \proj \intvar \delta f {\cal E}_p^r Dp^{\langle
     \nu_1}p^{\nu_2}...p^{\nu_n\rangle}\\
     &\quad + \proj \intvar \delta f (D{\cal E}_p^r) p^{\langle
     \nu_1}p^{\nu_2}...p^{\nu_n\rangle}\\
     &= \proj \intvar (D\delta f){\cal E}_p^r p^{\langle \nu_1}p^{\nu_2}...p^{\nu_n\rangle}\\
     &\quad - n\intvar \delta f {\cal E}_p^{r+1} p^{\langle \mu_1}p^{\mu_2}...a^{\mu_n\rangle}\\
     &\quad - r\proj a_\sigma \intvar \delta f {\cal E}_p^{r-1} p^{\langle \sigma
     \rangle}p^{\langle \nu_1}p^{\nu_2}...p^{\nu_n\rangle}\\
\end{split}
\end{equation}
where we defined the fluid acceleration by $a^\mu = Du^\mu$, and used the fact that
$D{\cal E}_p = -a_\mu p^\mu = -a_\mu p^{\langle\mu\rangle}$, along with Eq.(\ref{id1}). 
Using Eq.(\ref{eq:Waldmann_recursion}),
we can expand the last term on the right-hand side 
\begin{equation}\label{eq1919}
\begin{split}
     \proj D \rho_r^{\nu_1...\nu_n} &= \proj \intvar (D\delta f){\cal E}_p^r p^{\langle
     \nu_1}p^{\nu_2}...p^{\nu_n\rangle}\\
     &\quad - n\intvar \delta f {\cal E}_p^{r+1} p^{\langle \mu_1}p^{\mu_2}...a^{\mu_n\rangle}\\
     &\quad - ra_\sigma \intvar \delta f {\cal E}_p^{r-1} p^{\langle
     \sigma}p^{\mu_1}p^{\mu_2}...p^{\mu_n\rangle}\\
     &\quad - r\frac{n}{2n+1}a_\sigma \intvar \delta f {\cal E}_p^{r-1} ({\cal E}_p^2 - m^2)p^{\langle
     \mu_1}p^{\mu_2}...\Delta^{\mu_n\rangle \sigma}\\
\end{split}
\end{equation}
To express $D\delta f$ in terms of $\delta f$, we can use the following form of the
Boltzmann equation
\begin{equation}\label{boltz}
\begin{split}
    p^\mu \partial_\mu f_0 + {\cal E}_pD\delta f + p^{\langle\mu\rangle}\nabla_\mu \delta f
    = C[f]
\end{split}
\end{equation}
in Eq.(\ref{eq1919}) where we used the decomposition
\begin{equation}\label{deriv_decom}
    \partial_\mu = g_\mu^\alpha \partial_\alpha = \left(-u_\mu u^\alpha +
    \Delta_\mu^\alpha\right)\partial_\alpha = -u_\mu D + \nabla_\mu
\end{equation}
This gives
\begin{equation}\label{eq241}
\begin{split}
     \proj D \rho_r^{\nu_1...\nu_n} &= - n\intvar \delta f {\cal E}_p^{r+1} p^{\langle
     \mu_1}p^{\mu_2}...a^{\mu_n\rangle}\\
     &\quad - r a_\sigma \intvar \delta f {\cal E}_p^{r-1} p^{\langle
     \sigma}p^{\mu_1}p^{\mu_2}...p^{\mu_n\rangle}\\
     &\quad - r\frac{n}{2n+1}\intvar \delta f {\cal E}_p^{r-1} ({\cal E}_p^2 -
     m^2)p^{\langle\mu_1}p^{\mu_2}...a^{\mu_n\rangle}\\
     &\quad + \proj \intvar C[f]{\cal E}_p^{r-1}p^{\langle \nu_1}p^{\nu_2}...p^{\nu_n\rangle}\\
     &\quad - \proj \intvar (\partial_\lambda f_0){\cal E}_p^{r-1}p^{\lambda}p^{\langle
     \nu_1}p^{\nu_2}...p^{\nu_n\rangle}\\
     &\quad - \proj \intvar (\nabla_\lambda \delta f){\cal E}_p^{r-1}p^{\langle \lambda
     \rangle}p^{\langle \nu_1}p^{\nu_2}...p^{\nu_n\rangle}\\
\end{split}
\end{equation}
Here, we define $\nabla_\mu = \Delta_\mu^{\nu}\partial_\nu$ as the
projected derivative, corresponding to the spatial gradient in the fluid
rest-frame.
Using the chain rule, we can pull $\nabla_\lambda$ in the last term on the right-hand
side of Eq.(\ref{eq241}) out of the integral:
\begin{equation}\label{eq243}
\begin{split}
    \proj D \rho_r^{\nu_1...\nu_n} &= - n\intvar \delta f {\cal E}_p^{r+1} p^{\langle
    \mu_1}p^{\mu_2}...a^{\mu_n\rangle}\\
     &\quad - r a_\sigma \intvar \delta f {\cal E}_p^{r-1} p^{\langle
     \sigma}p^{\mu_1}p^{\mu_2}...p^{\mu_n\rangle}\\
     &\quad - r\frac{n}{2n+1}\intvar \delta f {\cal E}_p^{r-1} ({\cal E}_p^2 -
     m^2)p^{\langle\mu_1}p^{\mu_2}...a^{\mu_n\rangle}\\
     &\quad + \proj \intvar C[f]{\cal E}_p^{r-1}p^{\langle \nu_1}p^{\nu_2}...p^{\nu_n\rangle}\\
     &\quad - \proj \intvar (\partial_\lambda f_0){\cal E}_p^{r-1}p^{\lambda}p^{\langle
     \nu_1}p^{\nu_2}...p^{\nu_n\rangle}\\
     &\quad - \proj \nabla_\lambda \intvar \delta f {\cal E}_p^{r-1}p^{\langle \lambda
     \rangle}p^{\langle \nu_1}p^{\nu_2}...p^{\nu_n\rangle}\\
     &\quad + \proj \intvar \delta f (\nabla_\lambda {\cal E}_p^{r-1})p^{\langle \lambda
     \rangle}p^{\langle \nu_1}p^{\nu_2}...p^{\nu_n\rangle}\\
     &\quad + \proj \intvar \delta f {\cal E}_p^{r-1}(\nabla_\lambda p^{\langle \lambda
     \rangle})p^{\langle \nu_1}p^{\nu_2}...p^{\nu_n\rangle}\\
     &\quad + \proj \intvar \delta f {\cal E}_p^{r-1}p^{\langle \lambda \rangle} (\nabla_\lambda
     p^{\langle \nu_1}p^{\nu_2}...p^{\nu_n\rangle})\\
\end{split}
\end{equation}
Now, note that the second-last term on the right-hand side can be simplified as
\begin{equation}
\begin{split}
    &\proj \intvar \delta f {\cal E}_p^{r-1}(\nabla_\lambda p^{\langle \lambda
    \rangle})p^{\langle \nu_1}p^{\nu_2}...p^{\nu_n\rangle}\\
    &\quad = \proj \intvar \delta f {\cal E}_p^{r-1}\nabla_\lambda (p^\lambda - {\cal E}_p
    u^\lambda)p^{\langle \nu_1}p^{\nu_2}...p^{\nu_n\rangle}\\
    &\quad = - \theta \intvar \delta f {\cal E}_p^r p^{\langle
    \mu_1}p^{\mu_2}...p^{\mu_n\rangle}
\end{split}
\end{equation}
since $\nabla_\lambda p^\lambda = 0$ and $u^\lambda \nabla_\lambda {\cal E}_p = u^\lambda
\Delta_\lambda^\alpha \partial_\alpha {\cal E}_p = 0$. Here, we define $\theta = \partial_\mu
u^\mu = \nabla_\mu u^\mu$, which represents the expansion rate of the fluid.

To briefly summarize, so far we have:
\begin{equation}\label{eq246}
\begin{split}
    \proj D \rho_r^{\nu_1...\nu_n} &= - n\intvar \delta f {\cal E}_p^{r+1} p^{\langle
    \mu_1}p^{\mu_2}...a^{\mu_n\rangle}\\
     &\quad - r a_\sigma \intvar \delta f {\cal E}_p^{r-1} p^{\langle
     \sigma}p^{\mu_1}p^{\mu_2}...p^{\mu_n\rangle}\\
     &\quad - r\frac{n}{2n+1}\intvar \delta f {\cal E}_p^{r-1} ({\cal E}_p^2 -
     m^2)p^{\langle\mu_1}p^{\mu_2}...a^{\mu_n\rangle}\\
     &\quad + \intvar C[f]{\cal E}_p^{r-1}p^{\langle \mu_1}p^{\mu_2}...p^{\mu_n\rangle}\\
     &\quad - \intvar (\partial_\lambda f_0){\cal E}_p^{r-1}p^{\lambda}p^{\langle
     \mu_1}p^{\mu_2}...p^{\mu_n\rangle}\\
     &\quad - \proj \nabla_\lambda \intvar \delta f {\cal E}_p^{r-1}p^{\langle \lambda
     \rangle}p^{\langle \nu_1}p^{\nu_2}...p^{\nu_n\rangle}\\
     &\quad + \proj \intvar \delta f {\cal E}_p^{r-1}p^{\langle \lambda \rangle} (\nabla_\lambda
     p^{\langle \nu_1}p^{\nu_2}...p^{\nu_n\rangle})\\
     &\quad - \theta \intvar {\cal E}_p^r \delta f p^{\langle
     \mu_1}p^{\mu_2}...p^{\mu_n\rangle}\\
     &\quad + \intvar \delta f (\nabla_\lambda {\cal E}_p^{r-1})p^{\langle \lambda
     \rangle}p^{\langle \mu_1}p^{\mu_2}...p^{\mu_n\rangle}
\end{split}
\end{equation}
We continue to simplify the last three terms by calculating the gradients. Observe that
\begin{equation}
\begin{split}
    \nabla_\lambda {\cal E}_p^{r-1} &= (r-1){\cal E}_p^{r-2}(\nabla_\lambda {\cal E}_p)\\
    &= -(r-1){\cal E}_p^{r-2}\nabla_\lambda (u_\alpha p^\alpha)\\
    &= -(r-1){\cal E}_p^{r-2}p^\alpha\nabla_\lambda u_\alpha\\
    &= -(r-1){\cal E}_p^{r-2}(p^{\langle\alpha\rangle} + {\cal E}_p u^\alpha)\nabla_\lambda u_\alpha\\
    &= -(r-1){\cal E}_p^{r-2}p^{\langle\alpha\rangle}\nabla_\lambda u_\alpha\\
\end{split}
\end{equation}
using the normalization condition $u_\alpha u^\alpha = -1$. Plugging this into
Eq.(\ref{eq246}) gives
\begin{equation}\label{eq248}
\begin{split}
    \proj D \rho_r^{\nu_1...\nu_n} &= - n\intvar \delta f {\cal E}_p^{r+1} p^{\langle
    \mu_1}p^{\mu_2}...a^{\mu_n\rangle}\\
     &\quad - r a_\sigma \intvar \delta f {\cal E}_p^{r-1} p^{\langle
     \sigma}p^{\mu_1}p^{\mu_2}...p^{\mu_n\rangle}\\
     &\quad - r\frac{n}{2n+1}\intvar \delta f {\cal E}_p^{r-1} ({\cal E}_p^2 -
     m^2)p^{\langle\mu_1}p^{\mu_2}...a^{\mu_n\rangle}\\
     &\quad + \intvar C[f]{\cal E}_p^{r-1}p^{\langle \mu_1}p^{\mu_2}...p^{\mu_n\rangle}\\
     &\quad - \intvar (\partial_\lambda f_0){\cal E}_p^{r-1}p^{\lambda}p^{\langle
     \mu_1}p^{\mu_2}...p^{\mu_n\rangle}\\
     &\quad - \proj \nabla_\lambda \intvar \delta f {\cal E}_p^{r-1}p^{\langle \lambda
     \rangle}p^{\langle \nu_1}p^{\nu_2}...p^{\nu_n\rangle}\\
     &\quad + \proj \intvar \delta f {\cal E}_p^{r-1}p^{\langle \lambda \rangle} (\nabla_\lambda
     p^{\langle \nu_1}p^{\nu_2}...p^{\nu_n\rangle})\\
     &\quad - \theta \intvar {\cal E}_p^r \delta f p^{\langle
     \mu_1}p^{\mu_2}...p^{\mu_n\rangle}\\
     &\quad - (r-1) \intvar \delta f {\cal E}_p^{r-2}(\nabla_\lambda u_\alpha)p^{\langle \alpha
     \rangle}p^{\langle \lambda \rangle}p^{\langle \mu_1}p^{\mu_2}...p^{\mu_n\rangle}\\
\end{split}
\end{equation}
Now, using Eq.(\ref{id3}) proven in Appendix \ref{app:Dpn}, the third-last term on the
right-hand side can be written as
\begin{equation}
\begin{split}
    &\proj \intvar \delta f {\cal E}_p^{r-1}p^{\langle \lambda \rangle} (\nabla_\lambda
    p^{\langle \nu_1}p^{\nu_2}...p^{\nu_n\rangle})\\
    &\quad = - n\intvar {\cal E}_p^r \delta f p^{\langle \lambda \rangle} p^{\langle
    \mu_1}p^{\mu_2}...\nabla_\lambda u^{\mu_n \rangle}\\
\end{split}
\end{equation}

Eq.(\ref{eq248}) now becomes
\begin{equation}\label{eq251}
\begin{split}
    \proj D \rho_r^{\nu_1...\nu_n} &= - n\intvar \delta f {\cal E}_p^{r+1} p^{\langle
    \mu_1}p^{\mu_2}...a^{\mu_n\rangle}\\
     &\quad - r a_\sigma \intvar \delta f {\cal E}_p^{r-1} p^{\langle
     \sigma}p^{\mu_1}p^{\mu_2}...p^{\mu_n\rangle}\\
     &\quad - r\frac{n}{2n+1}\intvar \delta f {\cal E}_p^{r-1} ({\cal E}_p^2 -
     m^2)p^{\langle\mu_1}p^{\mu_2}...a^{\mu_n\rangle}\\
     &\quad + \intvar C[f]{\cal E}_p^{r-1}p^{\langle \mu_1}p^{\mu_2}...p^{\mu_n\rangle}\\
     &\quad - \intvar (\partial_\lambda f_0){\cal E}_p^{r-1}p^{\lambda}p^{\langle
     \mu_1}p^{\mu_2}...p^{\mu_n\rangle}\\
     &\quad - \proj \nabla_\lambda \intvar \delta f {\cal E}_p^{r-1}p^{\langle \lambda
     \rangle}p^{\langle \nu_1}p^{\nu_2}...p^{\nu_n\rangle}\\
     &\quad - n\intvar {\cal E}_p^r \delta f p^{\langle \lambda \rangle} p^{\langle
     \mu_1}p^{\mu_2}...\nabla_\lambda u^{\mu_n \rangle}\\
     &\quad - \theta \intvar {\cal E}_p^r \delta f p^{\langle
     \mu_1}p^{\mu_2}...p^{\mu_n\rangle}\\
     &\quad - (r-1) \intvar \delta f {\cal E}_p^{r-2}(\nabla_\lambda u_\alpha)p^{\langle \alpha
     \rangle}p^{\langle \lambda \rangle}p^{\langle \mu_1}p^{\mu_2}...p^{\mu_n\rangle}\\
\end{split}
\end{equation}
Applying Eq.(\ref{eq:Waldmann_recursion}) again to the sixth term on the right-hand side, we get
\begin{equation}
\begin{split}
    & - \proj \nabla_\lambda \intvar \delta f {\cal E}_p^{r-1}p^{\langle \lambda
    \rangle}p^{\langle \nu_1}p^{\nu_2}...p^{\nu_n\rangle}\\
    &\quad = -\proj \nabla_\lambda \intvar \delta f {\cal E}_p^{r-1}p^{\langle
    \lambda}p^{\nu_1}p^{\nu_2}...p^{\nu_n\rangle}\\
    &\quad \quad -\frac{n}{2n+1} \proj \nabla_\lambda \intvar \delta f {\cal E}_p^{r-1}
    ({\cal E}_p^2 - m^2)p^{\langle \nu_1}p^{\nu_2}...\Delta^{\nu_n \rangle \lambda}\\
\end{split}
\end{equation}
Plugging this back into Eq.(\ref{eq251}) gives us
\begin{equation}
\begin{split}
    \proj D \rho_r^{\nu_1...\nu_n} &= \intvar C[f]{\cal E}_p^{r-1}p^{\langle
    \mu_1}p^{\mu_2}...p^{\mu_n\rangle}\\
    &\quad - \intvar (\partial_\lambda f_0){\cal E}_p^{r-1}p^{\lambda}p^{\langle
    \mu_1}p^{\mu_2}...p^{\mu_n\rangle}\\
    &\quad - n\intvar \delta f {\cal E}_p^{r+1} p^{\langle \mu_1}p^{\mu_2}...a^{\mu_n\rangle}\\
    &\quad - r\frac{n}{2n+1}\intvar \delta f {\cal E}_p^{r-1} ({\cal E}_p^2 -
    m^2)p^{\langle\mu_1}p^{\mu_2}...a^{\mu_n\rangle}\\
    &\quad - r a_\sigma \intvar \delta f {\cal E}_p^{r-1} p^{\langle
    \sigma}p^{\mu_1}p^{\mu_2}...p^{\mu_n\rangle}\\
    &\quad - \proj \nabla_\lambda \intvar \delta f {\cal E}_p^{r-1}p^{\langle
    \lambda}p^{\nu_1}p^{\nu_2}...p^{\nu_n\rangle}\\
    &\quad -\frac{n}{2n+1} \proj \nabla_\lambda \intvar \delta f {\cal E}_p^{r-1} ({\cal E}_p^2 -
    m^2)p^{\langle \nu_1}p^{\nu_2}...\Delta^{\nu_n \rangle \lambda}\\
    &\quad - n\intvar {\cal E}_p^r \delta f p^{\langle \lambda \rangle} p^{\langle
    \mu_1}p^{\mu_2}...\nabla_\lambda u^{\mu_n \rangle}\\
    &\quad - \theta \intvar {\cal E}_p^r \delta f p^{\langle
    \mu_1}p^{\mu_2}...p^{\mu_n\rangle}\\
    &\quad - (r-1) \intvar \delta f {\cal E}_p^{r-2}(\nabla_\lambda u_\alpha)p^{\langle \alpha
    \rangle}p^{\langle \lambda \rangle}p^{\langle \mu_1}p^{\mu_2}...p^{\mu_n\rangle}\\
\end{split}
\end{equation}
Using the definition of the moments, we get
\begin{equation}\label{generaleom}
\begin{split}
    \proj D \rho_r^{\nu_1...\nu_n} &= \intvar C[f]{\cal E}_p^{r-1}p^{\langle
    \mu_1}p^{\mu_2}...p^{\mu_n\rangle}\\
    &\quad - \intvar (\partial_\lambda f_0){\cal E}_p^{r-1}p^{\lambda}p^{\langle
    \mu_1}p^{\mu_2}...p^{\mu_n\rangle}\\
    &\quad -\theta \rho_r^{\mu_1 \cdots \mu_n} \\
    &\quad -\proj \nabla_\lambda \rho_{r-1}^{\lambda \nu_1 \cdots \nu_n} \\
    &\quad -\frac{n}{2 n+1}\left(\nabla^{\left\langle\mu_1\right.}
    \rho_{r+1}^{\left.\mu_2 \cdots \mu_n\right\rangle}-m^2
    \nabla^{\left\langle\mu_1\right.} \rho_{r-1}^{\left.\mu_2 \cdots
    \mu_n\right\rangle}\right) \\
    &\quad -r a_\alpha \rho_{r-1}^{\alpha \mu_1 \cdots \mu_n} \\
    &\quad +r \frac{n}{2 n+1} m^2 \rho_{r-1}^{\left\langle\mu_1 \cdots \mu_{n-1}\right.}
    a^{\left.\mu_n\right\rangle} \\
    &\quad -\frac{n(r+2 n+1)}{2 n+1} \rho_{r+1}^{\left\langle\mu_1 \cdots
    \mu_{n-1}\right.} a^{\left.\mu_n\right\rangle}\\
    &\quad - n\intvar {\cal E}_p^r \delta f p^{\langle \lambda \rangle} p^{\langle
    \mu_1}p^{\mu_2}...\nabla_\lambda u^{\mu_n \rangle}\\
    &\quad - (r-1) \intvar \delta f {\cal E}_p^{r-2}(\nabla_\lambda u_\alpha)p^{\langle \alpha
    \rangle}p^{\langle \lambda \rangle}p^{\langle \mu_1}p^{\mu_2}...p^{\mu_n\rangle}\\
\end{split}
\end{equation}
Now, we can further expand the term $- n\intvar {\cal E}_p^r \delta f p^{\langle \lambda
\rangle} p^{\langle \mu_1}p^{\mu_2}...\nabla_\lambda u^{\mu_n \rangle}$ as the following:
\begin{equation}\label{eq255}
\begin{split}
    &- n \intvar {\cal E}_p^r \delta f p^{\langle \lambda \rangle} p^{\langle
    \mu_1}p^{\mu_2}...\nabla_\lambda u^{\mu_n \rangle}\\
    & = -\intvar {\cal E}_p^r \delta f \bigg(\sum_{i=1}^n (\nabla_\lambda
    u^{\mu_i})p^{\langle\lambda\rangle}p^{\langle\mu_1...}p^{\mu_{i-1}}p^{\mu_{i+1}...}p^{\mu_n\rangle}\bigg)\\
    &\quad + \frac{2}{2n-1}\intvar {\cal E}_p^r \delta f
    \bigg(\sum_{i\neq j}^n \Delta^{\mu_i \mu_j}(\nabla_\lambda
    u_\alpha)p^{\langle\lambda\rangle}p^{\langle\alpha}p^{\mu_1...}p^{\mu_{i-1}}
    p^{\mu_{i+1}...}p^{\mu_{j-1}} p^{\mu_{j+1}...}p^{\mu_n\rangle}\bigg)
\end{split}
\end{equation}
where we used
\begin{equation}\label{eq:pnm1_an}
\begin{split}
    p^{\langle\mu_1...}p^{\mu_{n-1}}a^{\mu_n\rangle} &= \frac1n \sum_{i=1}^n
    a^{\langle\mu_i\rangle}p^{\langle\mu_1...}p^{\mu_{i-1}}p^{\mu_{i+1}...}p^{\mu_n\rangle}\\
    &\quad - \frac{2}{n(2n-1)}\sum_{i\neq j}^n \Delta^{\mu_i
    \mu_j}a_{\langle\lambda\rangle} 
    p^{\langle\lambda}p^{\mu_1...}p^{\mu_{i-1}} p^{\mu_{i+1}...}p^{\mu_{j-1}}
    p^{\mu_{j+1}...}p^{\mu_n\rangle}
\end{split}
\end{equation}
in which $a^{\langle\mu\rangle}$
is an arbitrary transverse vector.
This identity comes frmo Eq.(\ref{eq:q_P_n}).
Using Eq.(\ref{eq:Waldmann_recursion}) 
to combine the angular brackets,
we can further expand Eq.(\ref{eq255}) as
\begin{equation}
\begin{split}
    & - n\intvar {\cal E}_p^r \delta f p^{\langle \lambda \rangle} p^{\langle
    \mu_1}p^{\mu_2}...\nabla_\lambda u^{\mu_n \rangle}\\
    & = -\intvar {\cal E}_p^r \delta f \sum_{i=1}^n (\nabla_\lambda
    u^{\mu_i})p^{\langle\lambda}p^{\mu_1...}p^{\mu_{i-1}}p^{\mu_{i+1}...}p^{\mu_n\rangle}\\
    & \quad - \frac{n-1}{2n-1}\intvar {\cal E}_p^r \delta f \sum_{i=1}^n (\nabla_\lambda
    u^{\mu_i})({\cal E}_p^2 - m^2) p^{\langle\mu_1...} p^{\mu_{i-1}}p^{\mu_{i+1}...}
    \Delta^{\mu_n\rangle\lambda}\\
    & \quad + \frac{2}{2n-1}\intvar {\cal E}_p^r \delta f \sum_{i\neq j}^n
    \Delta^{\mu_i \mu_j}(\nabla_\lambda u_\alpha) p^{\langle\lambda}p^\alpha
    p^{\mu_1...}p^{\mu_{i-1}}p^{\mu_{i+1}...}p^{\mu_{j-1}}p^{\mu_{j+1}...}p^{\mu_n\rangle}\\
    & \quad + \frac{2(n-1)}{(2n-1)^2} \intvar {\cal E}_p^r \delta f \sum_{i\neq
    j}^n \Delta^{\mu_i \mu_j}(\nabla_\lambda u_\alpha)({\cal E}_p^2 -
    m^2)p^{\langle\alpha}p^{\mu_1...}p^{\mu_{i-1}}p^{\mu_{i+1}...}p^{\mu_{j-1}}p^{\mu_{j+1}...}\Delta^{\mu_n\rangle\lambda}
\end{split}
\end{equation}
which can be written in terms of the moments:
\begin{equation}
\begin{split}
    & - n\intvar {\cal E}_p^r \delta f p^{\langle \lambda \rangle} p^{\langle
    \mu_1}p^{\mu_2}...\nabla_\lambda u^{\mu_n \rangle}\\
    & = -\sum_{i=1}^n (\nabla_\lambda
    u^{\mu_i})\rho_r^{\lambda\mu_1...\mu_{i-1}\mu_{i+1}...\mu_n}\\
    & \quad + \frac{2}{2n-1}\sum_{i\neq j}^n \Delta^{\mu_i\mu_j}(\nabla_\lambda
    u_\alpha)\rho_r^{\lambda\alpha\mu_1...\mu_{i-1}\mu_{i+1}...\mu_{j-1}\mu_{j+1}...\mu_n}\\
    & \quad -\frac{n-1}{2n-1}\sum_{i=1}^n
    \rho_{r+2}^{\langle\mu_1...\mu_{i-1}\mu_{i+1}...\mu_{n-1}}\nabla^{\mu_n\rangle}u^{\mu_i}\\
    & \quad + \frac{2(n-1)}{(2n-1)^2}\sum_{i\neq j}^n
    \Delta^{\mu_i\mu_j}\rho_{r+2}^{\langle\alpha\mu_1...\mu_{i-1}\mu_{i+1}...\mu_{j-1}\mu_{j+1}...\mu_{n-1}}\nabla^{\mu_n\rangle}u_\alpha\\
    & \quad + \frac{m^2 (n-1)}{2n-1}\sum_{i=1}^n
    \rho_r^{\langle\mu_1...\mu_{i-1}\mu_{i+1}...\mu_{n-1}}\nabla^{\mu_n\rangle}u^{\mu_i}\\
    & \quad - \frac{2m^2 (n-1)}{(2n-1)^2}\sum_{i\neq j}^n
    \Delta^{\mu_i\mu_j}\rho_r^{\langle\alpha\mu_1...\mu_{i-1}\mu_{i+1}...\mu_{j-1}\mu_{j+1}...\mu_{n-1}}\nabla^{\mu_n\rangle}u_\alpha\\
    & = -\sum_{i=1}^n \nabla_\lambda u^{\langle
    \mu_i}\rho_r^{\mu_1...\mu_{i-1}\mu_{i+1}...\mu_n\rangle\lambda}\\
    &\quad - \frac{n-1}{2n-1}\sum_{i=1}^n
    \rho_{r+2}^{\langle\mu_1...\mu_{i-1}\mu_{i+1}...\mu_{n-1}}\sigma^{\mu_n\mu_i\rangle}\\
    &\quad + \frac{m^2 (n-1)}{2n-1}\sum_{i=1}^n
    \rho_r^{\langle\mu_1...\mu_{i-1}\mu_{i+1}...\mu_{n-1}}\sigma^{\mu_n\mu_i\rangle}\\
\end{split}
\end{equation}
where
\begin{equation}
    \sigma^{\mu\nu} = \nabla^{\langle\mu}u^{\nu\rangle}
\end{equation}
is the symmetric Navier-Stokes shear tensor. Since the angular bracket represents the
traceless and symmetric combination of the Lorentz indices, all permutations of the
Lorentz indices inside the bracket give the same term. Thus
\begin{equation}\label{eq260}
\begin{split}
    & - n\intvar {\cal E}_p^r \delta f p^{\langle \lambda \rangle} p^{\langle
    \mu_1}p^{\mu_2}...\nabla_\lambda u^{\mu_n \rangle}\\
    & = -n\rho_r^{\lambda\langle\mu_1...\mu_{n-1}}\nabla_\lambda u^{\mu_n\rangle} -
    \frac{n(n-1)}{2n-1}\rho_{r+2}^{\langle\mu_1...\mu_{n-2}}\sigma^{\mu_{n-1}\mu_n\rangle}\\
    &\quad + \frac{m^2
    (n-1)n}{2n-1}\rho_r^{\langle\mu_1...\mu_{n-2}}\sigma^{\mu_{n-1}\mu_n\rangle}
\end{split}
\end{equation}
Here, we can replace $\nabla_\lambda u^{\mu_n}$ using
\begin{equation}\label{vorticity_id}
    \nabla^\lambda u^\mu = \sigma^{\mu\nu} + \omega^{\mu\nu} +
    \frac{\theta}{3}\Delta^{\mu\nu}
\end{equation}
where
\begin{equation}
    \omega^{\mu\nu} = \frac12 \bigg(\nabla^\mu u^\nu - \nabla^\nu u^\mu\bigg)
\end{equation}
is the anti-symmetric vorticity tensor. Doing so gives us
\begin{equation}\label{eq262}
\begin{split}
    & - n\intvar {\cal E}_p^r \delta f p^{\langle \lambda \rangle} p^{\langle
    \mu_1}p^{\mu_2}...\nabla_\lambda u^{\mu_n \rangle}\\
    & = -n\rho_r^{\lambda\langle\mu_1...\mu_{n-1}}\sigma_\lambda^{\mu_n\rangle}
    -n\rho_r^{\lambda\langle\mu_1...\mu_{n-1}}\omega_\lambda^{\mu_n\rangle}
    -\frac{n}{3}\theta\rho_r^{\mu_1...\mu_n}\\
    &\quad -
    \frac{n(n-1)}{2n-1}\rho_{r+2}^{\langle\mu_1...\mu_{n-2}}\sigma^{\mu_{n-1}\mu_n\rangle}
    + \frac{m^2
    (n-1)n}{2n-1}\rho_r^{\langle\mu_1...\mu_{n-2}}\sigma^{\mu_{n-1}\mu_n\rangle}
\end{split}
\end{equation}
Now let's go back to the general moment equation Eq.(\ref{generaleom}) and take a look
at the term $- (r-1) \intvar \delta f {\cal E}_p^{r-2}(\nabla_\lambda u_\alpha)p^{\langle \alpha
\rangle}p^{\langle \lambda \rangle}p^{\langle \mu_1}p^{\mu_2}...p^{\mu_n\rangle}$. Using
Eq.(\ref{vorticity_id}), this term can be written as
\begin{equation}
\begin{split}
    &- (r-1) \intvar \delta f {\cal E}_p^{r-2}(\nabla_\lambda u_\alpha)p^{\langle \alpha
    \rangle}p^{\langle \lambda \rangle}p^{\langle \mu_1}p^{\mu_2}...p^{\mu_n\rangle} \\
    &= - (r-1)\sigma_{\lambda\alpha} \intvar \delta f {\cal E}_p^{r-2}p^{\langle \alpha
    \rangle}p^{\langle \lambda \rangle}p^{\langle \mu_1}p^{\mu_2}...p^{\mu_n\rangle}\\
    &\quad - \frac{(r-1)}{3}\theta \intvar \delta f {\cal E}_p^{r-2}({\cal E}_p^2 - m^2)p^{\langle
    \mu_1}p^{\mu_2}...p^{\mu_n\rangle}\\
    &= - (r-1)\sigma_{\lambda\alpha} \intvar \delta f {\cal E}_p^{r-2}p^{\langle \alpha
    \rangle}p^{\langle \lambda \rangle}p^{\langle \mu_1}p^{\mu_2}...p^{\mu_n\rangle}\\
    &\quad - \frac{(r-1)}{3}\theta \rho_r^{\mu_1...\mu_n} +
    \frac{(r-1)m^2}{3}\theta\rho_{r-2}^{\mu_1...\mu_n}
\end{split}
\end{equation}
Note that the term with $\omega_{\lambda\alpha}$ vanishes due to its anti-symmetric
property. We then proceed to expand the first term on the right-hand side using 
Eq.(\ref{eq:alpha_lambda_n})
\begin{equation}
\begin{split}
    &- (r-1)\sigma_{\lambda\alpha} \intvar \delta f {\cal E}_p^{r-2}p^{\langle \alpha
    \rangle}p^{\langle \lambda \rangle}p^{\langle \mu_1}p^{\mu_2}...p^{\mu_n\rangle}\\
    &= -(r-1)\sigma_{\lambda\alpha}\rho_{r-2}^{\alpha\lambda\mu_1...\mu_n} \\
    &\quad -\frac{2(r-1)}{2n+3}\sum_{i=1}^n \sigma_\alpha^{\mu_i}
    \rho_r^{\alpha\mu_1...\mu_{i-1}\mu_{i+1}...\mu_n}\\
    & \quad + \frac{4(r-1)}{(2n+3)(2n-1)}\sum_{i\neq
    j}^n \Delta^{\mu_i\mu_j} \sigma_{\lambda\alpha}
    \rho_r^{\alpha\lambda\mu_1...\mu_{i-1}\mu_{i+1}...\mu_{j-1}\mu_{j+1}...\mu_n}\\
    &\quad + \frac{2m^2 (r-1)}{2n+3}\sum_{i=1}^n \sigma_\alpha^{\mu_i}
    \rho_{r-2}^{\alpha\mu_1...\mu_{i-1}\mu_{i+1}...\mu_n}\\
    & \quad - \frac{4m^2(r-1)}{(2n+3)(2n-1)}\sum_{i\neq
    j}^n \Delta^{\mu_i\mu_j} \sigma_{\lambda\alpha}
    \rho_{r-2}^{\alpha\lambda\mu_1...\mu_{i-1}\mu_{i+1}...\mu_{j-1}\mu_{j+1}...\mu_n}\\
    &\quad -
    \frac{(r-1)(n-1)n}{(2n+1)(2n-1)}\rho_{r+2}^{\langle\mu_1...\mu_{n-2}}\sigma^{\mu_{n-1}\mu_n\rangle}\\
    &\quad +
    \frac{2m^2(r-1)(n-1)n}{(2n+1)(2n-1)}\rho_r^{\langle\mu_1...\mu_{n-2}}\sigma^{\mu_{n-1}\mu_n\rangle}\\
    &\quad -
    \frac{m^4(r-1)(n-1)n}{(2n+1)(2n-1)}\rho_{r-2}^{\langle\mu_1...\mu_{n-2}}\sigma^{\mu_{n-1}\mu_n\rangle}
\end{split}
\end{equation}
Note that each pair of summations give the traceless and symmetric combination of
$\sigma_\alpha^{\mu_i}$ and $\rho_r^{\alpha\mu_1...\mu_{i-1}\mu_{i+1}...\mu_n}$. Thus
this reduces to
\begin{equation}
\begin{split}
    &- (r-1)\sigma_{\lambda\alpha} \intvar \delta f {\cal E}_p^{r-2}p^{\langle \alpha
    \rangle}p^{\langle \lambda \rangle}p^{\langle \mu_1}p^{\mu_2}...p^{\mu_n\rangle}\\
    &= -(r-1)\sigma_{\lambda\alpha}\rho_{r-2}^{\alpha\lambda\mu_1...\mu_n} \\
    &\quad -\frac{2(r-1)}{2n+3}\sum_{i=1}^n \sigma_\alpha^{\langle\mu_i}
    \rho_r^{\mu_1...\mu_{i-1}\mu_{i+1}...\mu_n\rangle\alpha}\\
    &\quad + \frac{2m^2 (r-1)}{2n+3}\sum_{i=1}^n \sigma_\alpha^{\langle\mu_i}
    \rho_{r-2}^{\mu_1...\mu_{i-1}\mu_{i+1}...\mu_n\rangle\alpha}\\
    &\quad -
    \frac{(r-1)(n-1)n}{(2n+1)(2n-1)}\rho_{r+2}^{\langle\mu_1...\mu_{n-2}}\sigma^{\mu_{n-1}\mu_n\rangle}\\
    &\quad +
    \frac{2m^2(r-1)(n-1)n}{(2n+1)(2n-1)}\rho_r^{\langle\mu_1...\mu_{n-2}}\sigma^{\mu_{n-1}\mu_n\rangle}\\
    &\quad -
    \frac{m^4(r-1)(n-1)n}{(2n+1)(2n-1)}\rho_{r-2}^{\langle\mu_1...\mu_{n-2}}\sigma^{\mu_{n-1}\mu_n\rangle}
\end{split}
\end{equation}
Since all permutations of the Lorentz indices inside the angular brackets give the
same term, this can be simplified to
\begin{equation}\label{eq264}
\begin{split}
    &- (r-1)\sigma_{\lambda\alpha} \intvar \delta f {\cal E}_p^{r-2}p^{\langle \alpha
    \rangle}p^{\langle \lambda \rangle}p^{\langle \mu_1}p^{\mu_2}...p^{\mu_n\rangle}\\
    &= -(r-1)\sigma_{\lambda\alpha}\rho_{r-2}^{\alpha\lambda\mu_1...\mu_n} \\
    &\quad
    -\frac{2(r-1)n}{2n+3}\rho_r^{\alpha\langle\mu_1...\mu_{n-1}}\sigma_\alpha^{\mu_n\rangle}\\
    &\quad + \frac{2m^2 (r-1)n}{2n+3}
    \rho_{r-2}^{\alpha\langle\mu_1...\mu_{n-1}}\sigma_\alpha^{\mu_n\rangle}\\
    &\quad -
    \frac{(r-1)(n-1)n}{(2n+1)(2n-1)}\rho_{r+2}^{\langle\mu_1...\mu_{n-2}}\sigma^{\mu_{n-1}\mu_n\rangle}\\
    &\quad +
    \frac{2m^2(r-1)(n-1)n}{(2n+1)(2n-1)}\rho_r^{\langle\mu_1...\mu_{n-2}}\sigma^{\mu_{n-1}\mu_n\rangle}\\
    &\quad -
    \frac{m^4(r-1)(n-1)n}{(2n+1)(2n-1)}\rho_{r-2}^{\langle\mu_1...\mu_{n-2}}\sigma^{\mu_{n-1}\mu_n\rangle}
\end{split}
\end{equation}
Plugging all the above results back into Eq.(\ref{generaleom}), and expressing
everything in terms of the moments, we arrive at the final form of the general moment
equation:
\begin{equation}
\begin{split}
    \proj D \rho_r^{\nu_1...\nu_n} &= \intvar C[f]{\cal E}_p^{r-1}p^{\langle
    \mu_1}p^{\mu_2}...p^{\mu_n\rangle}\\
    &\quad - \intvar (\partial_\lambda f_0){\cal E}_p^{r-1}p^{\lambda}p^{\langle
    \mu_1}p^{\mu_2}...p^{\mu_n\rangle}\\
    &\quad -\frac{n(2n+r+1)}{2n+1}\rho_{r+1}^{\langle\mu_1...\mu_{n-1}}a^{\mu_n\rangle}\\
    &\quad + rm^2\frac{n}{2n+1}\rho_{r-1}^{\langle\mu_1...\mu_{n-1}}a^{\mu_n\rangle}\\
    &\quad -ra_\lambda \rho_{r-1}^{\lambda\mu_1...\mu_n}\\
    &\quad - \Delta_{\nu_1...\nu_n}^{\mu_1...\mu_n} \nabla_\lambda
    \rho_{r-1}^{\lambda\nu_1...\nu_n}\\
    &\quad - \frac{n}{2n+1}\nabla^{\langle\mu_1}\rho_{r+1}^{\mu_2...\mu_n\rangle}\\
    &\quad + m^2\frac{n}{2n+1}\nabla^{\langle\mu_1}\rho_{r-1}^{\mu_2...\mu_n\rangle}\\
    &\quad - \frac{n+r+2}{3}\theta\rho_r^{\mu_1...\mu_n}\\
    &\quad - (r-1)\sigma_{\lambda\alpha}\rho_{r-2}^{\alpha\lambda\mu_1...\mu_n}\\
    &\quad + \frac{(r-1)m^2}{3}\theta\rho_{r-2}^{\mu_1...\mu_n}\\
    &\quad
    -\frac{n(2n+2r+1)}{2n+3}\rho_r^{\lambda\langle\mu_1...\mu_{n-1}}\sigma_\lambda^{\mu_n\rangle}\\
    &\quad -n\rho_r^{\lambda\langle\mu_1...\mu_{n-1}}\omega_\lambda^{\mu_n\rangle}\\
    &\quad -
    \frac{(2n+r)(n-1)n}{(2n-1)(2n+1)}\rho_{r+2}^{\langle\mu_1...\mu_{n-2}}\sigma^{\mu_{n-1}\mu_n\rangle}\\
    &\quad +
    2m^2\frac{(r-1)n}{2n+3}\rho_{r-2}^{\lambda\langle\mu_1...\mu_{n-1}}\sigma_\lambda^{\mu_n\rangle}\\
    &\quad -
    m^4\frac{(r-1)(n-1)n}{(2n+1)(2n-1)}\rho_{r-2}^{\langle\mu_1...\mu_{n-2}}\sigma^{\mu_{n-1}\mu_n\rangle}\\
    &\quad +
    m^2\frac{(2n+2r-1)(n-1)n}{(2n+1)(2n-1)}\rho_r^{\langle\mu_1...\mu_{n-2}}\sigma^{\mu_{n-1}\mu_n\rangle}
\end{split}
\end{equation}

\section{$F$ integrals and $\phi, \varphi, \psi$ coefficients}
\label{app:Fintegrals}
To evaluate the $F$ integrals, we first need to know the conservation laws. 
The stress-energy tensor is
\begin{equation}
    T^{\mu\nu} = \int{d^3p\over (2\pi)^3 E_p} f_0 p^\mu p^\nu + \pi^{\mu\nu} + \Pi \Delta^{\mu\nu}
\end{equation}
The energy-momentum conservation law is
\begin{equation}
\begin{split}
    0 & = \partial_\mu T^{\mu\nu}\\
    & = \int{d^3p\over (2\pi)^3 E_p} (\partial_\mu f_0) p^\mu p^\nu + \partial_\mu \pi^{\mu\nu} +
    (\nabla^\nu\Pi) + \Pi (u^\nu\theta + a^\nu)
\end{split}
\end{equation}
where we used
\begin{eqnarray}
\partial_\mu \Delta^{\mu\nu}
& = &
\partial_\mu (u^\mu u^\nu)
\nonumber\\
& = &
u^\nu\theta + a^\nu
\end{eqnarray}
and we specify $f_0 = e^{-\beta {\cal E}_p}$.

In the time direction $u_\nu \partial_\mu T^{\mu\nu} = 0$ yields
\begin{equation}
    0 = 
    -\pi^{\mu\nu}\sigma_{\mu\nu} 
    - \theta \Pi + I_{3,0}  D\beta 
    - {\beta \over 3}\theta I_{3,1}
\end{equation}
where we defined
\begin{equation}
    I_{n,k} = \intvar f_0 {\cal E}_p^{n-2k}({\cal E}_p^2-m^2)^k
\end{equation}
which can be evaluated in the local rest frame where ${\cal E}_p \to E_p$ and 
$({\cal E}_p^2-m^2) \to p^2$. 
This integral is always finite when $m\ne 0$ and $k \ge 0$.
In the $m\to 0$ limit, the integral behaves as $\log(m\beta)$ for $n = -2$,
and $I_{n,k} \sim T^{n+2}$.

Using integration by part, it can be shown that
\begin{eqnarray}
\beta I_{n,k} 
\ampeq
\beta\int {d^3p\over (2\pi)^3 E_p} E_p^{n-2k}p^{2k} e^{-\beta E_p}
\non
\ampeq
-\int {d^3p\over (2\pi)^3 E_p} E_p^{n-2k+1}p^{2k-1} \partial_p e^{-\beta E_p}
\non
\ampeq
(2k+1)I_{n-1,k-1} + (n-2k)I_{n-1,k} 
\label{eq:I_id}
\end{eqnarray}
as long as all intergrals are finite.
In particular
\begin{equation}
\begin{split}
    \beta I_{3,1} &= 3 I_{2,0} + I_{2,1}\\
    & = 3(\varepsilon + P)
\end{split}
\end{equation}

In the spatial direction
$\Delta_{\nu}^\rho \partial_\mu T^{\mu\nu} = 0$ yields
\begin{equation}
    0 = \Delta_{\nu}^\rho
\partial_\mu \pi^{\mu\nu}
+
(\nabla^\rho  \Pi)
+
a^\rho \Pi
- I_{3,1} {\nabla^\rho\beta\over 3} 
+ {\beta a^\rho\over 3} I_{3,1}
\end{equation}

Solving for the time derivatives $D\beta$ and $a^\rho = Du^\rho$, we obtain
\begin{equation}\label{eq:Dbeta}
D\beta
=
\chi_{\beta|0}\theta
+
\chi_{\beta|1}^{\pi\Pi} \left(\pi^{\gamma\rho}\sigma_{\gamma\rho} + \Pi\theta\right)
\end{equation}
where
\begin{equation}
    \chi_{\beta|0} =
    {\beta\over 3}{I_{3,1}\over I_{3,0}}
    \ \ \hbox{and}\ \ 
    \chi_{\beta|1}^{\pi\Pi}
    = {1\over I_{3,0}}
\end{equation}
From Eq.(\ref{eq:momentum_conserv}), we get
\begin{eqnarray}
\label{eq:asigma}
 a^\rho 
&=&
{1\over \Pi + (\varepsilon + P)}
\left(
-\nabla^\rho P
- (\nabla^\rho  \Pi)
-\Delta_{\nu}^\rho \partial_\mu \pi^{\mu\nu}
\right)
\nonumber\\
&\approx &
{1\over \varepsilon + P}
\left(
-\nabla^\rho P
- (\nabla^\rho  \Pi)
-\Delta_{\nu}^\rho \partial_\mu \pi^{\mu\nu}
\right)
-
{\Pi\over (\varepsilon + P)^2}
\left(
-\nabla^\rho P
\right)
\end{eqnarray}
where we used
\begin{equation}
{\nabla\beta\over 3} I_{3,1} 
=
-\nabla P
\end{equation}
The 0-th order acceleration is
\begin{equation}
    a_{|0}^{\rho}
=
-{\nabla^{\rho}P\over \varepsilon + P}
=
\nabla^\rho\beta {I_{3,1}\over 3(\varepsilon + P)}
= {\nabla^\rho\beta\over \beta}
\end{equation}
and the 1st order one satisfies
\begin{equation}
Q_\sigma^\rho    a_{|1}^{\sigma}
=
{1\over \varepsilon + P}
\left(
- \nabla^\rho \Pi 
- \Delta_{\nu}^{\rho}\nabla_\mu \pi^{\mu\nu}
\right)
-
{1\over (\varepsilon + P)^2} \Pi \left( -\nabla^\rho P \right)
\end{equation}
where
\begin{eqnarray}
Q_{\sigma}^{\rho} = g_\sigma^\rho + {1\over \varepsilon + P}\pi_\sigma^\rho
\end{eqnarray}

Now, observe that the only non-zero $F$ integrals are the spin 0, 1, and 2 integrals. 
The scalar integral is
\begin{equation}
\begin{split}
    F_{r}
& =
\intvar {\cal E}_p^r p^\lambda (\partial_\lambda f_0)
\\
& =
\intvar {\cal E}_p^r f_0
\left(
-{\cal E}_p^2 D \beta
+ \beta {\theta\over 3} ({\cal E}_p^2-m^2)
\right)
\\ & 
=
-I_{r+2,0} D\beta
+ {\beta\over 3}\theta I_{r+2, 1}
\\ &
=
\phi_{r|0}\theta
+\phi_{r|1}^{\pi\Pi}\left(\pi^{\rho\gamma}\sigma_{\rho\gamma} + \theta\Pi\right)
\end{split}
\end{equation}
where
\begin{equation}
    \phi_{r|0}
=
{\beta \over 3}\left( I_{r+2,1} - {I_{r+2,0}I_{3,1}\over I_{3,0}}\right)
\end{equation}
and
\begin{equation}
    \phi_{r|1}^{\pi\Pi}
= -{ I_{r+2,0} \over I_{3,0}}
\end{equation}
using Eq.(\ref{eq:Dbeta}).
Note that $\phi_{1|0} = 0$ and $\phi_{1|1}^{\pi\Pi} = -1$.
In the massless limit, we have
\begin{equation}
    I_{r,k} = \int \frac{d^3 p}{(2\pi)^3} p^{r-1} e^{-p/T} = {T^{r+2}\over 2\pi^2} (r+1)!
    \label{eq:I_rk}
\end{equation}
For the 14 moments, we need $F_{-1}$ whose coefficients are
\begin{equation}
    \phi_{-1|0} = -4 {T^2\over 2\pi^2}
\end{equation}
and
\begin{equation}
    \phi_{-1|1}^{\pi\Pi} = -{1\over 6}\beta^2
\end{equation}
The vector integral is
\begin{equation}
\begin{split}
    F_r^\sigma
& =
\intvar {\cal E}_p^r p^\lambda (\partial_\lambda f_0) p^{\langle\sigma\rangle}
\\
& =
\intvar {\cal E}_p^r f_0
\left(
-{\cal E}_p p^{\langle\lambda\rangle} \nabla_\lambda \beta
+ {\cal E}_p \beta p^{\langle\lambda\rangle} a_\lambda
\right) p^{\langle\sigma\rangle}
\\
& =
\psi_{r|1}
\left(
\Delta_{\nu}^\rho \partial_\mu \pi^{\mu\nu}
+
\nabla^\sigma  \Pi
+
a^\sigma \Pi
\right)
\end{split}
\end{equation}
where we used a slight different form of Eq.(\ref{eq:asigma})
\begin{equation}
    \beta a^\rho - \nabla^\rho\beta
=
-{3\over I_{3,1}}
\left(
\Delta_{\nu}^\rho \partial_\mu \pi^{\mu\nu}
+
(\nabla^\rho  \Pi)
+
a^\rho \Pi
\right)
\end{equation}
The coefficient is
\begin{equation}
    \psi_{r|1} = -{I_{r+3,1}\over I_{3,1}}
\end{equation}
Note that $\psi_{0|1} = -1$.
Here, $I_{3,1} = 3(\varepsilon+P)T$ and $I_{2,1} = 3P$
can be used if needed. With $r= -1$ and $m = 0$,
\begin{equation}
    \psi_{-1|1} = -{1\over 4} \beta
\end{equation}
The spin-2 integral is relatively simple 
\begin{equation}
\begin{split}
    F_r^{\sigma\gamma}
& =
\intvar {\cal E}_p^r p^\lambda (\partial_\lambda f_0)
p^{\langle\sigma}p^{\gamma\rangle}
\\
& =
\intvar {\cal E}_p^r f_0
\left(
\beta p^{\langle\lambda} p^{\alpha\rangle}\nabla_\lambda u_\alpha
\right)
p^{\langle\sigma}p^{\gamma\rangle}
\\
& =
\varphi_{r|0} \sigma^{\sigma\gamma}
\end{split}
\end{equation}
where
\begin{equation}
    \varphi_{r|0} = {2\over 15} \beta I_{r+4,2}
    = {2\over 15}\left(
    5I_{r+3,1} + r I_{r+3,2}
    \right)
\label{eq:varphi_r0}
\end{equation}
is obtained with the help of the normalization condition 
Eq.(\ref{eq:p_poly_ortho}) 
(See also Refs.~\cite{Denicol:2012es,degroot}),
and Eq.(\ref{eq:I_id}).
With $r = -1$ and $m = 0$,
\begin{equation}\label{phi_10}
    \varphi_{-1|0} = {16\over 5} {T^4\over 2\pi^2}
    = {8\over 15}\varepsilon
\end{equation}

Let's check whether Landau conditions $\rho_2 = 0$ and $\rho_1^\mu = 0$ are 
consistent with the $F$-integrals.
Setting $r = 2$ in Eq.(\ref{eq:rank0_eq}), we get
\begin{eqnarray}
D\rho_2
&=&
-{\rho_2\over \tau_R}
-F_{1} 
+m^2 \frac{\theta}{3} \rho_{0} 
-{4\over 3}\theta\rho_2 
\nonumber\\ && {}
-\nabla_\lambda\rho^\lambda_1 - ra_\lambda\rho^\lambda_1
-\sigma_{\lambda \alpha} \rho_{0}^{\alpha \lambda}
\end{eqnarray}
Setting $r = 1$ in Eq.(\ref{eq:rank1_eq}), we get
\begin{eqnarray}
\Delta_{\nu_1}^{\mu_1}D\rho_1^{\nu_1}
& = &
-\frac{\rho_1^{\mu_1}}{\tau_R}-F_{0}^{\mu_1}
- {4\over 3}\theta \rho_1^{\mu_1} 
\nonumber\\ && {}
- a_\alpha \rho_{0}^{\alpha \mu_1} 
-\Delta_{\nu_1}^{\mu_1} \nabla_\lambda \rho_{0}^{\lambda \nu_1}
-\omega_\lambda^{\mu_1} \rho_1^\lambda
\nonumber\\ && {}
+ {1\over 3}
\left( m^2 \rho_{0} -4 \rho_{2} \right) a^{\mu_1}
\nonumber\\ && {}
-\frac{1}{3}\left(\nabla^{\mu_1} \rho_{2}-m^2 \nabla^{\mu_1} \rho_{0}\right) 
\nonumber\\ && {}
- \rho_1^\lambda \sigma_\lambda^{\mu_1} 
\end{eqnarray}
Using $\pi^{\mu\nu} = \rho_0^{\mu\nu}$, $\Pi = -m^2\rho_0/3$ as well as 
\begin{equation}
F_1 = -(\pi^{\rho\sigma}\sigma_{\rho\sigma} + \theta\Pi)
\end{equation}
and
\begin{equation}
F_0^{\mu}
=
-
(\Delta^{\mu}_{\nu}\nabla_\lambda \pi^{\lambda\nu} + a_\lambda \pi^{\lambda\mu}
+ \nabla^\rho\Pi + a^\rho\Pi 
)
\end{equation}
these evolution equations become
\begin{eqnarray}
D\rho_2
&=&
-{\rho_2\over \tau_R}
-{4\over 3}\theta\rho_2 
-\nabla_\lambda\rho^\lambda_1 - ra_\lambda\rho^\lambda_1
\end{eqnarray}
and
\begin{eqnarray}
\Delta_{\nu_1}^{\mu_1}D\rho_1^{\nu_1}
& = &
-\frac{\rho_1^{\mu_1}}{\tau_R}
- {4\over 3}\theta \rho_1^{\mu_1} 
-\omega_\lambda^{\mu_1} \rho_1^\lambda
\nonumber\\ && {}
- {4\over 3} \rho_{2} a^{\mu_1}
-\frac{1}{3}\nabla^{\mu_1} \rho_{2}
- \rho_1^\lambda \sigma_\lambda^{\mu_1} 
\end{eqnarray}
Hence as long as the initial values for $\rho_2$ and $\rho_1^\mu$
all vanish, $\rho_2$ and $\rho_1^\mu$ remain zero.

\bibliography{reg_hydro}

\end{document}